\documentclass[prd,amsmath,groupedaddress,onecolumn,eqsecnum,nofootinbib,10pt]{revtex4-2}
\usepackage{natbib}
\usepackage{hyperref}
\usepackage{graphicx}
\usepackage{lipsum}
\usepackage{enumerate}
\usepackage{lipsum, babel}
\usepackage{graphicx}
\usepackage{dcolumn}
\usepackage{amsmath}
\usepackage{amssymb}
\usepackage{amsfonts}
\usepackage{orcidlink}
\usepackage{bm}
\usepackage{float}
\hypersetup{colorlinks=true, linkcolor=blue, citecolor=blue, urlcolor=blue}
\usepackage{caption}
\usepackage{subfigure}
\usepackage[hmargin={0.45in,0.45in},height=9.0in, top=0.7in
]{geometry}

\begin{document}
\begin{center}
{\Large \textbf{A Black Hole Solution in Kalb-Ramond Gravity with Quintessence Field: From Geodesic Dynamics to Thermal Criticality}}
\end{center}

\vspace{0.4cm}

\begin{center}

{\bf Ahmad Al-Badawi}\orcidlink{0000-0002-3127-3453}\\
Department of Physics, Al-Hussein Bin Talal University, 71111,
Ma'an, Jordan. \\
e-mail: ahmadbadawi@ahu.edu.jo
 \\

\vspace{0.2cm}

{\bf Faizuddin Ahmed}\orcidlink{0000-0003-2196-9622}\\Department of Physics, The Assam Royal Global University, Guwahati, 781035, Assam, India\\
e-mail: faizuddinahmed15@gmail.com\\

\vspace{0.2cm}
{\bf \.{I}zzet Sakall{\i}}\orcidlink{0000-0001-7827-9476}\\
Physics Department, Eastern Mediterranean University, Famagusta 99628, North Cyprus via Mersin 10, Turkey\\
e-mail: izzet.sakalli@emu.edu.tr 

\end{center}

\date{\today}

\begin{abstract}
We present a theoretical investigation of black hole solutions in Kalb-Ramond gravity embedded with quintessence fields. Our study examines how combined effects of Lorentz violation through the Kalb-Ramond field parameter $\eta$ and exotic matter contributions via quintessence parameters $(\mathrm{C}, w)$ systematically modify spacetime geometry, particle dynamics, and observational signatures compared to standard Schwarzschild black holes. The analysis encompasses geodesic motion for both photons and massive particles, revealing substantial modifications to effective potentials, photon sphere characteristics, and innermost stable circular orbit properties. We derive analytical expressions for black hole shadow radii across different quintessence states, demonstrating systematic parameter dependencies enabling observational discrimination between theoretical frameworks. Our perturbation analysis of scalar and electromagnetic fields shows how Lorentz violation and quintessence effects alter wave propagation and stability properties. Using Gauss-Bonnet theorem methodology, we calculate gravitational lensing deflection angles incorporating both modified gravity and exotic matter contributions. The thermodynamic investigation reveals complex phase structures with modified Hawking temperature evolution, Gibbs free energy characteristics, and specific heat capacity behavior significantly deviating from general relativity predictions. Lorentz violation amplifies gravitational effects, whereas quintessence exerts counteractive forces, generating complex parameter spaces allowing precise manipulation of observable quantities. 
\end{abstract}



\maketitle

\section{Introduction}\label{sec1}

The theoretical landscape of modern gravitational physics has undergone profound transformations in recent decades, driven by both observational discoveries and fundamental theoretical developments that challenge our understanding of spacetime geometry and gravitational dynamics. Among the most significant developments has been the emergence of modified gravity theories that extend Einstein's General Relativity (GR) to incorporate new physical phenomena and address longstanding theoretical puzzles \cite{isz01,isz02,isz03,isz03x1,isz03x2,isz03x3}. Simultaneously, the discovery of cosmic acceleration and the subsequent introduction of dark energy concepts have revolutionized our understanding of the universe's large-scale dynamics, necessitating new theoretical frameworks that can accommodate exotic matter with negative pressure characteristics \cite{isz04,isz05,isz06,isz06x1,isz06x2}.

Black holes (BHs) represent among the most extreme manifestations of gravitational physics, serving as natural laboratories for testing the fundamental principles of spacetime geometry under conditions where gravitational fields reach their maximum intensity. The recent breakthroughs in observational astronomy, particularly the direct imaging of BH shadows by the Event Horizon Telescope (EHT) collaboration and the detection of gravitational waves from merging BH systems by LIGO/Virgo networks, have provided unprecedented opportunities to probe strong-field gravity and test alternative theoretical frameworks \cite{isz07,isz08,isz09}. These observations have demonstrated that BH physics serves as a crucial testing ground for fundamental physics, where deviations from GR predictions could manifest in measurable observational signatures.

Kalb-Ramond gravity (KRG) emerges as a particularly compelling modification of Einstein's theory, incorporating antisymmetric tensor fields that naturally arise in string theory and other fundamental frameworks \cite{isz10,isz11,isz11x1}. The theoretical foundation of KRG rests on the inclusion of KRG fields, which are antisymmetric tensor fields that couple non-minimally to spacetime curvature, leading to the spontaneous breaking of Lorentz symmetry through the development of non-zero vacuum expectation values (VEVs) \cite{isz12,isz13}. This Lorentz violation (LV) introduces systematic modifications to gravitational dynamics that can significantly alter BH properties, geodesic structure, and observational characteristics compared to standard GR predictions.

The incorporation of quintessence fields (QFs) represents another crucial theoretical development addressing the dark energy problem in modern cosmology. QFs are scalar fields with negative pressure characteristics that can drive cosmic acceleration while maintaining dynamic evolution, in contrast to the cosmological constant approach \cite{isz14,isz15,isz16}. When QFs surround BH systems, they introduce additional modifications to spacetime geometry through their stress-energy contributions, creating complex interplays between exotic matter effects and gravitational dynamics. The quintessence state parameter $w$, which characterizes the equation of state relating pressure to energy density, provides a fundamental parameter for controlling the strength and nature of these exotic matter effects.

The combination of KRG and QF environments creates a rich theoretical framework where both modified gravitational dynamics and exotic matter contributions simultaneously influence BH properties. This KRG-QF framework represents a natural extension of previous investigations that have separately examined either modified gravity effects or exotic matter influences on BH physics. The systematic study of KRG-QF BH systems enables comprehensive understanding of how multiple beyond-GR effects interact and combine to produce observable signatures that could distinguish these scenarios from standard gravitational physics \cite{isz17,isz18}.

Recent theoretical developments have established the mathematical foundations for constructing BH solutions within KRG frameworks, revealing how LV parameters systematically modify horizon structures, geodesic properties, and thermodynamic characteristics \cite{isz19,isz20}. Similarly, extensive investigations of BH systems surrounded by various exotic matter configurations, including quintessence environments, have demonstrated significant modifications to observable properties including shadows, gravitational lensing, and electromagnetic signatures \cite{isz21,isz22,isz23,isz23x1, isz23x2}. However, the combined investigation of KRG-QF systems represents a largely unexplored theoretical territory with significant potential for revealing new physical phenomena and observational consequences.

The motivation for our investigation stems from several converging theoretical and observational considerations. First, the fundamental question of whether Lorentz symmetry remains exact at all energy scales continues to drive theoretical exploration, with potential violations expected in Quantum Gravity frameworks including Loop Quantum Gravity and theories incorporating the Generalized Uncertainty Principle (GUP) \cite{isz24,isz25}. Second, the persistent dark energy problem in cosmology necessitates continued investigation of alternative mechanisms, including dynamic QF scenarios that could provide observational discrimination from cosmological constant models. Third, the remarkable precision of current and planned observational facilities creates unprecedented opportunities for detecting subtle deviations from GR predictions in strong gravitational field regimes. In this regard, our investigation addresses several fundamental questions regarding the interplay between modified gravity and exotic matter effects in determining BH properties and observational signatures. We systematically examine how LV effects through KRG dynamics and exotic matter contributions through QF environments combine to modify spacetime geometry, particle dynamics, wave propagation, and thermodynamic properties compared to standard Schwarzschild BH systems. The comprehensive scope of our analysis encompasses multiple physical phenomena including geodesic motion, photon sphere characteristics, innermost stable circular orbits (ISCO), BH shadows, scalar and electromagnetic perturbations, gravitational lensing through Gauss-Bonnet Theorem (GBT) methodology, and complete thermodynamic analysis including phase transitions and stability properties \cite{isz25x1}.

The geodesic investigation provides fundamental insights into how test particle motion is modified by the combined KRG-QF effects, with particular emphasis on photon trajectories that determine observable shadow characteristics and massive particle orbits that influence accretion disk dynamics and gravitational wave emission from inspiraling compact objects. Our analysis systematically characterizes how the LV parameter $\eta$ and QF characteristics including the normalization constant $\mathrm{C}$ and state parameter $w$ collectively influence effective potentials, orbital stability through Lyapunov exponent analysis, and critical impact parameters for photon capture. The shadow analysis establishes direct connections between theoretical parameters and observable quantities accessible to high-resolution imaging campaigns. By deriving comprehensive analytical expressions for shadow radii across different QF states and examining shadow morphology through celestial coordinate systems, we provide essential theoretical foundations for interpreting EHT observations and constraining KRG-QF parameters through precision measurements of BH shadow characteristics. Our perturbation analysis investigates both scalar and electromagnetic field fluctuations in KRG-QF spacetimes, revealing how modifications to effective potentials directly affect quasinormal mode (QNM) spectra and stability properties. These investigations provide crucial insights for interpreting gravitational wave observations and electromagnetic signatures from BH environments, establishing theoretical frameworks for detecting KRG-QF effects through precision measurements of perturbation dynamics. The gravitational lensing study employs advanced GBT methodology to derive deflection angle expressions that incorporate both LV and QF contributions. This analysis might provide crucial tools for using gravitational lensing observations to constrain modified gravity parameters and test alternative theoretical frameworks through precision astrometric measurements of light deflection around compact objects. Finally, our thermodynamic analysis uncovers intricate phase structures and critical phenomena that markedly differ from conventional black hole thermodynamics. By examining Hawking temperature evolution, Gibbs free energy characteristics, and specific heat capacity behavior across the KRG-QF parameter space, we identify potential observational signatures in thermal emission spectra and establish theoretical foundations for constraining fundamental physics through precision measurements of BH thermal properties \cite{isz26,isz27,isz28}. 

This paper is organized as follows: Section \ref{isec2} constructs the KRG-QF BH spacetime through systematic solution of modified Einstein field equations and analyzes fundamental geometric properties including horizon structures and curvature characteristics. Section \ref{isec3} investigates geodesic dynamics for both photons and massive particles, examining effective potentials, orbital stability, and ISCO properties. Section \ref{isec4} analyzes BH shadow phenomenology and establishes observational constraints through comprehensive shadow radius calculations and morphology analysis. Sections \ref{isec5} and \ref{isec6} examine scalar and electromagnetic perturbations respectively, deriving effective potentials and analyzing wave propagation characteristics. Section \ref{isec7} investigates gravitational lensing through GBT methodology, deriving comprehensive deflection angle expressions. Section \ref{isec8} presents complete thermodynamic analysis including temperature evolution, phase transitions, and stability properties. Finally, we conclude with a summary of our findings and outline future research directions emerging from this investigation.

\section{KRG-QF BH Spacetime}\label{isec2}

In KRG theory, the gravitational action incorporates contributions from the KRG field \cite{sec2is01}, the BH's gravitational field, and the surrounding QF, yielding a comprehensive framework that extends beyond standard GR. The total action can be decomposed as:
\begin{equation}
S = S_{GR} + S_{KRG} + S_{QF}.
\end{equation}

The complete action takes the explicit form:
\begin{equation}\label{action}
S = \int d^4x\sqrt{-g}\left[\frac{1}{2\kappa}\left(R+\varepsilon\, B^{\mu\lambda}B^\nu\, _\lambda R_{\mu\nu}\right)-\frac{1}{12}H_{\lambda\mu\nu}H^{\lambda\mu\nu}-V(B_{\mu\nu}B^{\mu\nu}\pm b^2)+\mathcal{L}_{QF}\right],
\end{equation}
where $\kappa=8\pi G$ represents the Einstein gravitational coupling with $G$ being the Newtonian gravitational constant, $\varepsilon$ denotes the coupling strength between gravitation and the KRG field, and $b^2$ is a positive real constant characterizing the symmetry breaking scale \cite{sec2is02}. The KRG field strength tensor is defined as $H_{\mu\nu\rho}\equiv \partial_{[\mu}B_{\nu\rho]}$, while $\mathcal{L}_{QF}$ represents the Lagrangian density governing the QF dynamics.

The self-interacting potential $V(B_{\mu\nu}B^{\mu\nu}\pm b^2)$ plays a crucial role in spontaneous LV, generating a non-zero VEV of the KRG field such that $\langle B_{\mu\nu} \rangle=b_{\mu\nu}$ satisfies the constant norm condition $b_{\mu\nu}b^{\mu\nu}=\mp b^2$ \cite{sec2is03}. This mechanism leads to the null field strength configuration of the KRG field, fundamentally altering the spacetime geometry from its GR counterpart.

Variation of the action with respect to the metric tensor $g_{\mu \nu}$ yields the modified Einstein field equations:
\begin{equation}
R_{\mu \nu }-\frac{1}{2}g_{\mu \nu }R=\kappa ( T^{\text{KRG}}_{\mu\nu}+T^{\text{QF}}_{\mu\nu}),
\label{fe}    
\end{equation}
where $R_{\mu\nu}$ denotes the Ricci tensor, $T^{\text{KRG}}_{\mu\nu}$ represents the energy-momentum tensor of the KRG field, and $T^{\text{QF}}_{\mu\nu}$ corresponds to the QF contribution \cite{sec2is04}.

The KRG field energy-momentum tensor exhibits a complex structure:
\begin{eqnarray}
\kappa T^{\text{KRG}}_{\mu\nu} &=&\frac{1}{2} H_{\mu \alpha \beta } H_{\nu }{}^{\alpha \beta } - \frac{1}{12} g_{\mu \nu } H^{\alpha \beta \rho } H_{\alpha \beta \rho }+2V'(X) B_{\alpha\mu}B^{\alpha}{}_\nu - g_{\mu\nu}V(X) \nonumber \\
&+& \varepsilon \left[\frac{1}{2} g_{\mu \nu } B^{\alpha \gamma } B^{\beta }{}_{\gamma }R_{\alpha \beta } - B^{\alpha }{}_{\mu } B^{\beta }{}_{\nu }R_{\alpha \beta }- B^{\alpha \beta } B_{\nu \beta } R_{\mu \alpha } - B^{\alpha \beta } B_{\mu \beta } R_{\nu \alpha }\right. \nonumber \\
&+&\left.\frac{1}{2} \nabla _{\alpha }\nabla _{\mu }\left(B^{\alpha \beta } B_{\nu \beta }\right) +\frac{1}{2} \nabla _{\alpha }\nabla _{\nu }\left(B^{\alpha \beta } B_{\mu \beta }\right)-\frac{1}{2}\nabla ^{\alpha }\nabla _{\alpha }\left(B_{\mu }{}^{\gamma }B_{\nu \gamma } \right) - \frac{1}{2} g_{\mu \nu } \nabla _{\alpha }\nabla _{\beta }\left(B^{\alpha \gamma } B^{\beta }{}_{\gamma }\right)\right],
\label{tkr}
\end{eqnarray}
where the prime indicates differentiation with respect to the argument.

The QF contribution manifests through its energy-momentum tensor:
\begin{equation}
T^{t}_{t}=T^{r}_{r}=\rho_q,\quad T^{\theta}_{\theta}=T^{\phi}_{\phi}=-\frac{1}{2}\,\rho_q\,(3\,w+1),\label{pp3}
\end{equation}
where $\rho_q$ denotes the QF energy density, and the pressure relates to density via the equation of state $p_q = w \rho_q$, with $w$ being the quintessence state parameter characterizing the dark energy-like behavior \cite{sec2is05}. The Bianchi identities ensure conservation of the combined tensor $T^{\text{KRG}}_{\mu\nu}+T^{\text{QF}}_{\mu\nu}$.

To derive static, spherically symmetric solutions, we employ the metric ansatz:
\begin{equation}
ds^2=-G(r)dt^2+F(r)dr^2+r^2 d\theta^2+r^2 \sin^2\theta d\phi^2.
\label{trial}
\end{equation}

We consider a pseudoelectric KRG field configuration where only $b_{01}$ and $b_{10}$ components are non-zero. The constant norm condition yields:
\begin{equation}
b_{01}=-b_{10}=|b| \sqrt{\frac{G(r)F(r)}{2}}.
\end{equation}

Assuming the KRG field remains frozen at its VEV and employing the diagonal ansatz $G(r)=1/F(r)=A(r)$, the field equations reduce to:
\begin{equation}
A''(r)+\frac{2}{r} A'(r)-\frac{\mathrm{C}}{4}\left(3w+1\right)\left(\frac{3w}{r^{3(w+1)}}\right)=0,\label{n11}
\end{equation}
\begin{equation}
A''(r)+\frac{1+\eta}{\eta\,r}A'(r)-\frac{1}{\eta\,r^2}+\frac{1-\eta}{\eta\,r^2}A(r)=0,\label{n12}
\end{equation}
where $\eta=\varepsilon b^2/2$ parameterizes the LV strength \cite{sec2is06}.

Solving these coupled differential equations yields the desired spacetime metric:
\begin{equation}
ds^2=-A(r)dt^2+\frac{1}{A(r)}dr^2+r^2 d\theta^2+r^2 \sin^2\theta d\phi^2,
\label{aa1}
\end{equation}
with the metric function:
\begin{equation}
A(r)=\frac{1}{1-\eta}-\frac{2\,M}{r}-\frac{\mathrm{C}}{r^{3\,w+1}},\label{bb2}
\end{equation}
where $(\mathrm{C}, w)$ represent the quintessential parameters characterizing the QF properties.

\begin{figure}
\begin{center}
\includegraphics[scale=1]{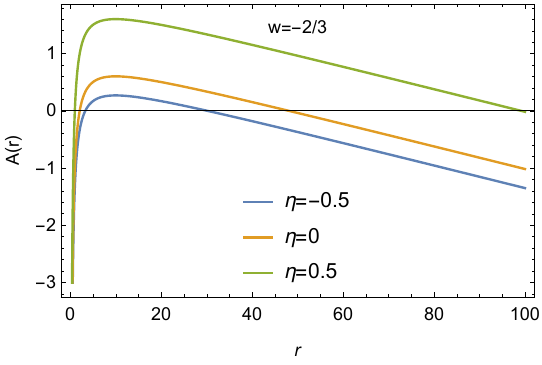}
\includegraphics[scale=1]{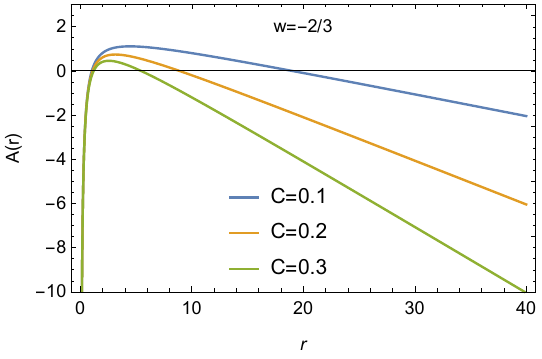}
\end{center}
\caption{\footnotesize Plot of the metric function $A(r)$ vs $r$ for different values of $\eta$ (left), and $C$ (right). Here, $M=1$.}\label{hor12}
\end{figure}

This solution exhibits rich limiting behavior: setting $\mathrm{C}=0$ recovers the Schwarzschild BH in KRG \cite{sec2is01}, while $\eta=0$ yields the standard Schwarzschild BH surrounded by QF \cite{sec2is07}. The metric function behavior, illustrated in Figure \ref{hor12}, reveals the presence of two horizons: the event horizon $r_h$ and cosmological horizon $r_c$.

The horizon locations are determined by solving $A(r_\text{h}) = 0$:
\begin{equation}
\frac{1}{1-\eta}-\frac{2\,M}{r}-\frac{\mathrm{C}}{r^{3\,w+1}}=0. \label{cc9} 
\end{equation}

For the specific case $w=-2/3$, the analytical horizon solutions are:
\begin{equation}
r_h=\frac{-1+\sqrt{1-8\,\mathrm{C}(\eta-1)^2M}}{2\,\mathrm{C}(\eta-1)}, \quad r_c=\frac{-1-\sqrt{1-8\,\mathrm{C}(\eta-1)^2M}}{2\,\mathrm{C}(\eta-1)}.
\end{equation}

\begin{figure}
\begin{center}
\includegraphics[scale=0.8]{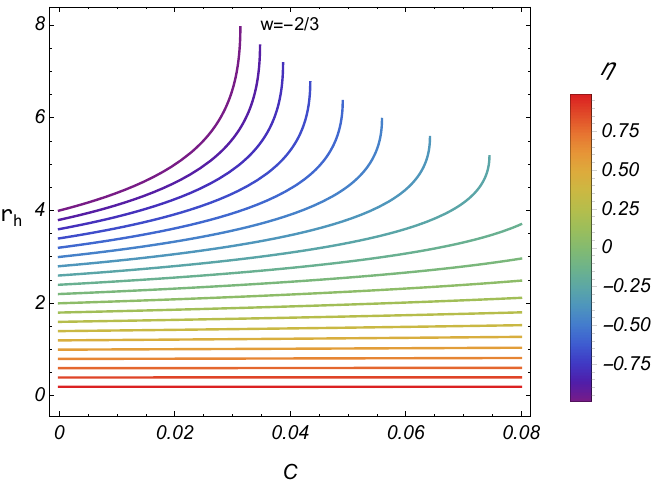}\includegraphics[scale=0.8]{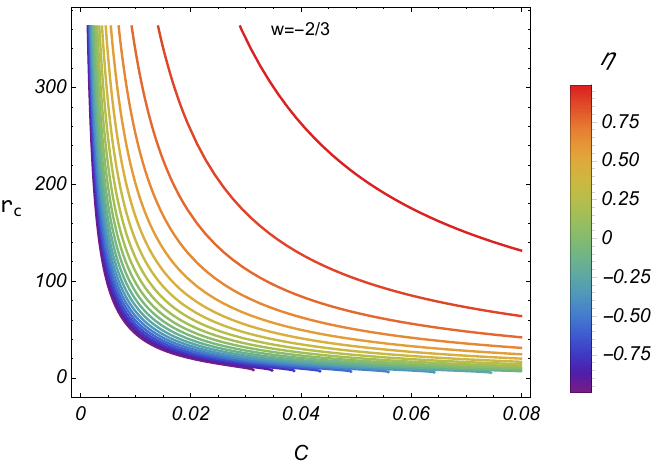}
\end{center}
\caption{\footnotesize Plot of the horizons $r_h$ (left) and $r_c$ (right) vs $\mathrm{C}$ for different values of $\eta$. Here, $M=1$ and $w=-2/3$.}\label{hor123}
\end{figure}

Figure \ref{hor123} demonstrates that the event horizon increases with $\mathrm{C}$ but decreases with $\eta$, while the cosmological horizon exhibits opposite behavior. This reflects the complex interplay between LV and QF effects in determining the causal structure.

The spacetime curvature properties are characterized by the Ricci and Kretschmann scalars:
\begin{eqnarray}
R &=& \frac{1}{r^3}\left( \frac{2\,\eta\,r}{\eta-1}+3\,\mathrm{C}\,w\,r^{-3w}(3w-1) \right), \\
\mathcal{K} &=& \frac{1}{r^6}\left( 4\left( 2\,M +\frac{2\,\eta\,r}{\eta-1}+\mathrm{C}\,r^{-3w}\right)^2+4r^{-6w}\left(\mathrm{C}+ 2\,M\,r^{3w} +3\,\mathrm{C}\,w\right)^2+r^{-6w}\left(4\,M\,r^{3w}+\mathrm{C}(2+9w(1+w))\right)^2 \right).
\end{eqnarray}

To analyze the specific influence of different QF states on the curvature structure, we examine three physically relevant values of the quintessence state parameter $w$ as presented in Table \ref{tableiz1}. The case $w=-1/3$ corresponds to a QF with intermediate dark energy characteristics, exhibiting $r^{-2}$ scaling in both curvature invariants, while $w=-2/3$ represents a phantom-like QF regime where the Ricci scalar develops linear radial dependence $r^{-1}$ alongside the standard $r^{-2}$ KRG contribution. The cosmological constant limit $w=-1$ yields constant Ricci scalar contributions from the QF term, demonstrating how different dark energy equations of state fundamentally alter the near-horizon curvature behavior and the interplay between LV and quintessence effects.

\begin{table}[h]
\begin{center}
\begin{tabular}{|c|c|c|}
\hline 
$w$ & Ricci Scalar & Kretschmann scalar \\ \hline
$-1/3$ & $\frac{2(1+\mathrm{C})}{r^2}+\frac{2}{(\eta-1)r^2}$ & $\frac{4\left( 8\,M^2+\left(2\,M+(1+\mathrm{C}+\frac{1}{\eta-1})r \right)^2\right)}{r^6}$ \\
\hline
$-2/3$ & $\frac{6\,\mathrm{C}}{r}+\frac{2}{(\eta-1)r^2}$ & $\frac{4\left( 4\,M^2+\mathrm{C}\,r^2-2\,M)^2+\left(2\,M+r(\mathrm{C}\,r+\frac{1}{\eta-1}) \right)^2\right)}{r^6}$ \\
\hline
$-1$ & $12\,\mathrm{C}+\frac{2\,\eta}{(\eta-1)r^2}$ & $24\,\mathrm{C}^2+\frac{8\mathrm{C}\,\eta}{(\eta-1)r^2}+ \frac{4}{r^6}\left( 12\,M^2+\frac{4\eta Mr}{\eta-1}+\frac{\eta^2 r^2}{(\eta-1)^2} \right)$ \\ 
\hline
\end{tabular}
\caption{\footnotesize The Ricci and Kretschmann scalars for various QF parameters.}\label{tableiz1}
\end{center}
\end{table}

The Kretschmann scalar $\mathcal{K}$ represents a fundamental curvature invariant in differential geometry, formally defined as the contraction of the Riemann curvature tensor with itself:

\begin{equation}
\mathcal{K} = R_{\mu\nu\rho\sigma}R^{\mu\nu\rho\sigma},
\end{equation}

where $R_{\mu\nu\rho\sigma}$ denotes the Riemann curvature tensor components. This scalar quantity provides a coordinate-independent measure of spacetime curvature and serves as a crucial diagnostic tool for identifying genuine curvature singularities in gravitational theories \cite{sec2is09}.

The Kretschmann scalar behavior at the origin confirms the presence of a true curvature singularity in our KRG-QF BH solution:

\begin{equation}
\lim_{r\to 0} \mathcal{K}\approx\infty.
\end{equation}

This divergent behavior indicates that the spacetime curvature becomes arbitrarily large as one approaches $r=0$, establishing the existence of a physical singularity that cannot be removed by coordinate transformations. The divergence of $\mathcal{K}$ at the origin represents a fundamental breakdown of the classical geometric description of spacetime, where tidal forces become infinite and the deterministic evolution of classical physics ceases to apply \cite{sec2is10}. The presence of this curvature singularity in our modified gravitational framework demonstrates several important physical implications. First, it confirms that despite the modifications introduced by LV through the KRG field and the exotic matter content represented by the QF, the central singularity characteristic of classical BH solutions persists \cite{sec2is11}. This suggests that neither the spontaneous breaking of Lorentz symmetry nor the presence of quintessence dark energy is sufficient to resolve the fundamental singularity problem that plagues classical GR. Furthermore, the invariant nature of the Kretschmann scalar ensures that this singularity diagnosis is coordinate-independent and physically meaningful. Unlike coordinate singularities that can be eliminated through appropriate coordinate choices, the divergence of $\mathcal{K}$ represents a genuine pathology in the spacetime manifold \cite{sec2is12}. This analysis establishes that the BH solution possesses an inherent singularity at $r=0$, consistent with the expectation for classical BH solutions in modified gravity theories incorporating both LV and dark energy effects.

\section{Geodesic Dynamics in KRG-QF BH Spacetime}\label{isec3}

This section presents a comprehensive investigation of geodesic motion for both photons and massive particles within the modified gravitational framework of KRG theory coupled with QF. We systematically analyze how the geometric modifications introduced by LV through the KRG field parameter $\eta$ and the exotic matter contributions from QF parameters $(\mathrm{C}, w)$ collectively influence particle trajectories, photon sphere characteristics, effective radial forces, and orbital stability properties.

Given the static and spherically symmetric nature of our spacetime, we restrict the geodesic analysis to the equatorial plane, defined by $\theta = \frac{\pi}{2}$ and $\dot{\theta} = 0$, without loss of generality. Employing the Lagrangian formalism with the Lagrangian density $\mathcal{L}=\frac{1}{2}\,g_{\mu\nu}\,\dot{x}^{\mu}\,\dot{x}^{\nu}$, where the dot denotes differentiation with respect to an affine parameter, we express the Lagrangian for the KRG-QF metric given in Eq.~(\ref{aa1}) as:

\begin{equation}
\mathcal{L}=\frac{1}{2}\,\left[-A(r, \eta)\,\dot{t}^2+\frac{\dot{r}^2}{A(r, \eta)}+r^2\,\dot{\phi}^2\right].\label{b1}
\end{equation}

The temporal and azimuthal Killing vectors of the spacetime yield two conserved quantities:
\begin{align}
\mathrm{E}&=A(r, \eta)\,\dot{t},\label{b3}\\
\mathrm{L}&=r^2\,\dot{\phi},\label{b4}
\end{align}
representing the specific energy and angular momentum per unit rest mass of the test particle, respectively.

These conservation laws enable us to express the radial geodesic equation as an energy conservation relation:
\begin{equation}
\dot{r}^2+V_\text{eff}(r)=\mathrm{E}^2,\label{b5}
\end{equation}
where the effective potential governing particle dynamics takes the form:
\begin{equation}
V_\text{eff}(r)=\left(-\epsilon+\frac{\mathrm{L}^2}{r^2}\right)\,A(r, \eta)=\left(-\epsilon+\frac{\mathrm{L}^2}{r^2}\right)\,\left(\frac{1}{1-\eta}-\frac{2\,M}{r}-\frac{\mathrm{C}}{r^{3\,w+1}}\right).\label{b6}
\end{equation}

Here, $\epsilon=0$ characterizes null geodesics (photons), while $\epsilon=-1$ corresponds to timelike geodesics (massive particles). The effective potential explicitly demonstrates how both the LV parameter $\eta$ and QF characteristics $(\mathrm{C}, w)$ fundamentally modify the gravitational dynamics compared to standard Schwarzschild geometry.

\begin{figure}[ht!]
\centering
\includegraphics[width=0.4\linewidth]{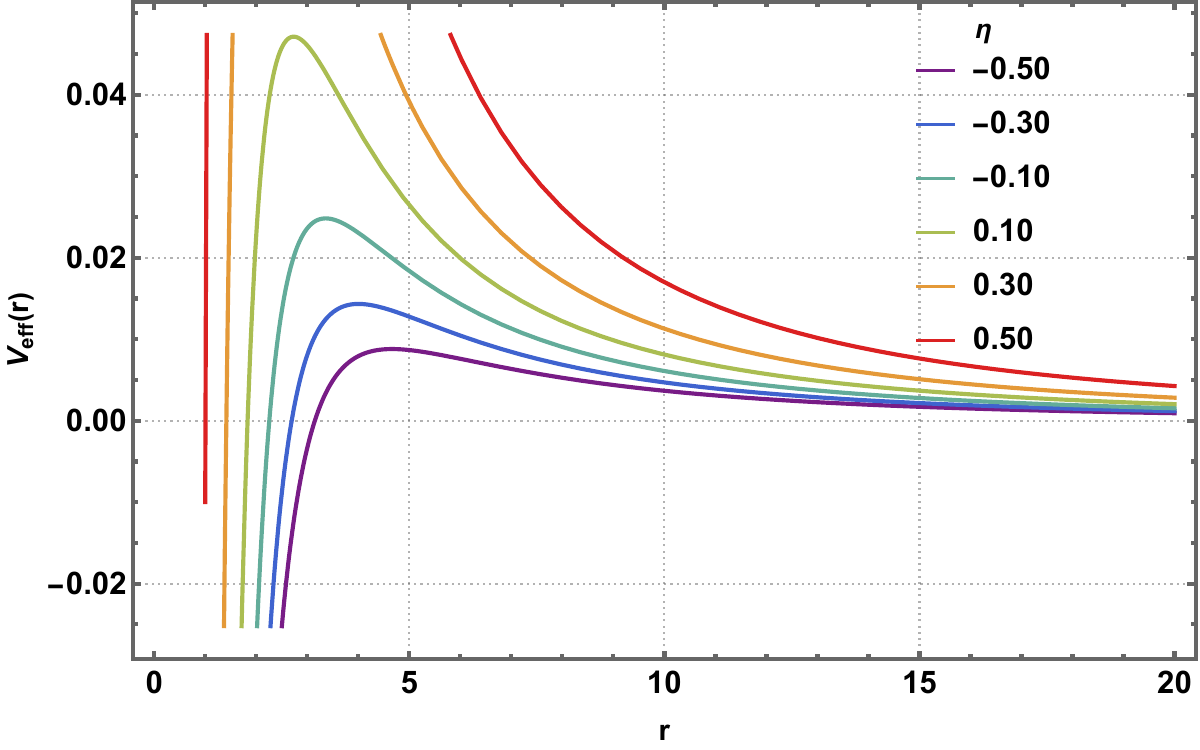}\quad\quad
\includegraphics[width=0.4\linewidth]{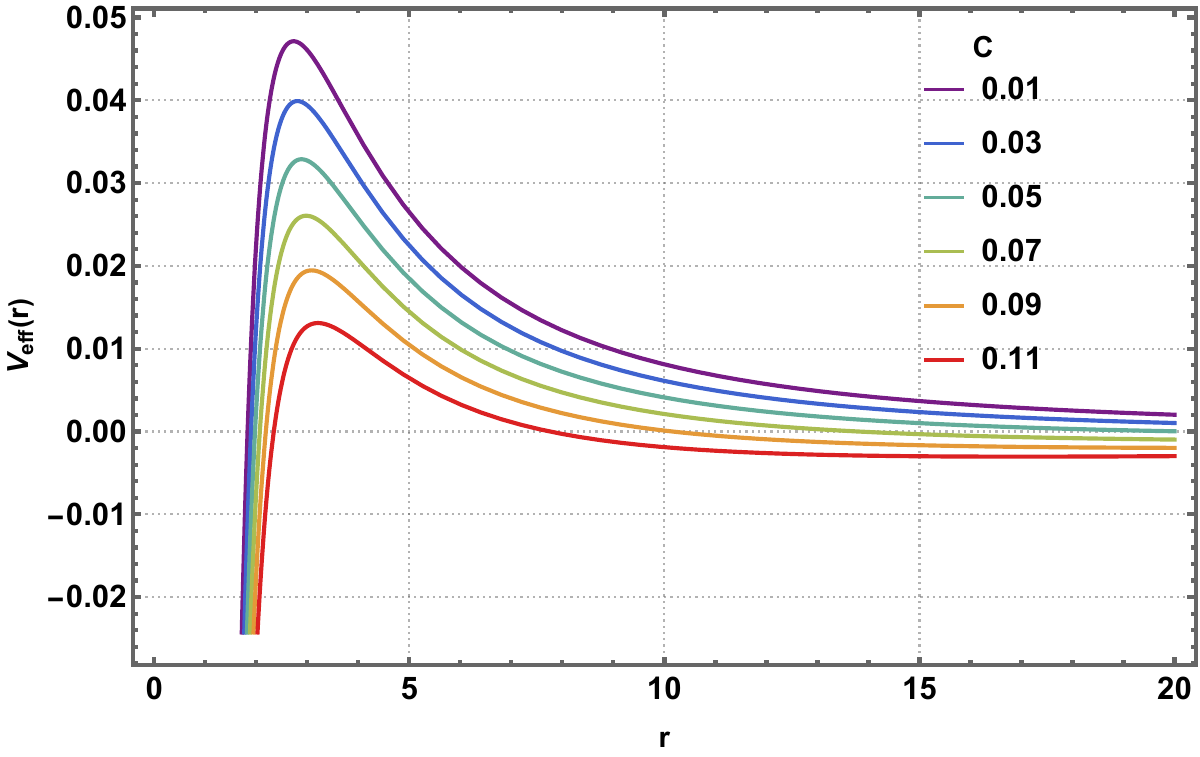}\\
(a) $\mathrm{C}=0.01,\,w=-2/3$ \hspace{6cm} (b) $\eta=0.1,\,w=-2/3$\\
\includegraphics[width=0.4\linewidth]{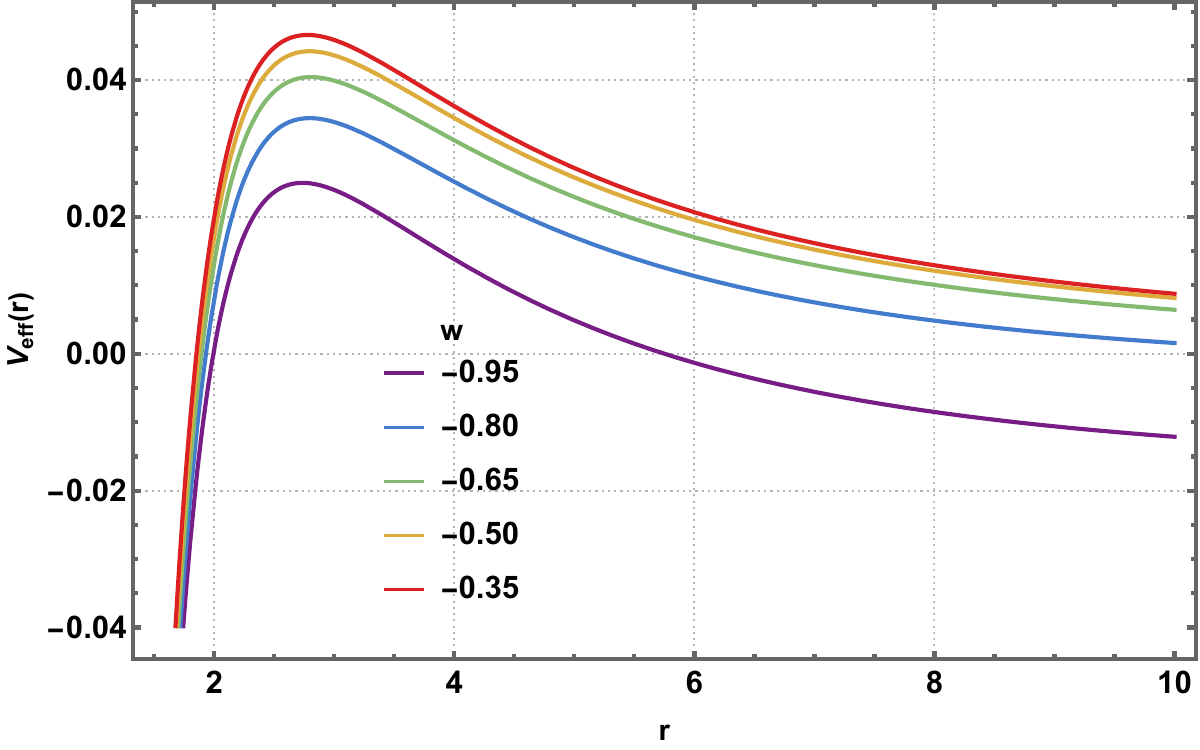}\\
(c) $\mathrm{C}=0.03,\,\eta=0.1$\\
\caption{\footnotesize Behavior of the effective potential governing photon dynamics as a function of $r$ for different values of KRG field parameter $\eta$ and QF normalization constant $\mathrm{C}$. Parameters: $M=1$, $\mathrm{L}=1$.}
\label{fig:potential-1}
\end{figure}

Figure \ref{fig:potential-1} illustrates the profound impact of KRG-QF modifications on the effective potential landscape. The left panel reveals that increasing the LV parameter $\eta$ systematically elevates the potential barrier, indicating enhanced gravitational binding effects due to spontaneous Lorentz symmetry breaking. Conversely, the right panel demonstrates that increasing the QF normalization constant $\mathrm{C}$ reduces the potential magnitude, reflecting the repulsive dark energy-like influence of quintessence matter. This complementary behavior between LV and QF effects establishes a rich parameter space for controlling particle dynamics and orbital characteristics.

\subsection{Null Geodesics: Photon Trajectories and Photon Sphere Analysis}

Null geodesics represent the fundamental trajectories of massless particles in curved spacetime and serve as crucial probes of strong gravitational fields near BH horizons. The analysis of photon motion involves studying an effective potential that encodes both spacetime curvature effects and conserved dynamical quantities. A central feature emerging from this investigation is the photon sphere—a critical spherical region where photons can maintain unstable circular orbits. Understanding these null geodesic properties is essential for interpreting observable phenomena including BH shadows, gravitational lensing effects, and light propagation in strongly curved geometries.

For null geodesics with $\epsilon=0$, the effective potential simplifies to:
\begin{equation}
V_\text{eff}(r)=\frac{\mathrm{L}^2}{r^2}\,A(r, \eta).\label{c1}
\end{equation}

The effective radial force experienced by photons is mathematically defined as $\mathrm{F}_\text{eff}(r)=-\frac{1}{2}\,\frac{dV_\text{eff}}{dr}$, yielding:
\begin{equation}
\mathrm{F}_\text{eff}(r)=\frac{\mathrm{L}^2}{r^3}\,\left(\frac{1}{1-\eta}-\frac{3\,M}{r}-\frac{\mathrm{C}\,(3\,w+3)/2}{r^{3\,w+1}}\right).\label{force-1}
\end{equation}

For the specific quintessence state parameter $w=-2/3$, this reduces to:
\begin{equation}
\mathrm{F}_\text{eff}(r)=\frac{\mathrm{L}^2}{r^3}\,\left(\frac{1}{1-\eta}-\frac{3\,M}{r}-\frac{\mathrm{C}}{2}\,r\right).\label{force-2}
\end{equation}

\begin{figure}[ht!]
\centering
\includegraphics[width=0.4\linewidth]{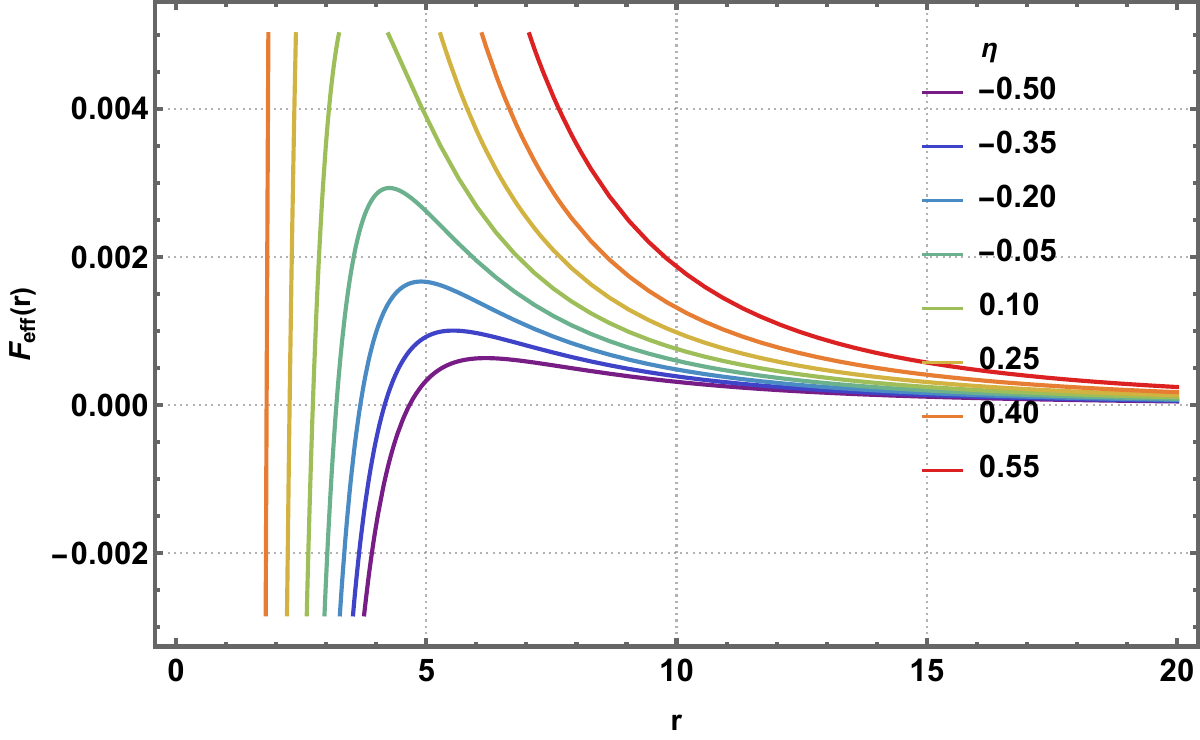}\quad\quad
\includegraphics[width=0.4\linewidth]{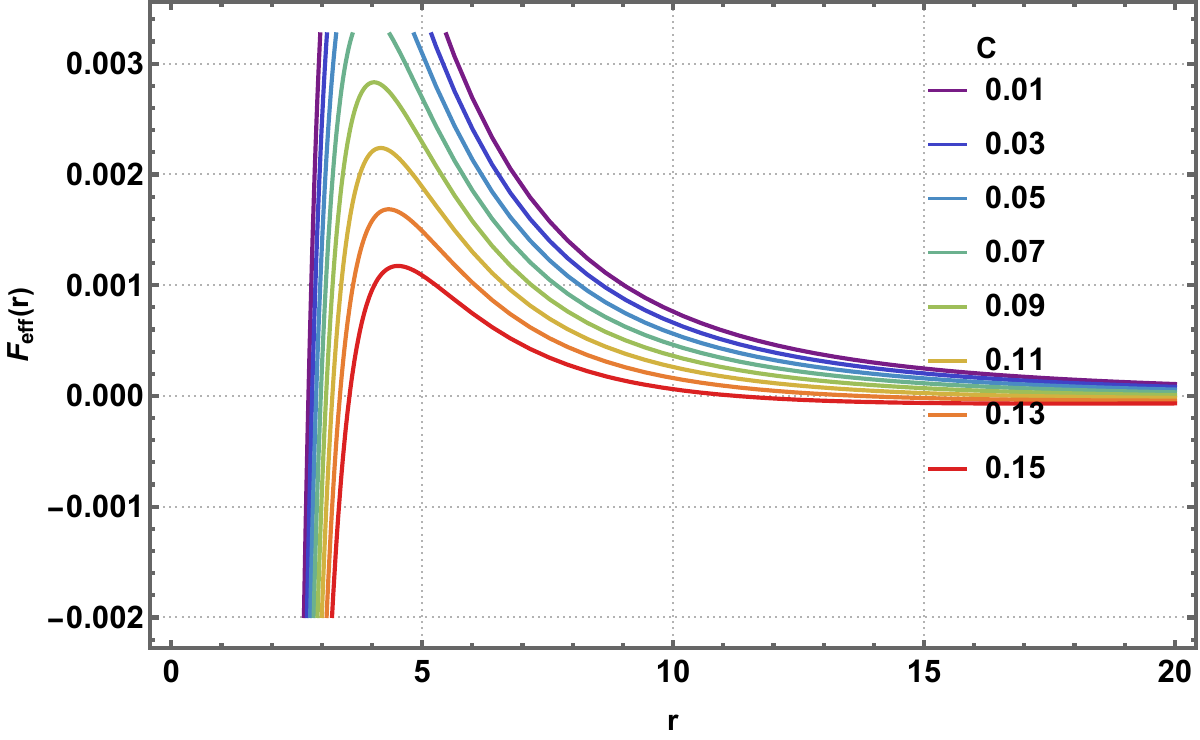}\\
(a) $\mathrm{C}=0.01,\,w=-2/3$ \hspace{6cm} (b) $\eta=0.1,\,w=-2/3$\\
\includegraphics[width=0.4\linewidth]{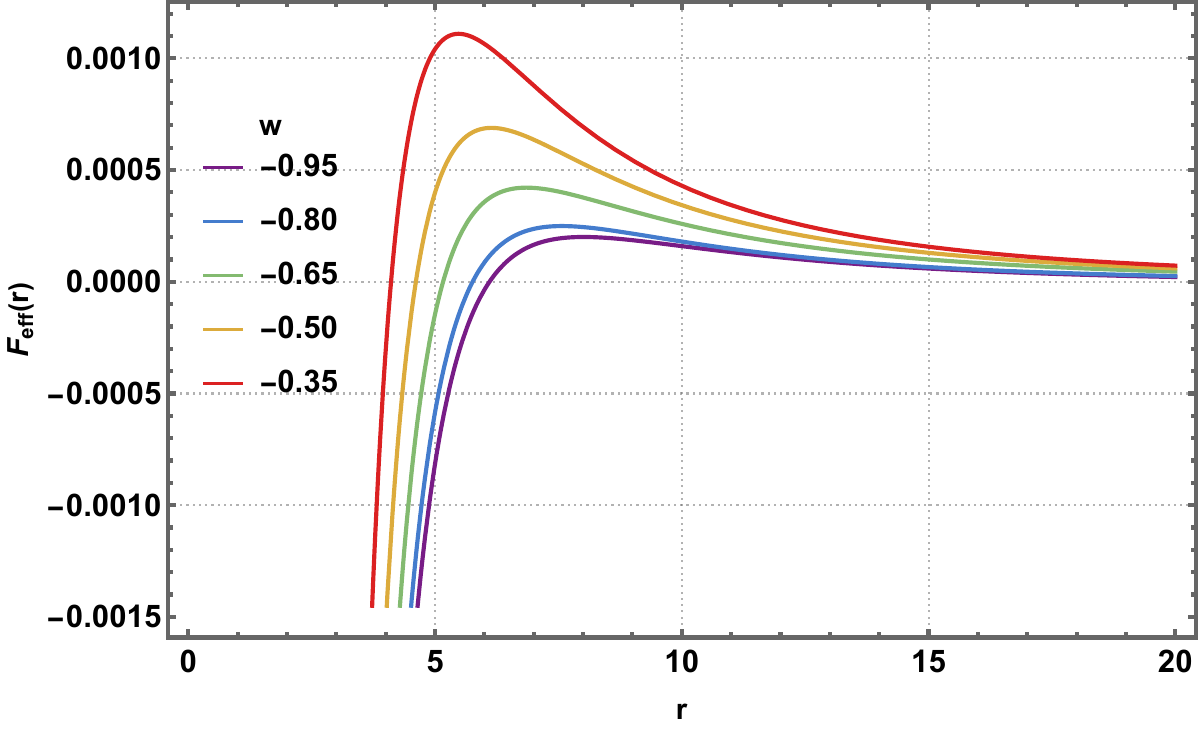}\\
(c) $\eta=0.1,\,\mathrm{C}=0.01$
\caption{Effective radial force on photons as a function of $r$ for varying KRG field parameter $\eta$ and QF normalization constant $\mathrm{C}$. Parameters: $M=1$, $\mathrm{L}=1$.}
\label{fig:force}
\end{figure}

Figure \ref{fig:force} provides crucial insights into the radial force dynamics governing photon motion. The systematic increase of force magnitude with growing $\eta$ values (left panel) demonstrates how LV enhances gravitational attraction, while the force reduction with increasing $\mathrm{C}$ (right panel) reflects the anti-gravitational influence of quintessence dark energy. These opposing trends establish the complex interplay between modified gravity and exotic matter effects in determining photon capture and deflection characteristics.

The photon orbital equation, derived from the conservation laws and effective potential, takes the form:
\begin{equation}
\left(\frac{1}{r^2}\,\frac{dr}{d\phi}\right)^2+\frac{1}{1-\eta}\,\frac{1}{r^2}=\frac{1}{\beta^2}+\frac{2\,M}{r^3}+\frac{\mathrm{C}}{r^{3\,w+3}},\label{c2}
\end{equation}
where $\beta = \mathrm{L}/\mathrm{E}$ represents the impact parameter.

Introducing the inverse radial coordinate $u = 1/r$ and differentiating with respect to $\phi$, we obtain the nonlinear differential equation governing photon trajectories:
\begin{equation}
\frac{d^2u}{d\phi^2}+\frac{u}{1-\eta}=3\,M\,u^2+\frac{1}{2}\,\mathrm{C}\,(3\,w+3)\,u^{3\,w+2}.\label{c4}
\end{equation}

For the case $w=-2/3$, this simplifies to:
\begin{equation}
\frac{d^2u}{d\phi^2}+\frac{u}{1-\eta}=3\,M\,u^2+\frac{1}{2}\,\mathrm{C}.\label{c4a}
\end{equation}

\begin{figure}[ht!]
\centering
\includegraphics[width=0.3\linewidth]{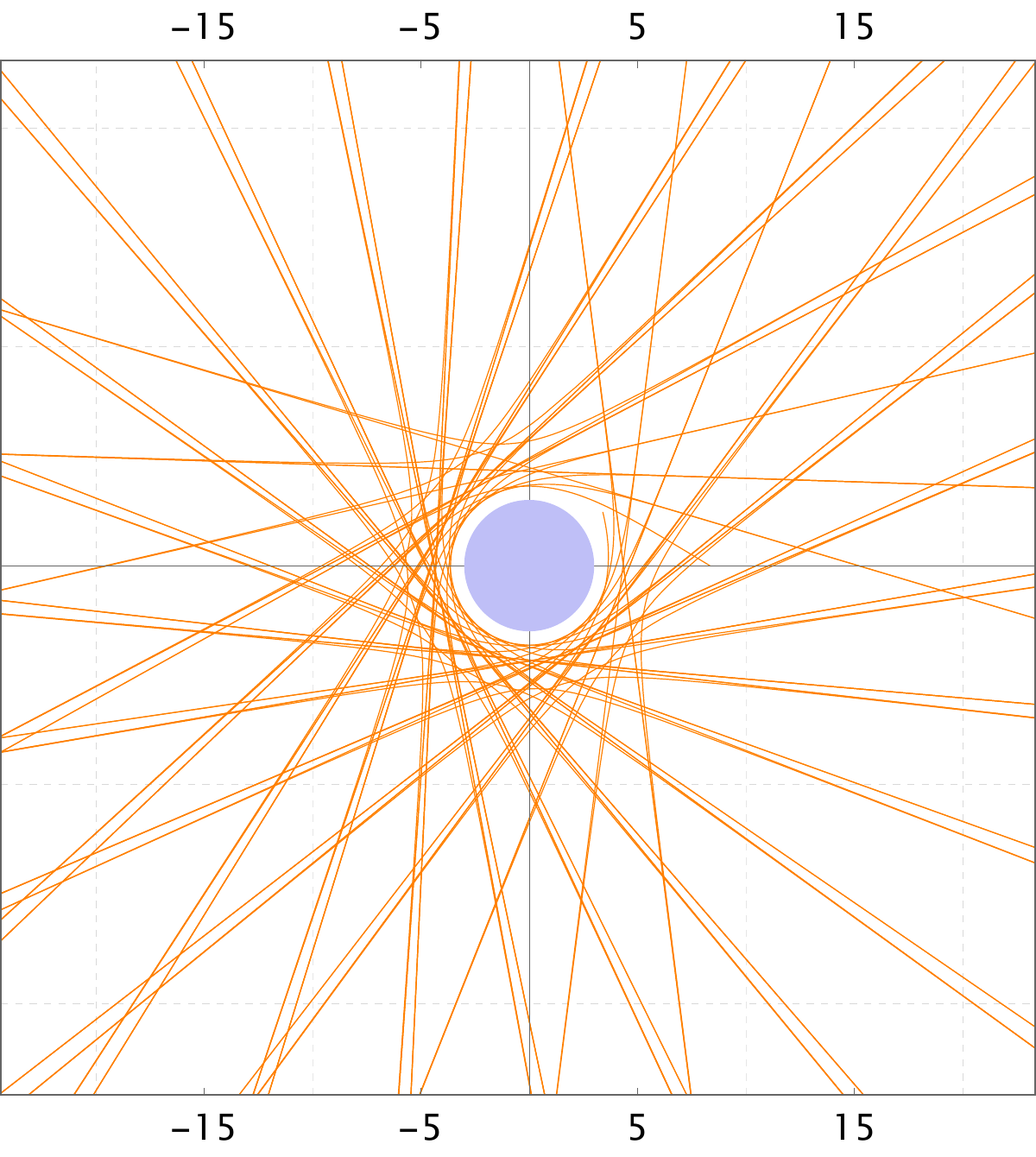}\quad
\includegraphics[width=0.3\linewidth]{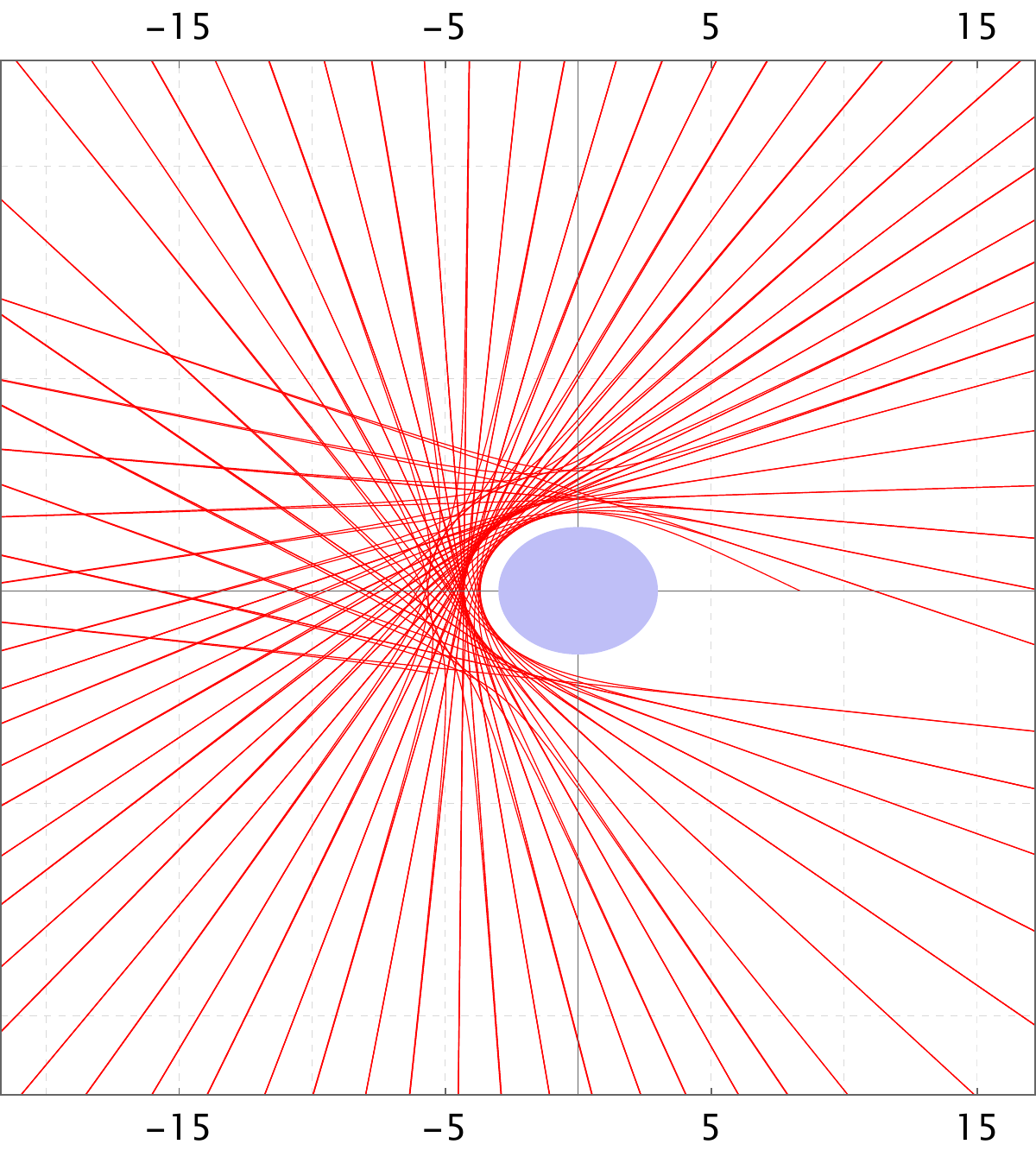}\quad
\includegraphics[width=0.3\linewidth]{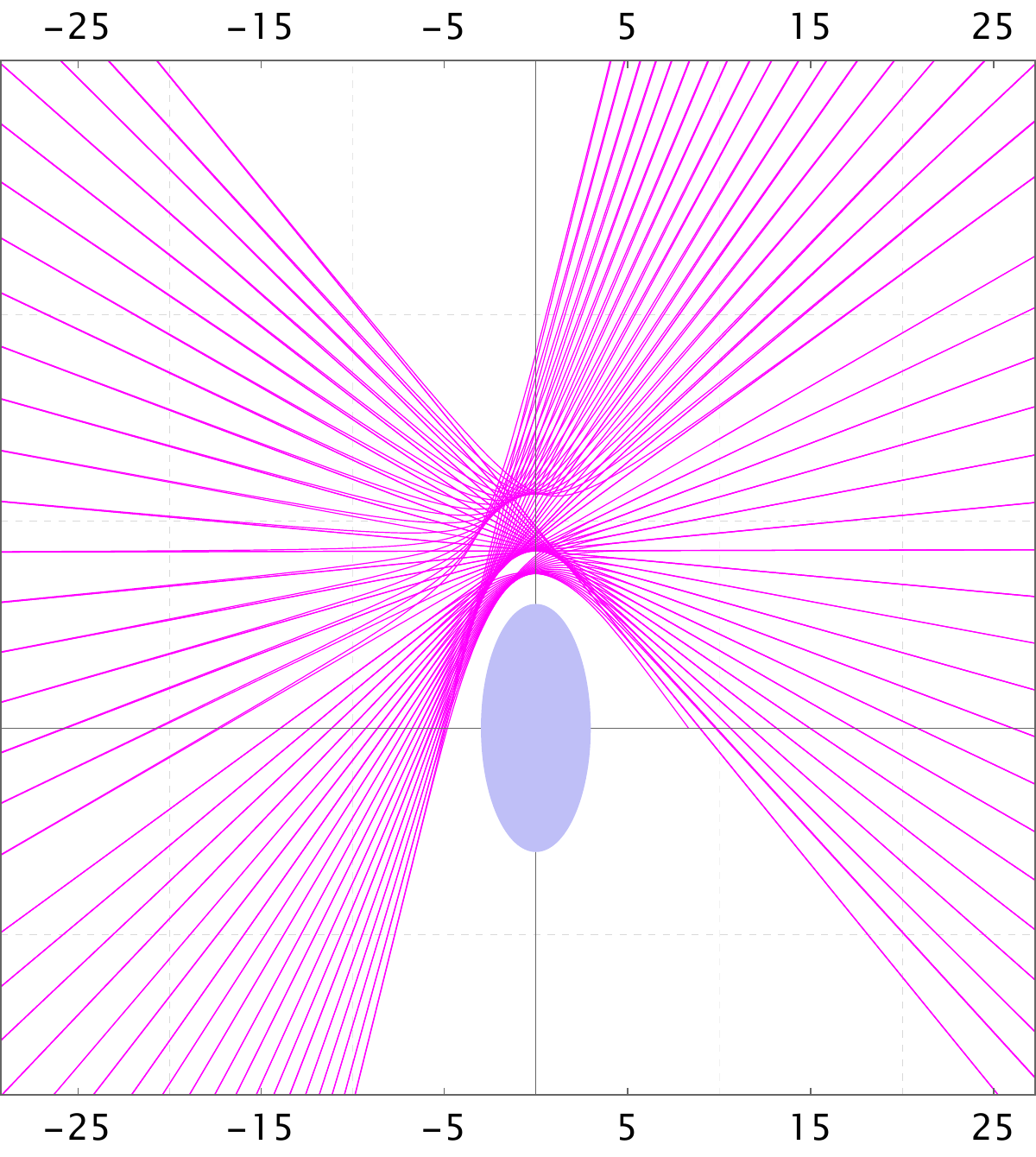}\\
(a) $\eta=0.23$ \hspace{4cm} (b) $\eta=0.24$ \hspace{4cm} (c) $\eta=0.25$\\
\includegraphics[width=0.3\linewidth]{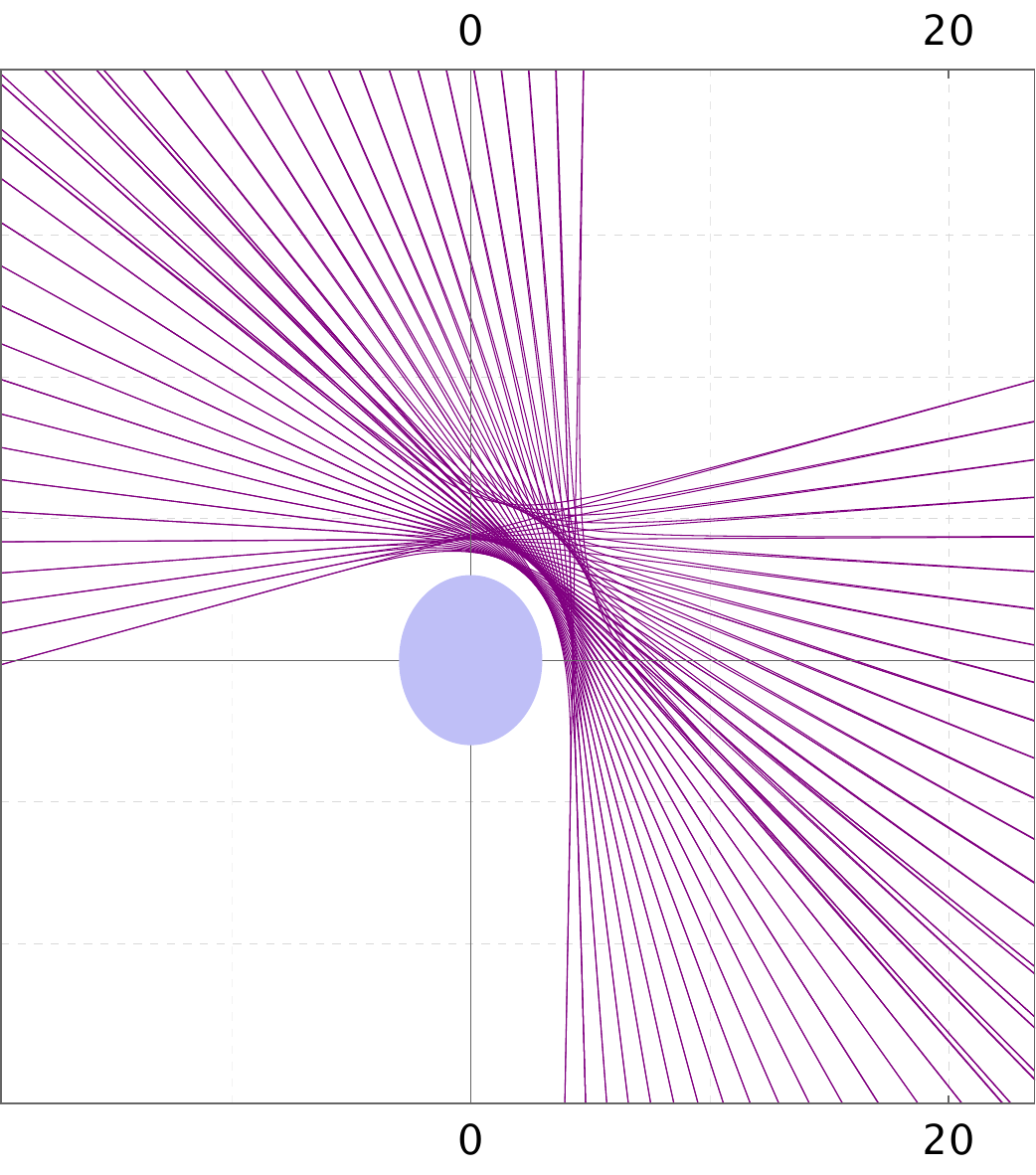}\quad
\includegraphics[width=0.3\linewidth]{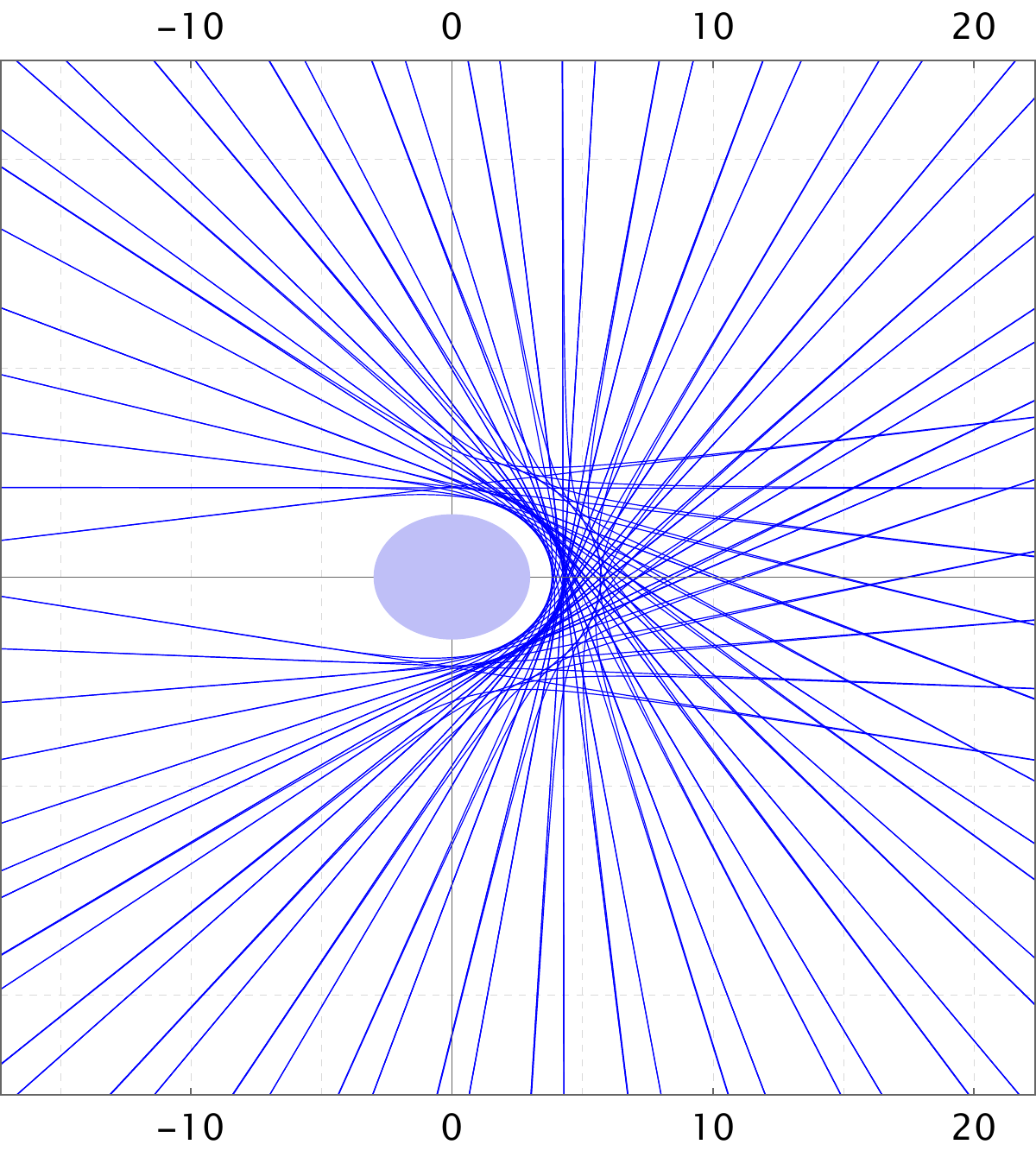}\quad
\includegraphics[width=0.3\linewidth]{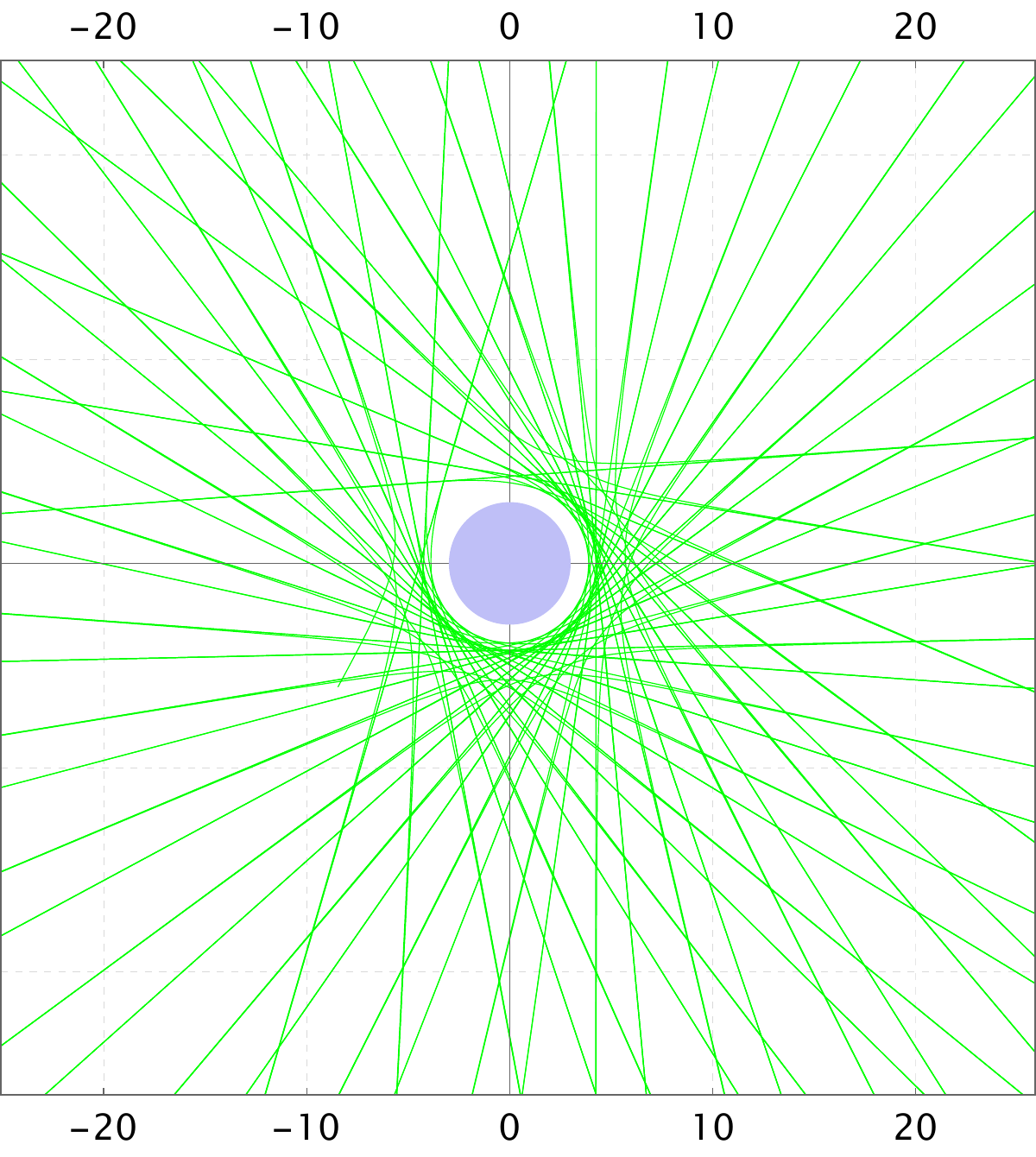}\\
(d) $\eta=0.26$ \hspace{4cm} (e) $\eta=0.27$ \hspace{4cm} (f) $\eta=0.28$
\caption{\footnotesize Parametric plots of photon geodesic paths $r(\phi)=\frac{1}{u(\phi)}$ demonstrating trajectory evolution with varying LV parameter $\eta$. Initial conditions: $u(0)=0.12$, $u'(0)=0.2$. Parameters: $M=1$, $\mathrm{C}=0.01$.}
\label{fig:parametric}
\end{figure}

Figure \ref{fig:parametric} presents a remarkable visualization of how LV systematically modifies photon geodesic structures. The progressive evolution from panels (a) through (f) reveals increasingly complex orbital patterns as $\eta$ increases from $0.23$ to $0.28$. The transition from relatively simple deflection patterns to intricate multi-loop structures demonstrates the sensitive dependence of photon dynamics on the strength of Lorentz symmetry breaking, highlighting potential observational signatures in gravitational lensing phenomena.

For circular photon orbits at radius $r_c$, the equilibrium conditions $\dot{r}=0$ and $\ddot{r}=0$ yield the critical impact parameter:
\begin{equation}
\beta_c=\frac{r_c}{\sqrt{\frac{1}{1-\eta}-\frac{2\,M}{r_c}-\frac{\mathrm{C}}{r^{3\,w+1}_c}}}.\label{c6}
\end{equation}

For $w=-2/3$, this becomes:
\begin{equation}
\beta_c=\frac{r_c}{\sqrt{\frac{1}{1-\eta}-\frac{2\,M}{r_c}-\mathrm{C}\,r_c}}.\label{c7}
\end{equation}

The orbital stability of circular photon paths is characterized by the Lyapunov exponent:
\begin{equation}
\lambda^\text{null}=\sqrt{-\frac{1}{2}\,\frac{1}{\dot{t}^2}\,\frac{d^2V_\text{eff}(r)}{dr^2}}\Bigg|_{r=r_c},\label{c8}
\end{equation}
which evaluates to:
\begin{equation}
\lambda^\text{null}=\frac{1}{r}\,\sqrt{\left(\frac{1}{1-\eta}-\frac{2\,M}{r}-\frac{\mathrm{C}}{r^{3\,w+1}}\right)\left(\frac{1}{1-\eta}+\frac{\mathrm{C}}{r^{3\,w+1}}\,\frac{(3\,w+1)(3\,w+2)-2}{2}\right)}\Bigg|_{r=r_c}.\label{c9}
\end{equation}

For $w=-2/3$:
\begin{equation}
\lambda^\text{null}=\frac{1}{r}\,\sqrt{\left(\frac{1}{1-\eta}-\frac{2\,M}{r}-\mathrm{C}\,r\right)\left(\frac{1}{1-\eta}-\mathrm{C}\,r\right)}\Bigg|_{r=r_c}.\label{c10}
\end{equation}

\begin{figure}[ht!]
\centering
\includegraphics[width=0.4\linewidth]{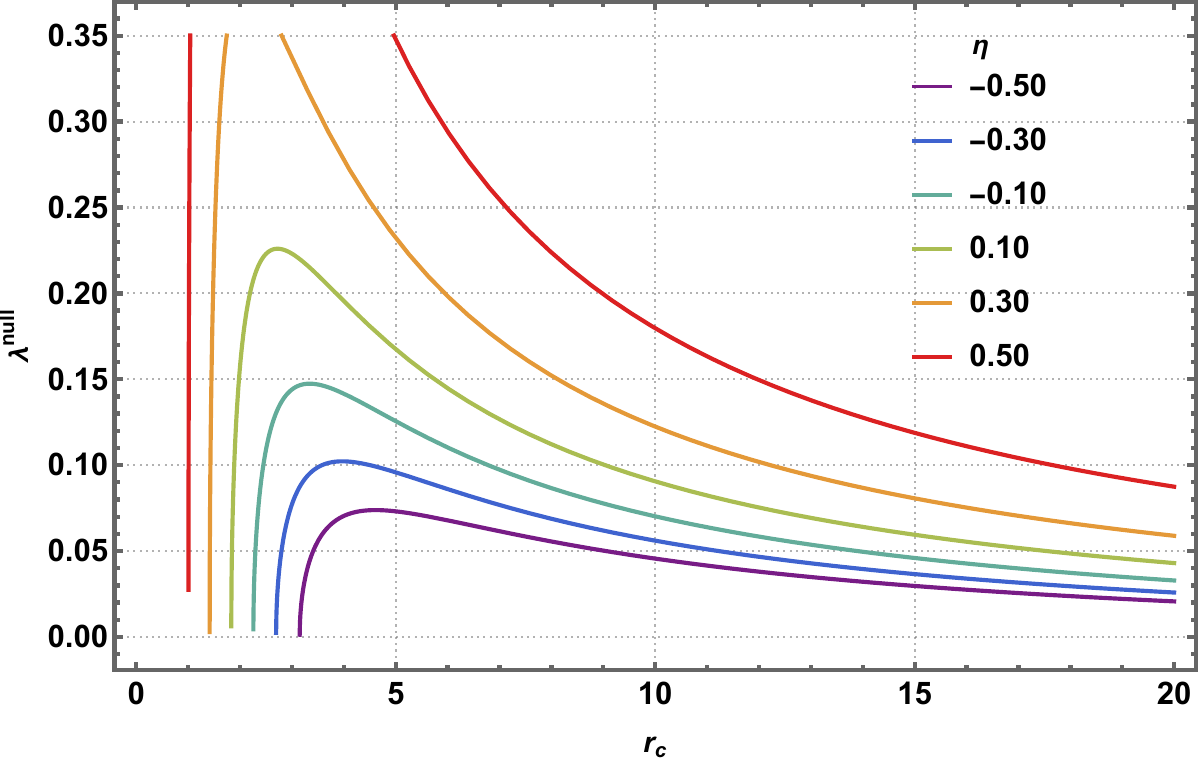}\quad\quad
\includegraphics[width=0.4\linewidth]{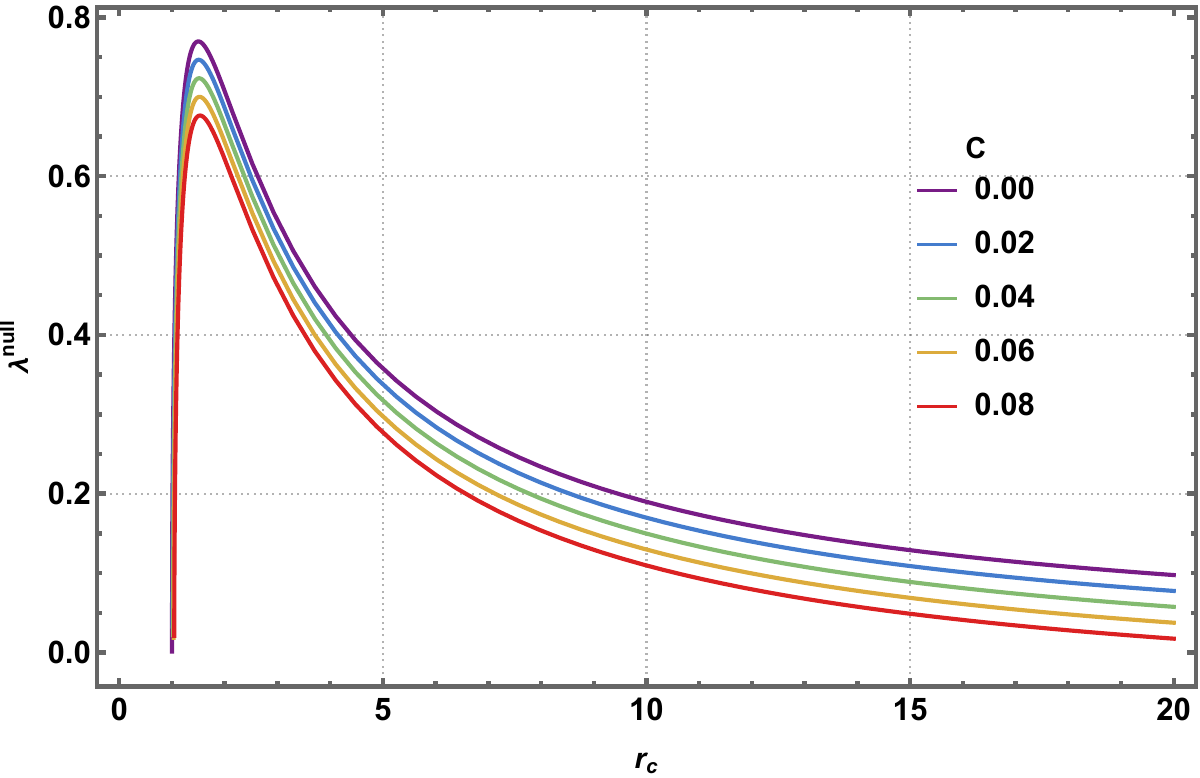}\\
(a) $\mathrm{C}=0.01,\,w=-2/3$ \hspace{4cm} (b) $\eta=0.5,\,w=-2/3$\\
\includegraphics[width=0.4\linewidth]{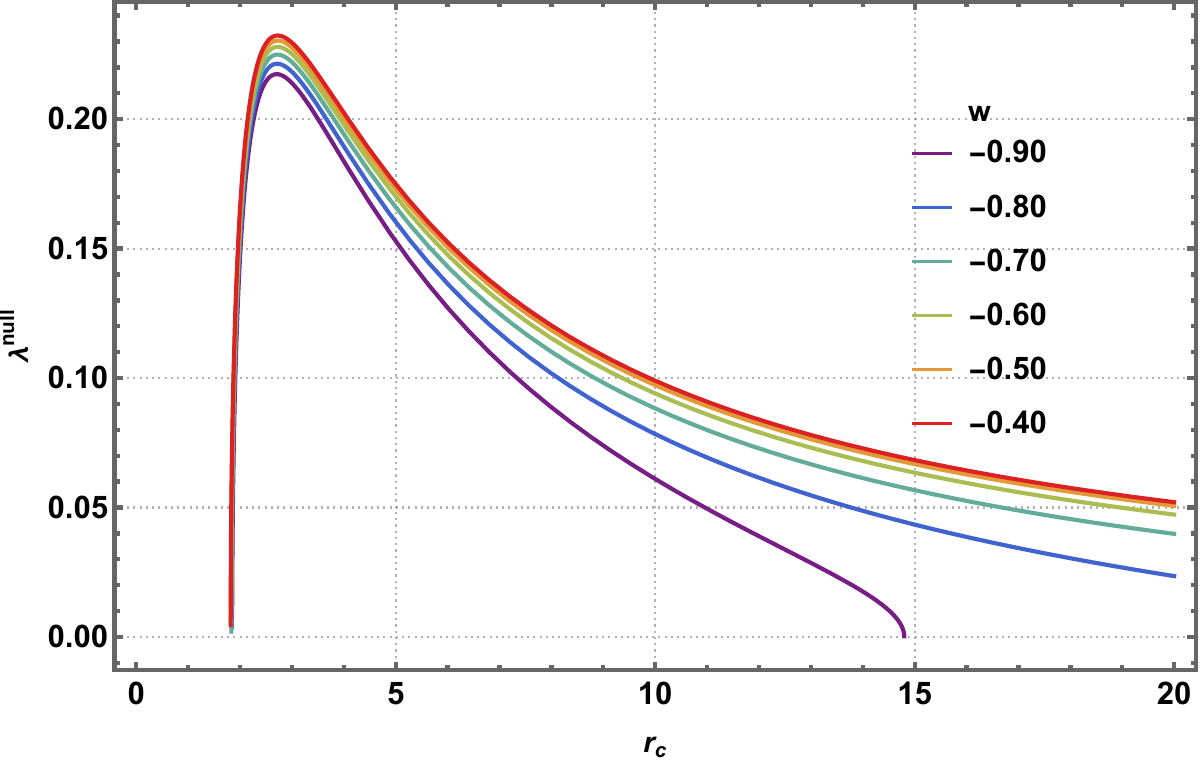}\\
(c) $\eta=0.1,\,\mathrm{C}=0.01$
\caption{\footnotesize Lyapunov exponent $\lambda^\text{null}$ evolution as a function of orbital radius $r_c$ for different KRG field parameter $\eta$ and QF normalization constant $\mathrm{C}$. Parameters: $M=1$.}
\label{fig:exponent}
\end{figure}

Figure \ref{fig:exponent} demonstrates the universal instability of circular photon orbits in the KRG-QF framework. The consistently positive Lyapunov exponent values across both parameter variations confirm that all circular null orbits remain fundamentally unstable, characteristic of photon sphere dynamics. However, the magnitude variations with $\eta$ and $\mathrm{C}$ indicate that the instability timescales are modulated by both LV and quintessence effects, potentially affecting observational signatures in BH imaging and shadow measurements.

The coordinate angular velocity for circular photon orbits is:
\begin{equation}
\omega^\text{null}=\frac{d\phi}{dt}=\frac{\dot{\phi}}{\dot{t}}=\frac{1}{r}\,\sqrt{\frac{1}{1-\eta}-\frac{2\,M}{r}-\frac{\mathrm{C}}{r^{3\,w+1}}}\Bigg|_{r=r_c}.\label{c11}
\end{equation}

\subsection{Timelike Geodesics: ISCO Analysis and Orbital Dynamics}

The investigation of massive particle dynamics within the KRG-QF gravitational framework provides crucial insights into the strong-field regime and potential astrophysical manifestations. Central to this analysis is the determination of ISCO properties, which represent the innermost stable circular trajectories for massive test particles. Understanding ISCO characteristics is fundamental for predicting accretion disk dynamics, gravitational wave emission from inspiraling compact objects, and other high-energy astrophysical phenomena that could provide observational tests of modified gravity theories.

For timelike geodesics with $\epsilon=-1$, the effective potential becomes:
\begin{equation}
V_\text{eff}(r)=\left(1+\frac{\mathrm{L}^2}{r^2}\right)\,\left(\frac{1}{1-\eta}-\frac{2\,M}{r}-\frac{\mathrm{C}}{r^{3\,w+1}}\right).\label{d1}
\end{equation}

\begin{figure}[ht!]
\centering
\includegraphics[width=0.4\linewidth]{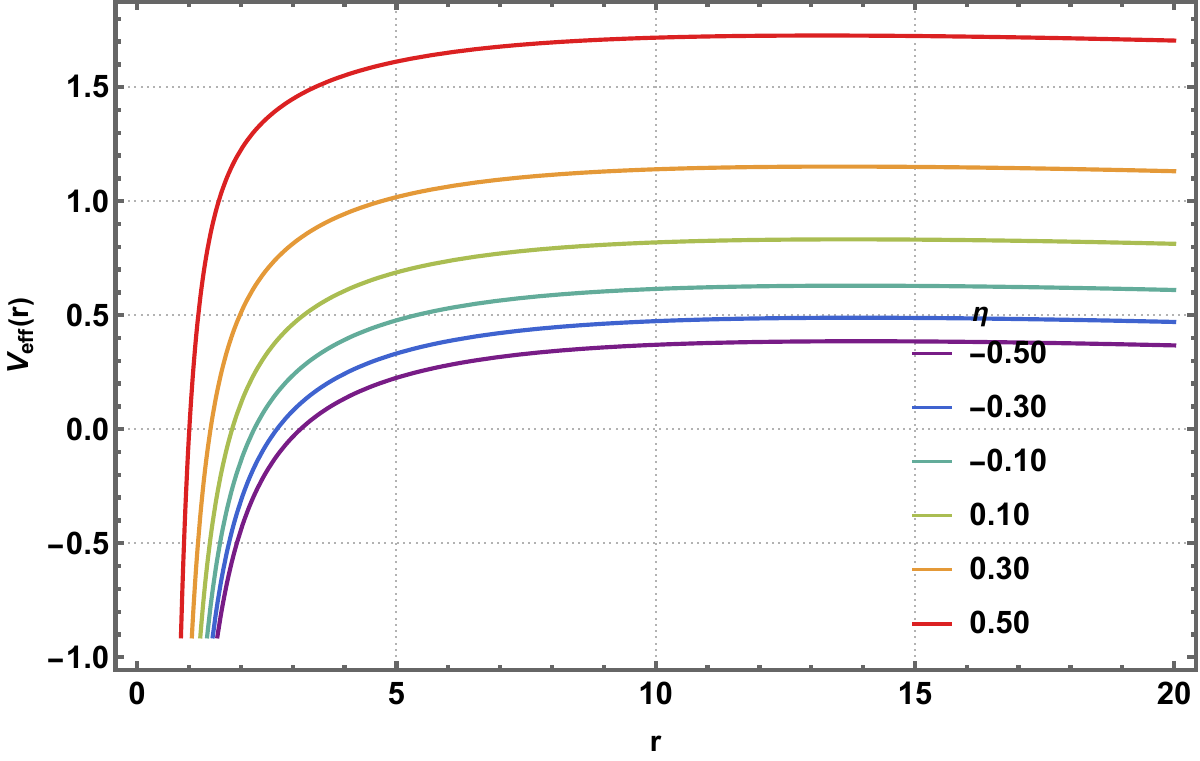}\quad\quad\quad
\includegraphics[width=0.4\linewidth]{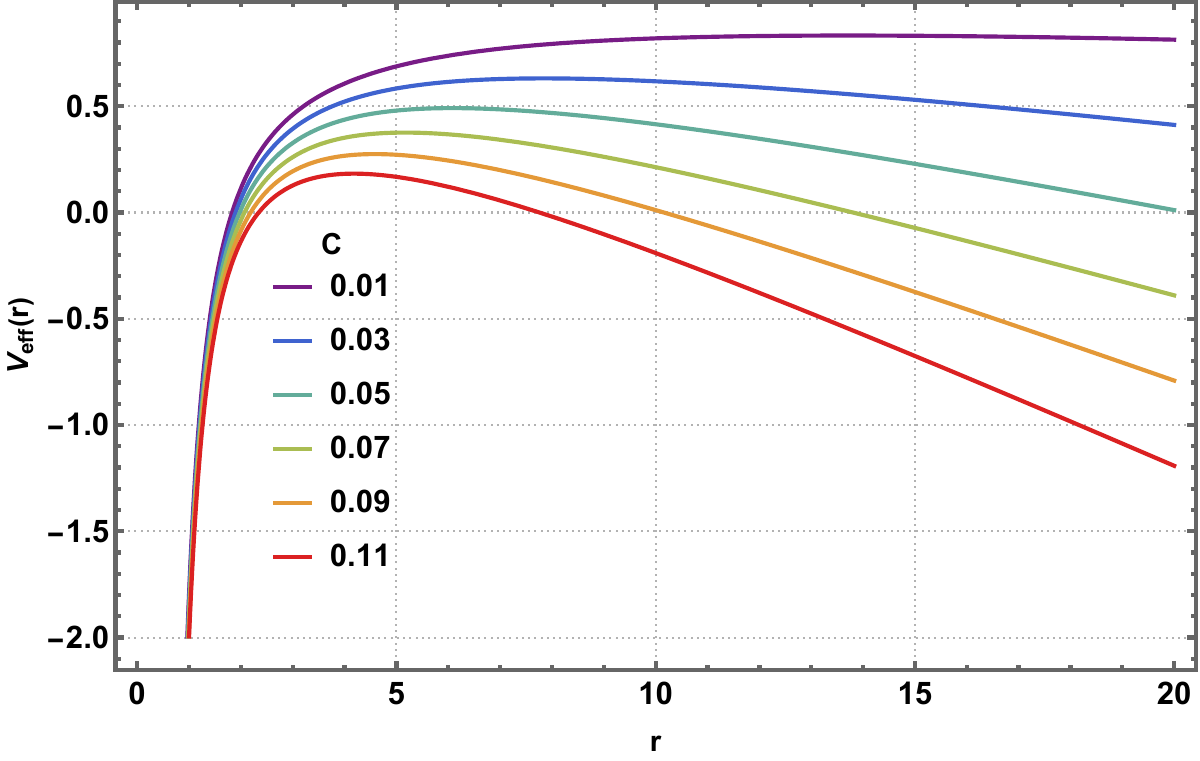}\\
(a) $\mathrm{C}=0.01,\,w=-2/3$ \hspace{6cm} (b) $\eta=0.1,\,w=-2/3$\\
\includegraphics[width=0.4\linewidth]{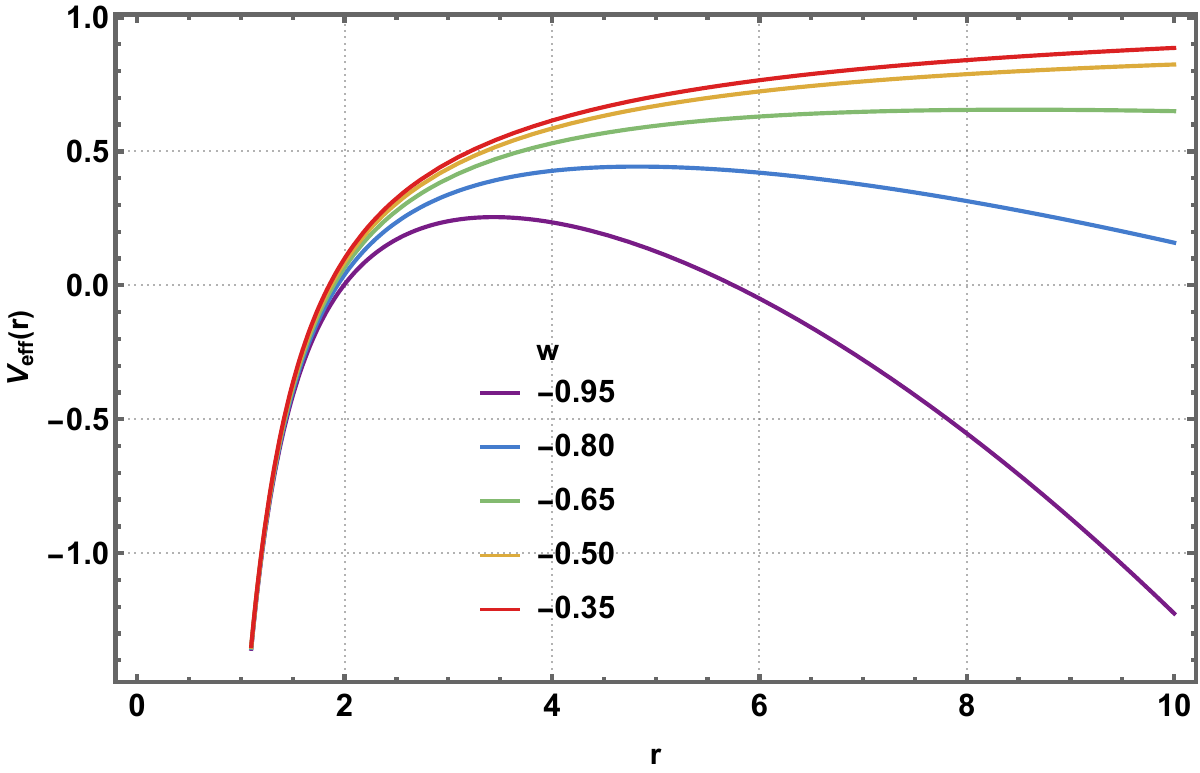}\\
(c) $\mathrm{C}=0.03,\,\eta=0.1$\\
\caption{\footnotesize Effective potential for massive particle dynamics showing parameter dependence on KRG field $\eta$ and QF constant $\mathrm{C}$. Parameters: $M=1$, $\mathrm{L}=1$.}
\label{fig:potential-2}
\end{figure}

Figure \ref{fig:potential-2} reveals the modified potential landscape governing massive particle motion. The enhanced potential barriers with increasing $\eta$ (left panel) and the reduced barriers with growing $\mathrm{C}$ (right panel) demonstrate how LV and QF effects oppositely influence particle binding and orbital characteristics, establishing a complex dynamical environment that significantly deviates from standard GR predictions.

The effective radial force on massive particles is:
\begin{equation}
\mathcal{F}_\text{eff}=-\frac{1}{2}\,\frac{dV_\text{eff}}{dr}=-\frac{1}{2\,r}\,\left(\frac{2\,M}{r}+\frac{\mathrm{C}\,(3\,w+1)}{r^{3\,w+1}}\right)+\frac{\mathrm{L}^2}{r^3}\,\left(\frac{1}{1-\eta}-\frac{3\,M}{r}-\frac{\mathrm{C}\,(3\,w+3)/2}{r^{3\,w+1}}\right).\label{d17}
\end{equation}

\begin{figure}[ht!]
\centering
\includegraphics[width=0.4\linewidth]{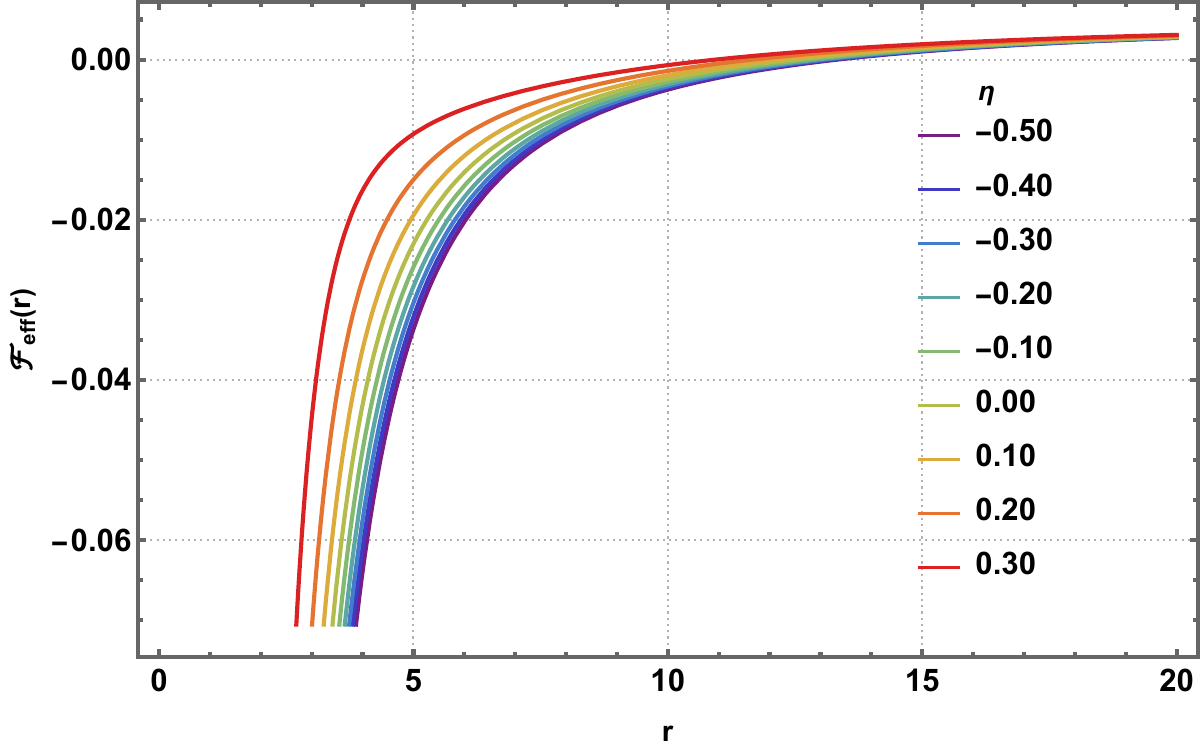}\quad\quad
\includegraphics[width=0.4\linewidth]{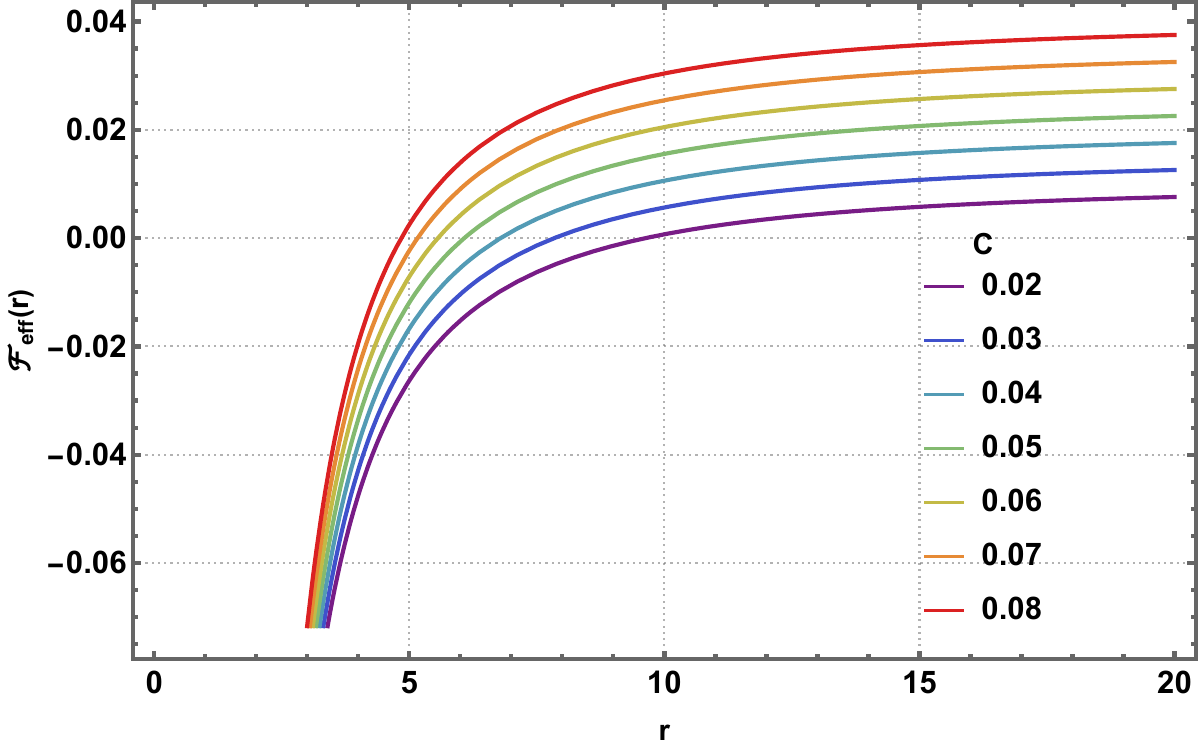}\\
(a) $\mathrm{C}=0.01,\,w=-2/3$ \hspace{6cm} (b) $\eta=0.1,\,w=-2/3$\\
\includegraphics[width=0.4\linewidth]{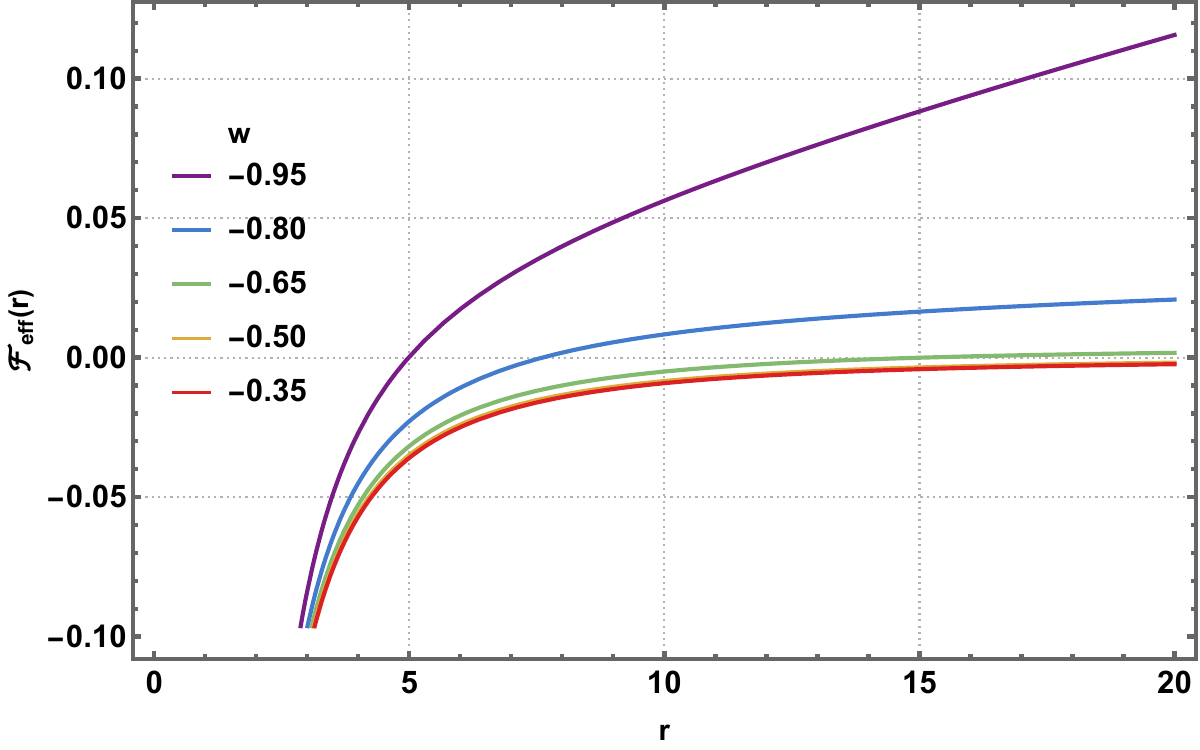}\\
(c) $\eta=0.1,\,\mathrm{C}=0.01$
\caption{\footnotesize Effective radial force on massive test particles demonstrating enhanced gravitational binding with increasing $\eta$ and $\mathrm{C}$. Parameters: $M=1$, $\mathrm{L}=2$.}
\label{fig:force-timelike}
\end{figure}

Figure \ref{fig:force-timelike} illustrates the systematic enhancement of gravitational binding forces as both $\eta$ and $\mathrm{C}$ increase. This behavior indicates that massive particles experience stronger confinement effects in the presence of both LV and quintessence modifications, suggesting potential observational consequences for particle acceleration processes and energy extraction mechanisms near BH horizons.

For circular orbits, the equilibrium conditions yield the specific energy and angular momentum:
\begin{equation}
\mathrm{L}_\text{sp}=r\,\sqrt{\frac{\left(\frac{M}{r} + \frac{\mathrm{C}\,(3\,w + 1)}{2}\,r^{-(3\,w+1)}\right)}{\frac{1}{1 - \eta} - \frac{3\,M}{r} -\frac{\mathrm{C}\,(3\,w + 3)}{2}\, r^{-(3w + 1)}}},\label{d3}
\end{equation}
\begin{equation}
\mathrm{E}_\text{sp}=\pm\,\frac{\left(\frac{1}{1 - \eta} - \frac{2\,M}{r} -\mathrm{C}\,r^{-(3w + 1)}\right)}{\sqrt{\frac{1}{1 - \eta} - \frac{3\,M}{r} -\frac{\mathrm{C}\,(3\,w + 3)}{2}\, r^{-(3w + 1)}}}.\label{d4}
\end{equation}

\begin{figure}[ht!]
\centering
\includegraphics[width=0.4\linewidth]{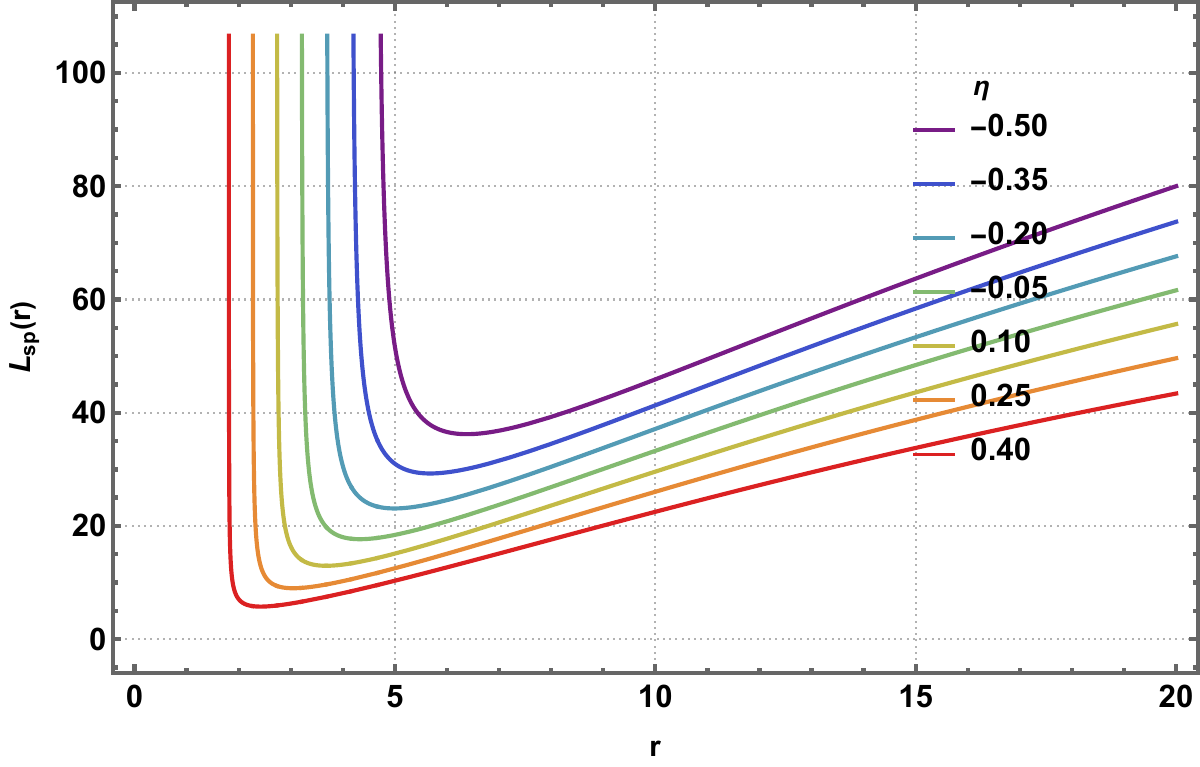}\qquad
\includegraphics[width=0.4\linewidth]{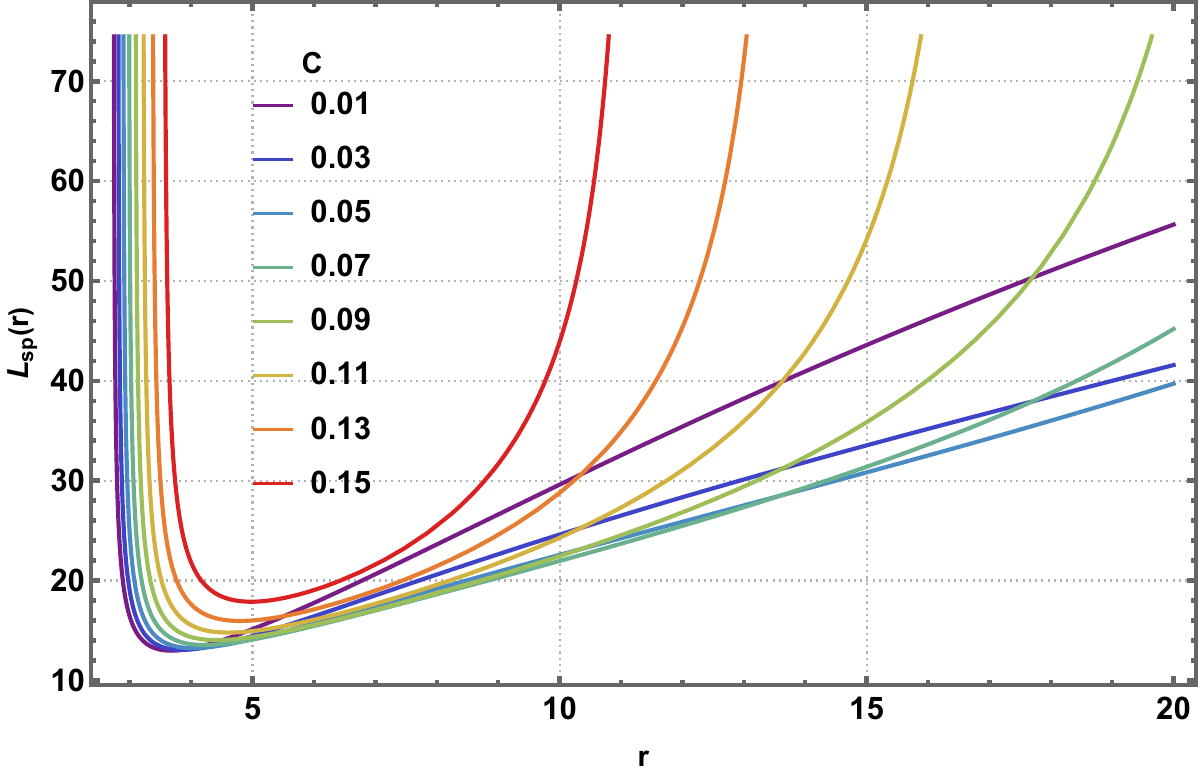}\\
(a) $\mathrm{C}=0.01,\,w=-2/3$ \hspace{4cm} (b) $\eta=0.1,w=-2/3$\\
\includegraphics[width=0.4\linewidth]{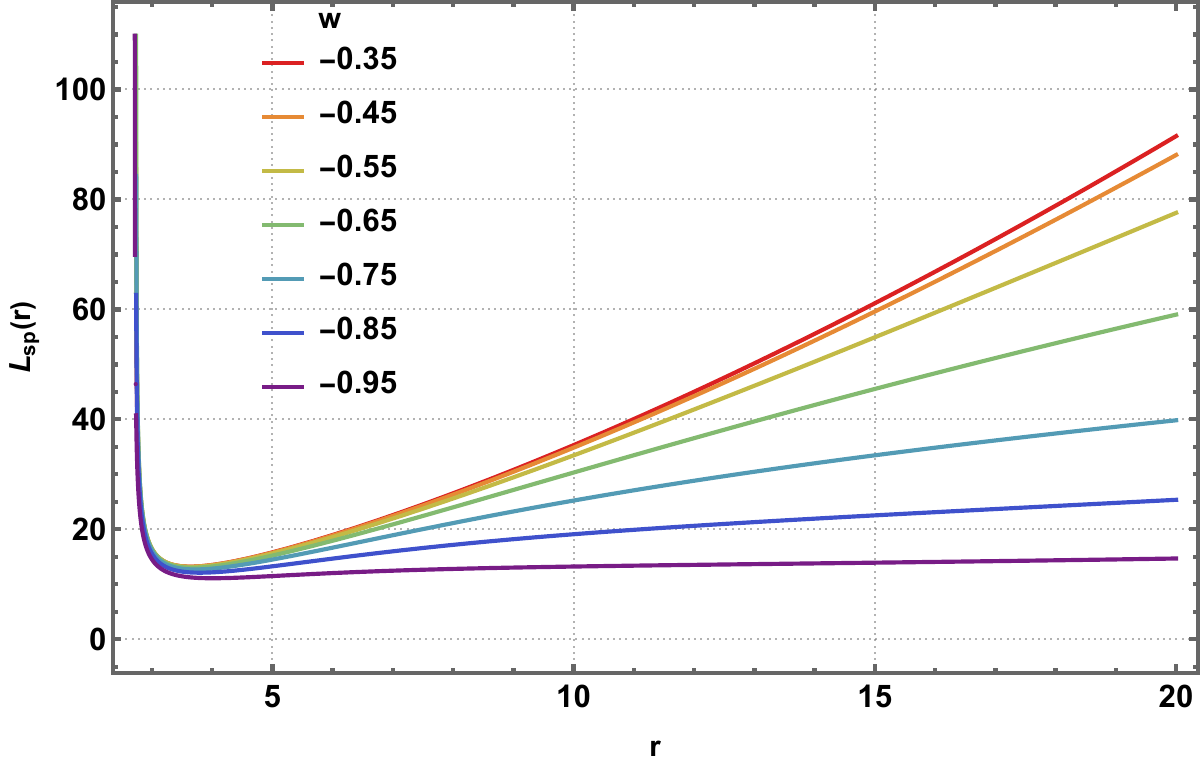}\\
(c) $\eta=0.1,\mathrm{C}=0.01$
\caption{\footnotesize Specific angular momentum $\mathrm{L}_\text{sp}$ evolution with radius for various parameter combinations. Here $M=1$.}
\label{fig:angular}
\end{figure}

\begin{figure}[ht!]
\centering
\includegraphics[width=0.4\linewidth]{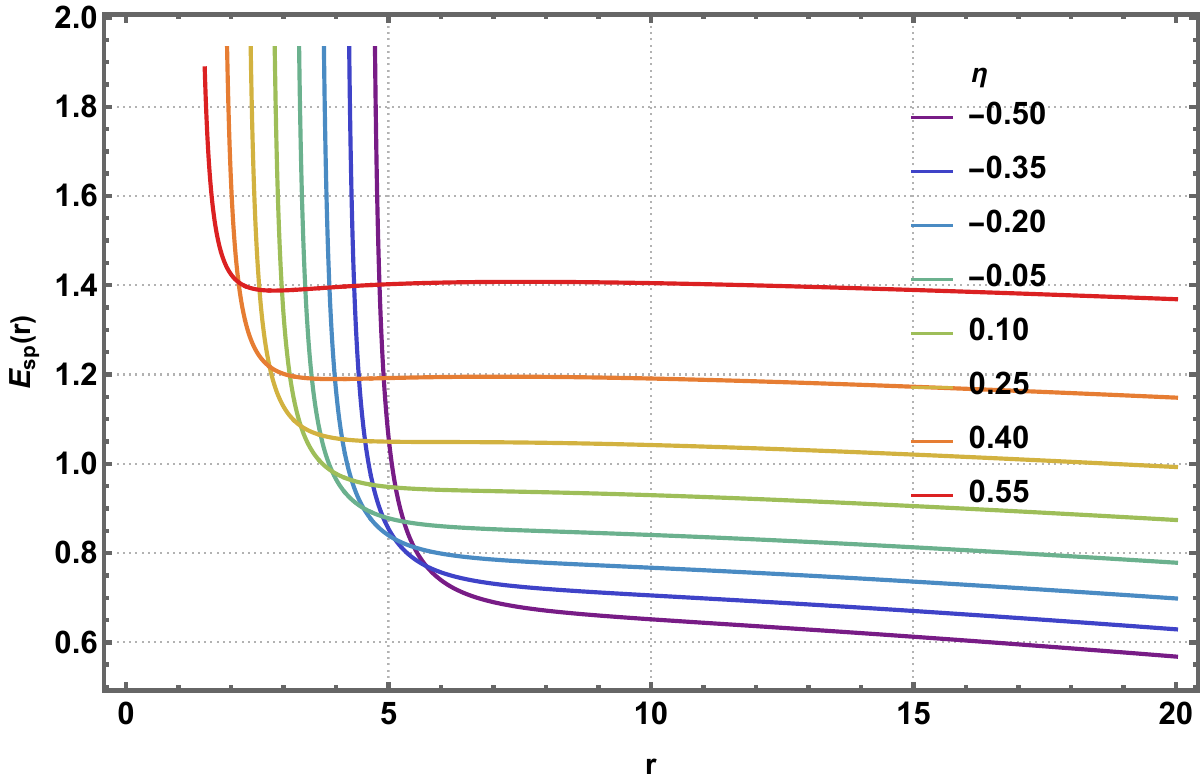}\qquad
\includegraphics[width=0.4\linewidth]{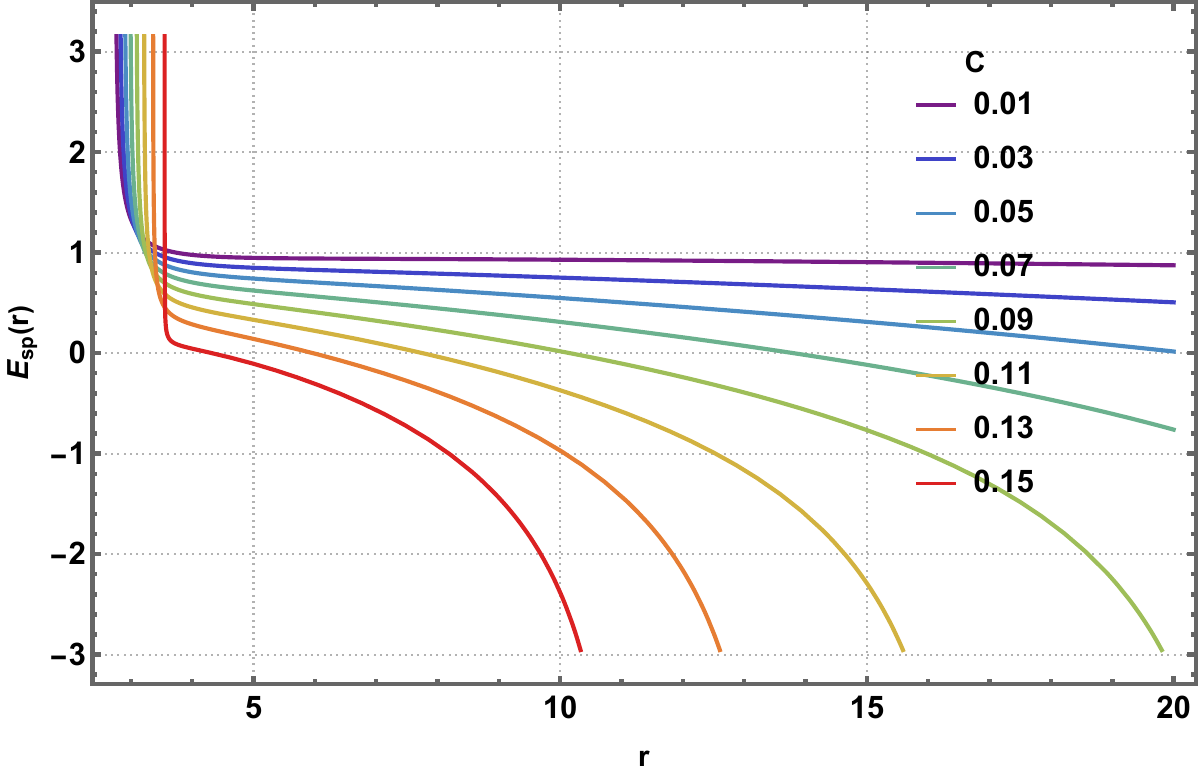}\\
(a) $\mathrm{C}=0.01,\,w=-2/3$ \hspace{4cm} (b) $\eta=0.1,w=-2/3$\\
\includegraphics[width=0.4\linewidth]{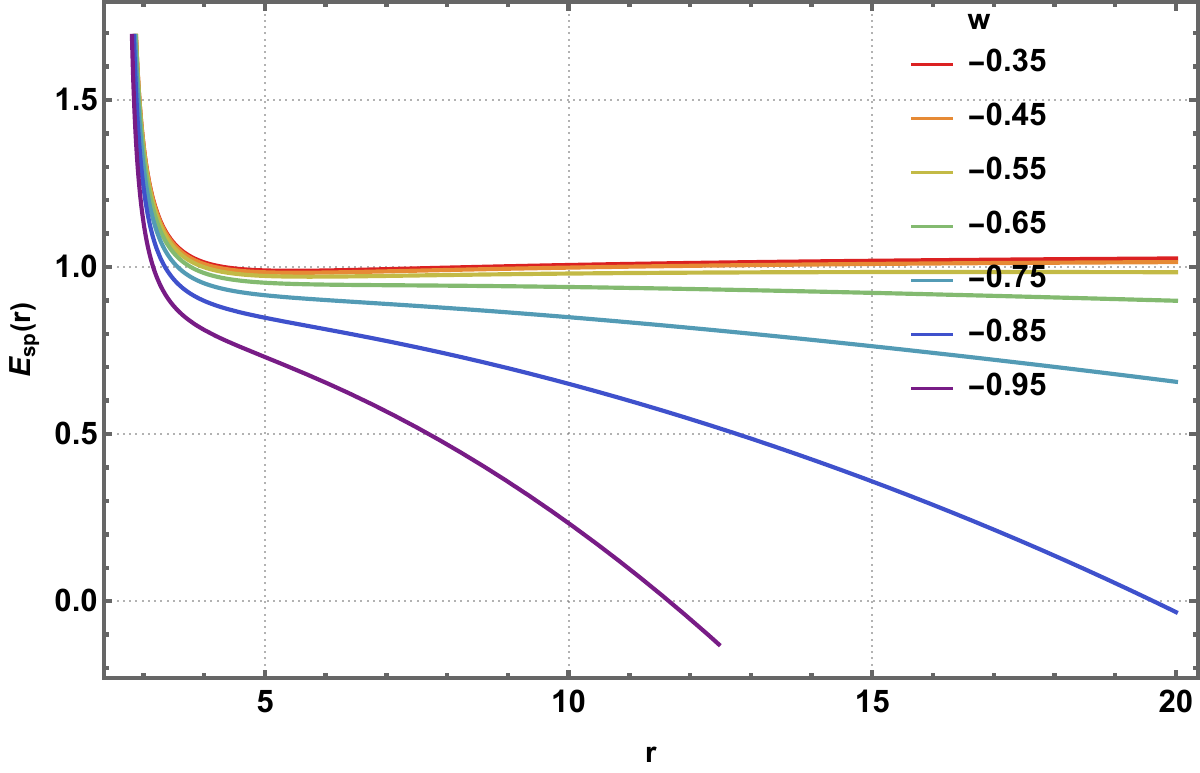}\\
(c) $\eta=0.1,\mathrm{C}=0.01$
\caption{\footnotesize Specific energy $\mathrm{E}_\text{sp}$ profiles demonstrating parameter-dependent modifications from GR predictions. Here $M=1$.}
\label{fig:energy}
\end{figure}

Figures \ref{fig:angular} and \ref{fig:energy} provide comprehensive visualization of how KRG-QF modifications systematically alter the fundamental orbital characteristics. The specific angular momentum profiles in Figure \ref{fig:angular} show pronounced parameter dependence, with varying $\eta$, $\mathrm{C}$, and $w$ values producing distinct evolutionary patterns that deviate significantly from standard Schwarzschild behavior. Similarly, Figure \ref{fig:energy} demonstrates how the specific energy requirements for circular orbits are modified by both LV and quintessence effects, establishing clear observational signatures for distinguishing between GR and KRG-QF scenarios.

The ISCO analysis requires satisfaction of three conditions: energy conservation, extremal condition, and marginal stability:
\begin{align}
\mathrm{E}^2&=V_\text{eff},\label{d10}\\
\frac{dV_{\text{eff}}}{dr}&= 0 \quad \text{(extremal condition)},\label{d11}\\ 
\frac{d^2V_{\text{eff}}}{dr^2}& \geq  0 \quad \text{(marginal stability)}.\label{d12}
\end{align}

The marginal stability condition yields:
\begin{align}
\left(\frac{3}{1 - \eta} - \frac{10M}{r} - \frac{C(6w + 5)}{r^{3w + 1}}\right)\,\left(\frac{2\,M}{r^2}+\frac{\mathrm{C}\,(3\,w+1)}{r^{3\,w+2}}\right)
+r\,\left(\frac{1}{1-\eta}-\frac{2\,M}{r}-\frac{\mathrm{C}}{r^{3\,w+1}}\right)\,\left(-\frac{4\,M}{r^3}-\frac{\mathrm{C}\,(3\,w+1)\,(3\,w+2)}{r^{3\,w+3}}\right)=0.\label{d13}
\end{align}

For $w=-2/3$, this reduces to a fourth-order polynomial:
\begin{align}
\mathrm{C}^2\, r^4 - \frac{3\,\mathrm{C}}{1 - \eta}\, r^3 + 12\,M\,\mathrm{C}\, r^2 + \frac{2\,M}{1 - \eta} r - 12\,M^2 = 0.\label{d14}
\end{align}

\begin{figure}[ht!]
\centering
\includegraphics[scale=1]{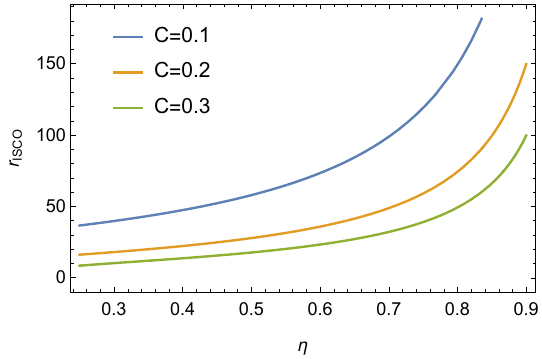}
\caption{\footnotesize ISCO radius variation with QF parameter $\mathrm{C}$ for different LV parameter $\eta$ values, demonstrating systematic modifications from GR predictions. Here, $M=1$.}
\label{figa92}
\end{figure}

Figure \ref{figa92} provides crucial insights into ISCO behavior in the KRG-QF framework. The systematic increase of ISCO radius with both $\mathrm{C}$ and $|\eta|$ demonstrates how modifications from GR lead to larger innermost stable orbits, potentially affecting accretion disk inner edges and gravitational wave emission characteristics from inspiraling compact objects.

\begin{table}[H]
\centering
\scalebox{1.25}{ 
\begin{tabular}{|c||c|c|c|c|c|c|}
\hline
$\eta$ & $C = 0.00$ & $C = 0.01$ & $C = 0.02$ & $C = 0.03$ & $C = 0.04$ & $C = 0.05$ \\
\hline
-0.4 & 8.4000 & 208.206 & 100.873 & 64.9415 & 46.8314 & 35.8150 \\
\hline
-0.3 & 7.8000 & 225.146 & 109.614 & 70.9862 & 51.5678 & 39.8136 \\
\hline
-0.2 & 7.2000 & 244.829 & 119.715 & 77.9233 & 56.9518 & 44.2976 \\
\hline
-0.1 & 6.6000 & 268.003 & 131.553 & 86.0061 & 63.1784 & 49.4325 \\
\hline
 0.0 & 6.0000 & 295.718 & 145.655 & 95.5874 & 70.5156 & 55.4388 \\
 \hline
 0.1 & 5.4000 & 329.490 & 162.778 & 107.175 & 79.3469 & 62.6274 \\
 \hline
 0.2 & 4.8000 & 371.592 & 184.061 & 121.528 & 90.2440 & 71.4587 \\
 \hline
 0.3 & 4.2000 & 425.596 & 211.289 & 139.839 & 104.103 & 82.6517 \\
 \hline
 0.4 & 3.6000 & 497.454 & 247.441 & 164.095 & 122.414 & 97.4006 \\
\hline
\end{tabular}
}
\caption{\footnotesize Numerical ISCO radius values $r_\text{ISCO}$ demonstrating systematic parameter dependence on both LV and QF effects, with $M = 1$.}
\label{tab:ISCO}
\end{table}

Table \ref{tab:ISCO} provides comprehensive numerical documentation of ISCO radius variations across the parameter space. The dramatic increase from the GR value of $6M$ (at $\eta=0$, $C=0$) to values exceeding $400M$ for moderate parameter combinations underscores the profound impact of KRG-QF modifications on orbital dynamics. This systematic parameter dependence offers potential observational discrimination between different theoretical frameworks through precise measurements of accretion disk properties and gravitational wave signals from compact binary systems.

\section{BH Shadow Phenomenology  in KRG-QF Framework}\label{isec4}

The BH shadow represents one of the most direct observational probes of strong gravitational fields and spacetime geometry near event horizons. This dark, approximately circular region emerges when a BH is surrounded by luminous matter, typically from an accretion disk, where photons are gravitationally captured or severely deflected by the extreme curvature. The shadow boundary is fundamentally determined by the photon sphere radius and critical impact parameters, making it distinct from but closely related to the event horizon itself \cite{sec2is14}. The landmark achievement of the EHT in 2019, capturing the first direct image of the supermassive BH in galaxy M87, marked a revolutionary moment in observational astronomy and provided unprecedented tests of GR under extreme conditions \cite{sec2is15}. These observations established BH shadow studies as powerful tools for measuring fundamental BH properties, investigating near-horizon physics, and discriminating between alternative gravitational theories \cite{sec2is16}.

In the context of our KRG-QF framework, the shadow characteristics are profoundly modified by both LV effects through the parameter $\eta$ and QF contributions characterized by $(\mathrm{C}, w)$. For a static spherically symmetric metric, the photon sphere radius $r_{\text{ph}}$ is determined by the condition:
\begin{equation}
r\,D'(r,\eta)=D(r,\eta),
\end{equation}
where $D=\sqrt{A(r,\eta)}$ encodes the metric modifications.

Applying this condition to our KRG-QF metric function from Eq.~(\ref{bb2}), the photon sphere equation becomes:
\begin{equation}
6\,M(\eta-1)+2 \,r+3\,\mathrm{C}\,r^{-3w}(1+w)(\eta-1)=0. \label{eps1}
\end{equation}

This equation demonstrates how both LV and QF parameters fundamentally alter the photon sphere location compared to the standard Schwarzschild result $r_{\text{ph}}=3M$. The analytical solutions depend critically on the QF state parameter $w$, as detailed in Table \ref{table:2}.

\begin{table}[ht!]
\centering
\scalebox{1}{ 
\begin{tabular}{|c|c|c|}
\hline 
$w$ & Photon Sphere Equation & $r_\text{ph}$ \\ 
\hline
$-1/3$ & $6\,M(\eta-1)+(2+2\,\mathrm{C}\,(\eta-1))r=0$ & $\dfrac{3M(1-\eta)}{1+\mathrm{C}(\eta-1)}$ \\
\hline
$-2/3$ & $6\,M(\eta-1)+2 \,r+\mathrm{C}\,r^{2}(\eta-1)=0$ & $\dfrac{1-\sqrt{1-6\,\mathrm{C}(\eta-1)^2M}}{C}$ \\
\hline
$-1$ & $3\,M(\eta-1)+r=0$ & $3(1-\eta)M$ \\ 
\hline
\end{tabular}
}
\caption{\footnotesize Analytical expressions for photon sphere radius across different QF state parameters, demonstrating the rich parameter dependence in KRG-QF spacetimes.}
\label{table:2}
\end{table}

Table \ref{table:2} reveals the sophisticated dependence of photon sphere characteristics on QF properties. The case $w=-1/3$ represents intermediate dark energy behavior with algebraic radius modifications, while $w=-2/3$ exhibits more complex square-root dependencies characteristic of phantom-like QF regimes. The cosmological constant limit $w=-1$ provides the simplest linear relationship, yet still incorporates significant deviations from GR through the LV parameter $\eta$.

The observable shadow radius $R_s$ for distant observers is computed using the critical impact parameter:
\begin{equation}
R_s=\beta_c=\frac{r_\text{ph}}{\sqrt{\frac{1}{1-\eta}-\frac{2\,M}{r_\text{ph}}-\frac{\mathrm{C}}{r_\text{ph}^{3\,w+1}}}}.\label{shadeq1}
\end{equation}

The analytical shadow radius expressions for different QF states are presented in Table \ref{table:3}.

\begin{table}[ht!]
\centering
\scalebox{1.2}{ 
\begin{tabular}{|c|c|c|}
\hline 
$w$ & Shadow Radius Equation & $R_{s}$ \\ \hline
$-1/3$ & $\frac{r_\text{ph}}{\sqrt{\frac{1}{1-\eta}-\frac{2\,M}{r_\text{ph}}-\mathrm{C}}}$ & $M\left(\frac{3(1-\eta)}{1+C(\eta-1)} \right)^{3/2}$ \\
\hline
$-2/3$ & $\frac{r_\text{ph}}{\sqrt{\frac{1}{1-\eta}-\frac{2\,M}{r_\text{ph}}-C\,r_\text{ph}}}$ & $ \frac{3\sqrt{3}(1-\eta)M}{\left(1+\mathrm{C}(\eta-1)\right)\sqrt{\frac{1}{1-\eta}+\mathrm{C}\left( 2+\frac{9(\eta-1)M}{1+\mathrm{C}(\eta-1)} \right)}}$ \\
\hline 
$-1$ & $\frac{r_\text{ph}}{\sqrt{\frac{1}{1-\eta}-\frac{2\,M}{r_\text{ph}}-\mathrm{C}\,r^2_\text{ph}}}$ & $\frac{3\sqrt{3}(1-\eta)M}{\left(1+\mathrm{C}(\eta-1)\right)\sqrt{\frac{1}{1-\eta}+\mathrm{C}\left( 2-\frac{27(\eta-1)^2M^2}{(1+\mathrm{C}(\eta-1))^2} \right)}}$ \\ 
\hline
\end{tabular}
}
\caption{\footnotesize Comprehensive shadow radius expressions demonstrating complex parameter interdependencies in KRG-QF modified spacetimes.} 
\label{table:3}
\end{table}

Table \ref{table:3} establishes the intricate mathematical structure governing shadow observables in modified gravity. The limiting case $\mathrm{C}=0$ recovers the pure KRG result $R_s=3\sqrt{3}M\sqrt{1-\eta}$ \cite{sec2is17}, while the combined KRG-QF effects produce significantly more complex expressions that encode rich observational signatures.

\begin{figure}[ht!]
\centering
\includegraphics[width=0.4\linewidth]{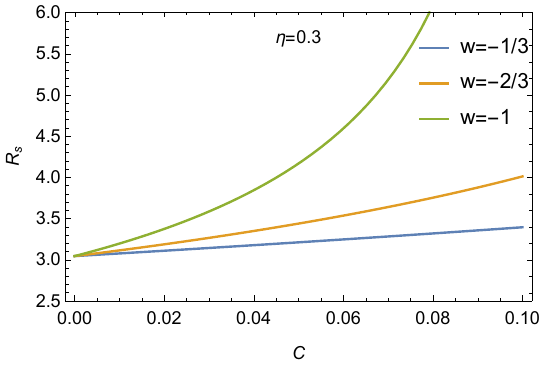}\qquad
\includegraphics[width=0.4\linewidth]{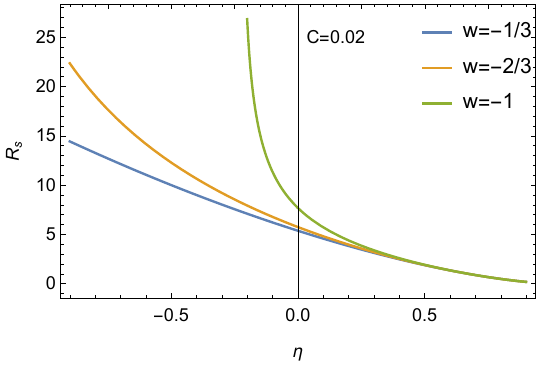}
\caption{\footnotesize Shadow radius evolution demonstrating QF state parameter effects (left panel) and LV parameter influence (right panel) on observable shadow characteristics. Parameters: $M=1$, $b=0.3$.}
\label{figa7}
\end{figure}

Figure \ref{figa7} provides crucial insights into the observational discrimination potential between different theoretical frameworks. The left panel demonstrates how varying QF state parameters $w$ systematically modify shadow radius scaling with the normalization constant $\mathrm{C}$, while the right panel reveals the universal suppression of shadow size with increasing LV parameter $\eta$ across all QF states. These complementary trends establish clear observational signatures for distinguishing KRG-QF models from standard GR and constraining theory parameters through precision shadow measurements \cite{sec2is18}.

\begin{figure}[ht!]
\centering
\includegraphics[scale=0.56]{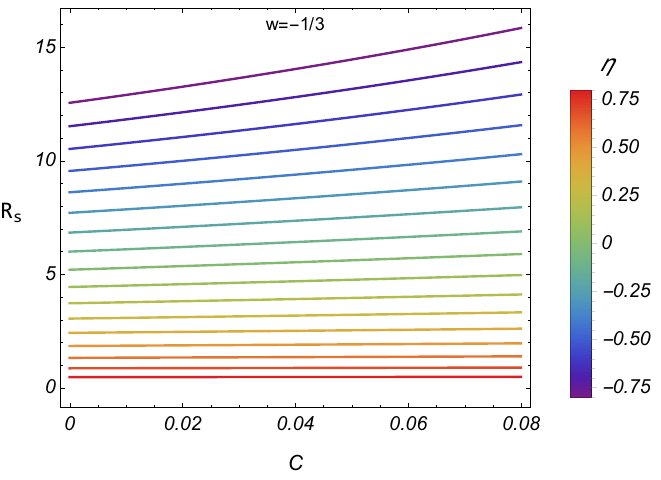}
\includegraphics[scale=0.56]{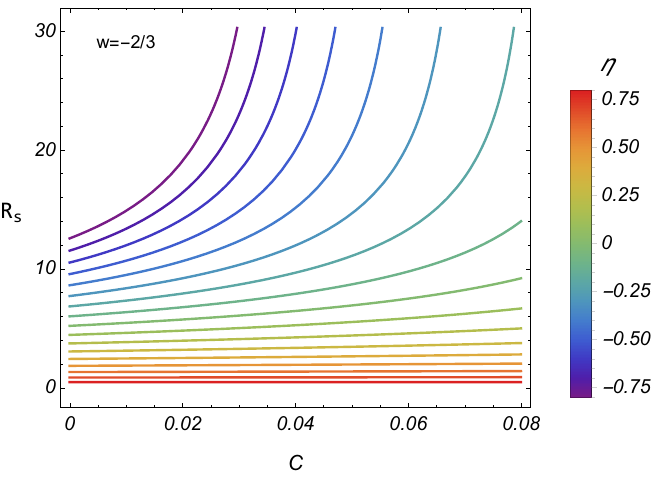}
\includegraphics[scale=0.56]{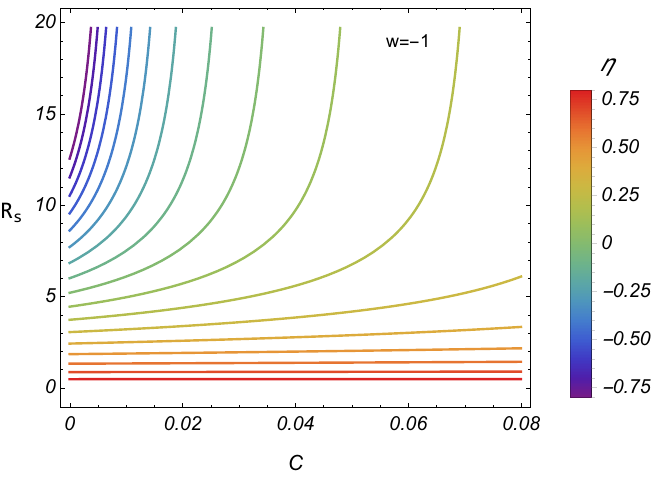}
\caption{\footnotesize Comprehensive parameter space exploration showing shadow radius dependence on QF normalization $\mathrm{C}$ across different QF state parameters $w$ and LV strengths $\eta$. Here, $M=1$.}
\label{figa2}
\end{figure}

Figure \ref{figa2} presents a systematic parameter space analysis across three representative QF regimes. The progression from $w=-1/3$ through $w=-2/3$ to $w=-1$ reveals increasingly pronounced parameter sensitivities and distinct scaling behaviors. The systematic enhancement of shadow radius with increasing $\mathrm{C}$ and the suppression with growing $|\eta|$ provide robust observational handles for constraining both LV and dark energy effects simultaneously \cite{sec2is19}.

\begin{table}[ht!]
\centering
\scalebox{1.2}{ 
\begin{tabular}{|c|c|ccc|ccc|ccc|}\hline \multicolumn{11}{|c|}{Shadow radius $R_s$}\\
\hline
& &\multicolumn{3}{|c|}{$w=-1/3$} &\multicolumn{3}{|c|}{$w=-2/3$}&\multicolumn{3}{|c|}{$w=-1$} \\ \hline 
$\mathrm{C}$& $\eta=0.0$ & $0.3$ & $0.6$ & $0.9$ & $0.3$ & $0.6$ & $0.9$ & $0.3$ & $0.6$ & $0.9$ \\ \hline
$0$ & $3\sqrt{3}$ & $3.04319$ & $1.31453$ & $0.164317$ & $3.04319$ & $1.31453$ & $0.164317$ & $3.04319$ & $1.31453$ & $0.164317$ \\ 
\hline
$0.02$ & $5.72976$ & $3.37259$ & $1.33799$ & $0.164362$ & $3.18584$ & $1.33386$ & $0.164466$ & $3.10823$ & $1.33047$ & $0.164811$ \\ 
\hline
$0.06$ & $7.48418$ & $4.59198$ & $1.3898$ & $0.164459$ & $3.53437$ & $1.37501$ & $0.164768$ & $3.24549$ & $1.36332$ & $0.165807$ \\ 
\hline
$0.1$ & $12.9099$ & $12.4356$ & $1.44943$ & $0.164564$ & $4.01026$ & $1.4199$ & $0.165079$ & $3.39316$ & $1.39754$ & $0.166813$\\
\hline
\end{tabular}
}
\caption{\footnotesize Numerical shadow radius values across the KRG-QF parameter space, providing quantitative benchmarks for observational constraints and model discrimination.} \label{table1a}
\end{table}

Table \ref{table1a} provides comprehensive numerical documentation of shadow radius variations, revealing dramatic parameter sensitivities that exceed factor-of-ten modifications from GR predictions. The systematic trends enable precise observational constraints on both LV and QF parameters through future high-resolution BH imaging campaigns \cite{sec2is20}.

To characterize the complete shadow morphology, we introduce celestial coordinates $(X,Y)$ representing the BH shadow as observed by distant observers:
\begin{equation}
X=\lim_{r_{\mathrm{o}}\rightarrow \infty }\left( -r_{\mathrm{o}}^{2}\sin\theta _{\mathrm{o}}\frac{d\varphi }{dr}\right) ,\quad Y=\lim_{r_{\mathrm{o}}\rightarrow \infty }\left( r_{\mathrm{o}}^{2}\frac{d\theta }{dr}\right).
\end{equation}

For equatorial observers at large distances, these coordinates yield the circular shadow relation:
\begin{equation}
X^{2}+Y^{2}=R_{s}^{2}.
\end{equation}

\begin{figure}[ht!]
\begin{center}
\includegraphics[scale=0.48]{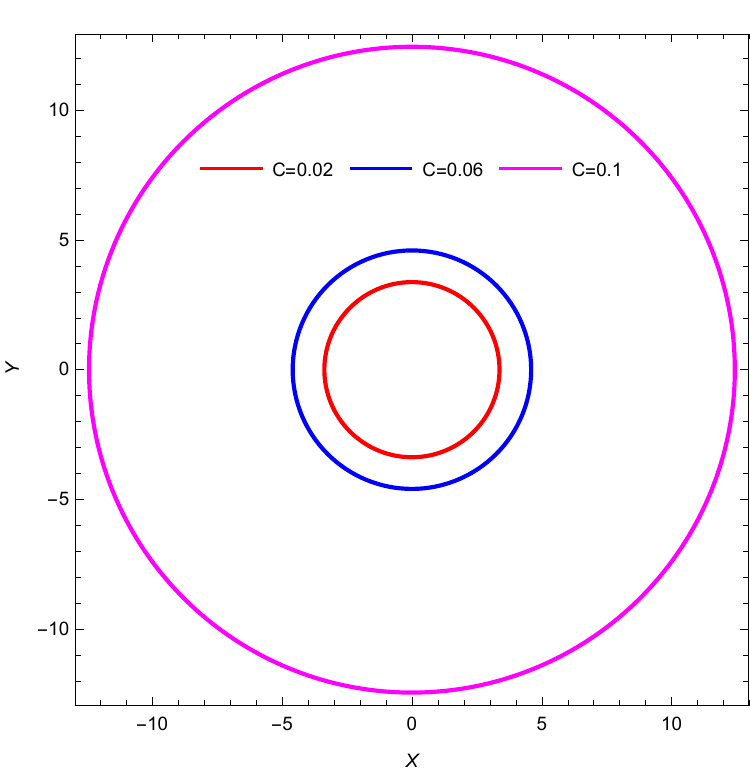}\quad
\includegraphics[scale=0.48]{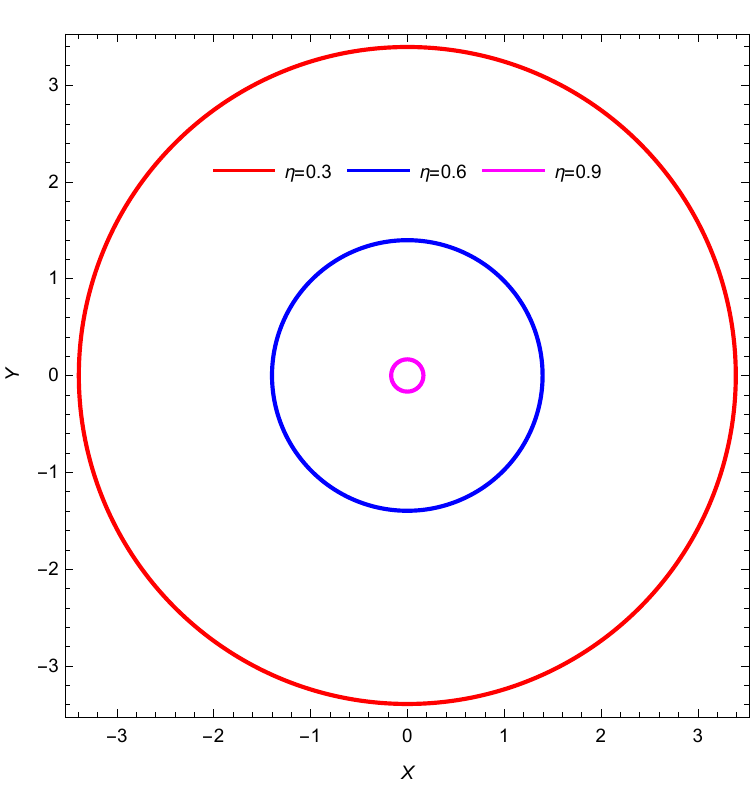}\quad
\includegraphics[scale=0.48]{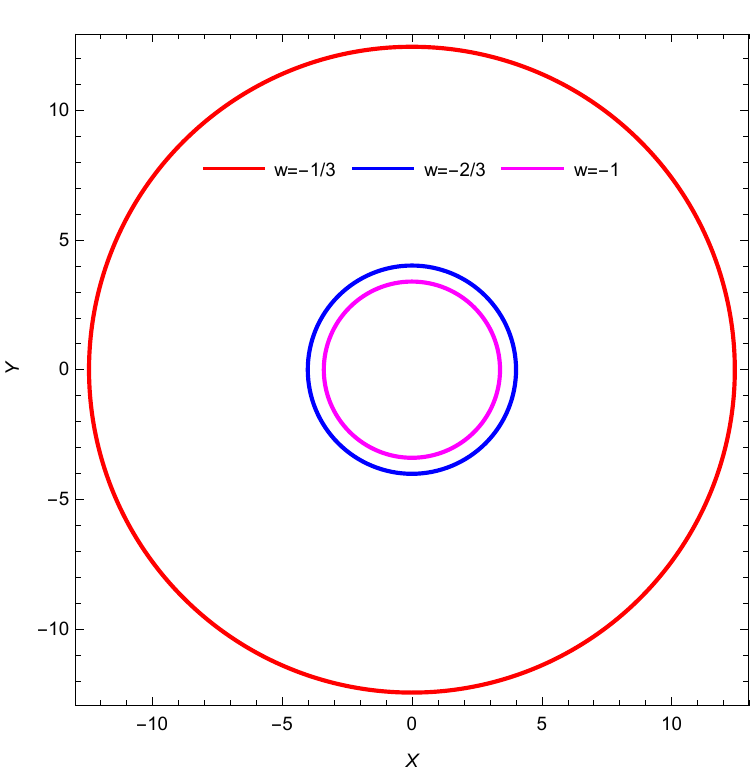}
\end{center}
\caption{\footnotesize Shadow morphology evolution demonstrating parameter-dependent modifications in celestial coordinates: QF normalization effects (left), LV parameter influence (middle), and QF state parameter variations (right).}\label{figa21}
\end{figure}

Figure \ref{figa21} provides direct visualization of shadow morphology modifications across the parameter space. The systematic expansion with increasing $\mathrm{C}$ (left panel), contraction with growing $\eta$ (middle panel), and state-dependent scaling with varying $w$ (right panel) establish clear observational signatures that could be detected by next-generation interferometric arrays and space-based telescopes \cite{sec2is21}.

\section{Scalar Field Perturbations in KRG-QF BH Spacetime}\label{isec5}

Scalar field perturbations constitute a fundamental probe for investigating the dynamical stability and wave propagation characteristics of BH spacetimes under small field fluctuations. These perturbations have been extensively studied across diverse BH solutions in both GR and modified gravity theories, providing crucial insights into stability properties, QNM spectra, and scalar field dynamics in curved geometries \cite{sec2is22,sec2is23,sec2is24,sec2is25,sec2is26}. In the context of our KRG-QF framework, scalar perturbations offer a particularly sensitive diagnostic tool for understanding how LV effects and quintessence matter collectively influence wave propagation and stability characteristics near BH horizons.

This section presents a comprehensive investigation of massless scalar field perturbations in the KRG-QF BH spacetime defined by Eq.~(\ref{bb2}). We employ standard BH perturbation theory, beginning with the Klein-Gordon equation for massless scalar fields and applying established techniques to derive the corresponding Schrödinger-like wave equation. The resulting effective potential encodes the combined influence of LV through parameter $\eta$ and QF characteristics through parameters $(\mathrm{C}, w)$, enabling systematic analysis of their impact on scalar field dynamics and spacetime stability properties \cite{sec2is27}.

The massless scalar field dynamics are governed by the Klein-Gordon equation:
\begin{equation}
\frac{1}{\sqrt{-g}}\,\partial_{\mu}\left(\sqrt{-g}\,g^{\mu\nu}\,\partial_{\nu}\Psi\right)=0\quad\quad (\mu,\nu=0,\cdots,3), \label{ff1}    
\end{equation}
where $\Psi$ represents the scalar field wave function, $g_{\mu\nu}$ denotes the covariant metric tensor, $g=\det(g_{\mu\nu})$ is the metric determinant, $g^{\mu\nu}$ corresponds to the contravariant metric components, and $\partial_{\mu}$ represents coordinate partial derivatives.

For our KRG-QF spacetime geometry given in Eq.~(\ref{aa1}), the metric components are:
\begin{align}
g_{\mu\nu}=\left(-A,\,A^{-1},\,r^2,\,r^2\,\sin^2 \theta\right),\quad
g^{\mu\nu}=\left(-1/A,\,A,\,1/r^{2},\,1/(r^{2}\,\sin^{2} \theta)\right),\quad 
g=\det (g_{\mu\nu})=-r^4\,\sin^2 \theta.\label{ff2}
\end{align}

We employ the standard separable ansatz for the scalar field wave function:
\begin{equation}
\Psi(t, r,\theta, \phi)=\exp(-i\,\omega\,t)\,Y^{m}_{\ell} (\theta,\phi)\,\frac{\psi(r)}{r},\label{ff3}
\end{equation}
where $\omega$ represents the (possibly complex) temporal frequency characterizing QNM behavior, $\psi(r)$ denotes the radial wave function, and $Y^{m}_{\ell}(\theta,\phi)$ are the spherical harmonics satisfying the eigenvalue equation:
\begin{equation}
\left[ \frac{1}{\sin\theta} \frac{\partial}{\partial\theta} \left( \sin\theta \frac{\partial}{\partial\theta} \right) + \frac{1}{\sin^2\theta} \frac{\partial^2}{\partial\phi^2} \right] Y_{\ell m}(\theta, \phi) = -\ell(\ell+1) Y_{\ell m}(\theta, \phi).\label{ff4}
\end{equation}

Substituting the separable ansatz into the Klein-Gordon equation yields the radial wave equation:
\begin{equation}
A\, \psi''(r) + A'\, \psi'(r) + \left( \frac{\omega^2}{A} - \frac{A'}{r} - \frac{\ell\,(\ell + 1)}{r^2} \right)\, \psi(r) = 0,\label{ff5}
\end{equation}
where the prime denotes differentiation with respect to the radial coordinate $r$.

To eliminate the first-order derivative term and obtain a Schrödinger-like wave equation, we implement the tortoise coordinate transformation:
\begin{eqnarray}
r_*=\int\,\frac{dr}{A}\quad,\quad \partial_{r_{*}}=A\,\partial_r\quad,\quad \partial^2_{r_{*}}=A^2\,\partial^2_r+A\,A'\,\partial_r.\label{ff6}
\end{eqnarray}

This coordinate transformation converts the radial equation into the standard Schrödinger form:
\begin{equation}
\frac{\partial^2 \psi(r_*)}{\partial r^2_{*}}+\left(\omega^2-V_\text{scalar}\right)\,\psi(r_*)=0,\label{ff7}
\end{equation}
where the effective scalar perturbation potential is given by:
\begin{eqnarray}
V_\text{scalar}(r)&=&\left(\frac{\ell\,(\ell+1)}{r^2}+\frac{A'}{r}\right)\,A\nonumber\\
&=&\left(\frac{1}{1-\eta}-\frac{2\,M}{r}-\frac{\mathrm{C}}{r^{3\,w+1}}\right)\,\left\{\frac{\ell\,(\ell+1)}{r^2}+\frac{2\,M}{r^3} + \mathrm{C}\,(3\,w + 1)\, r^{-(3w + 3)}\right\},\quad \ell\geq 0.\label{ff8}
\end{eqnarray}

Equation (\ref{ff8}) represents the fundamental scalar perturbation potential for our KRG-QF BH system. This expression explicitly demonstrates how both LV effects through parameter $\eta$ and QF contributions via parameters $(\mathrm{C}, w)$ systematically modify the potential landscape compared to standard Schwarzschild geometry. The potential structure reveals the intricate interplay between modified gravitational dynamics and exotic matter effects in determining wave propagation characteristics \cite{sec2is28}.

For the specific case of phantom-like quintessence with $w=-2/3$, the potential simplifies to:
\begin{eqnarray}
V_\text{scalar}(r)=\left(\frac{1}{1-\eta}-\frac{2\,M}{r}-\mathrm{C}\,r\right)\,\left\{\frac{\ell\,(\ell+1)}{r^2}+\frac{2\,M}{r^3} - \frac{\mathrm{C}}{r}\right\},\quad \ell\geq 0.\label{ff9}
\end{eqnarray}

Introducing dimensionless variables $x=r/M$ and $y=M\,\mathrm{C}$, we obtain the normalized potential:
\begin{eqnarray}
M^2\,V_\text{scalar}=\left(\frac{1}{1-\eta}-\frac{2}{x}-x\,y\right)\,\left\{\frac{\ell\,(\ell+1)}{x^2}+\frac{2}{x^3} - \frac{y}{x}\right\}.\label{ff10}
\end{eqnarray}

\begin{figure}[ht!]
\centering
\includegraphics[width=0.45\linewidth]{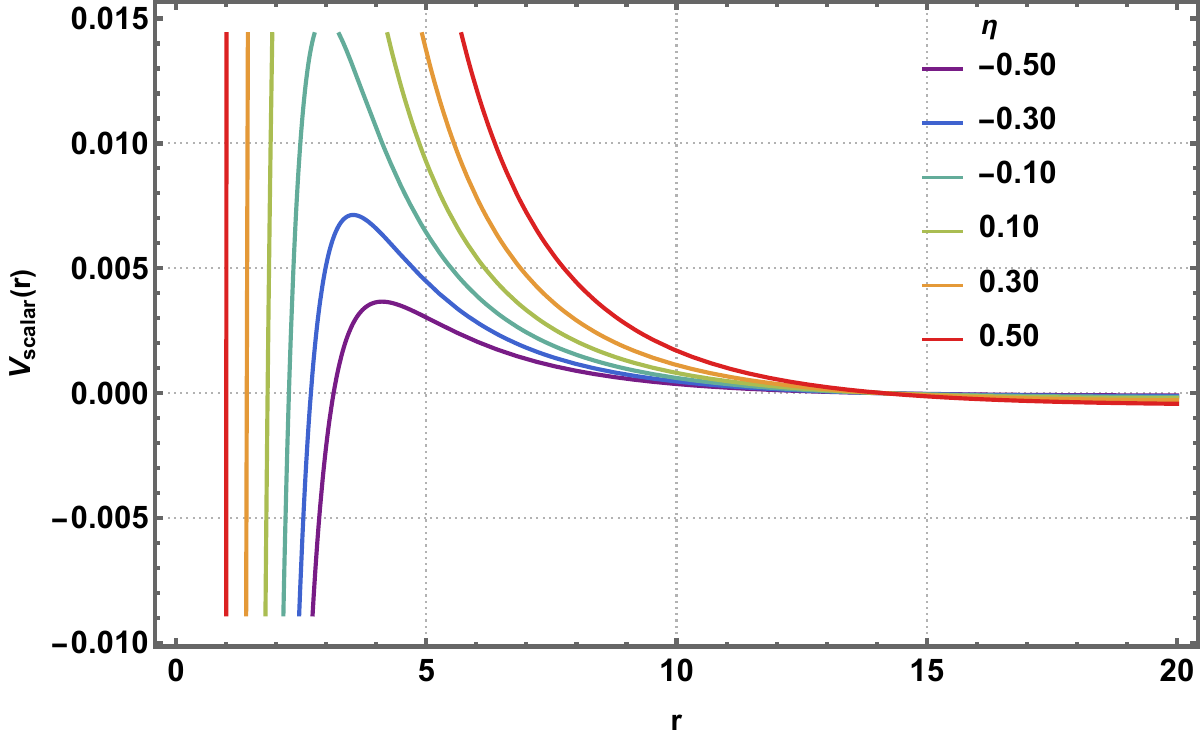}\quad\quad
\includegraphics[width=0.45\linewidth]{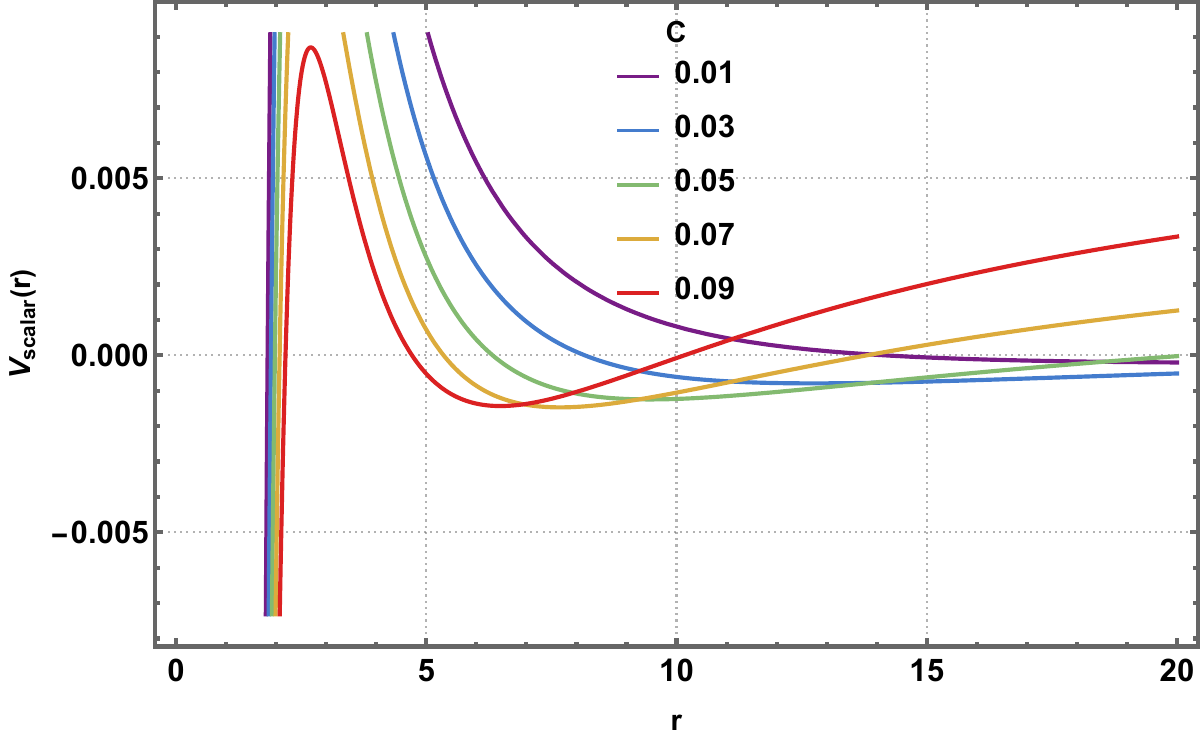}\\
(a) $\mathrm{C}=0.01$ \hspace{6cm} (b) $\eta=0.1$
\caption{\footnotesize Scalar perturbation potential $V_\text{scalar}(r)$ evolution for monopole mode $\ell=0$, demonstrating systematic parameter dependence on LV strength $\eta$ and QF normalization $\mathrm{C}$. Parameters: $M=1$, $w=-2/3$.}
\label{fig:scalar-potential}
\end{figure}

Figure \ref{fig:scalar-potential} provides fundamental insights into how KRG-QF modifications alter scalar wave propagation characteristics. The left panel demonstrates that increasing LV parameter $\eta$ systematically enhances the potential magnitude, indicating stronger wave scattering and modified QNM frequencies. Conversely, the right panel reveals that increasing QF parameter $\mathrm{C}$ reduces the potential strength, reflecting the influence of quintessence matter on wave dynamics. These opposing trends establish the complex parameter-dependent landscape governing scalar perturbation behavior in modified gravity theories \cite{sec2is29}.

\begin{figure}[ht!]
\centering
\includegraphics[width=0.4\linewidth]{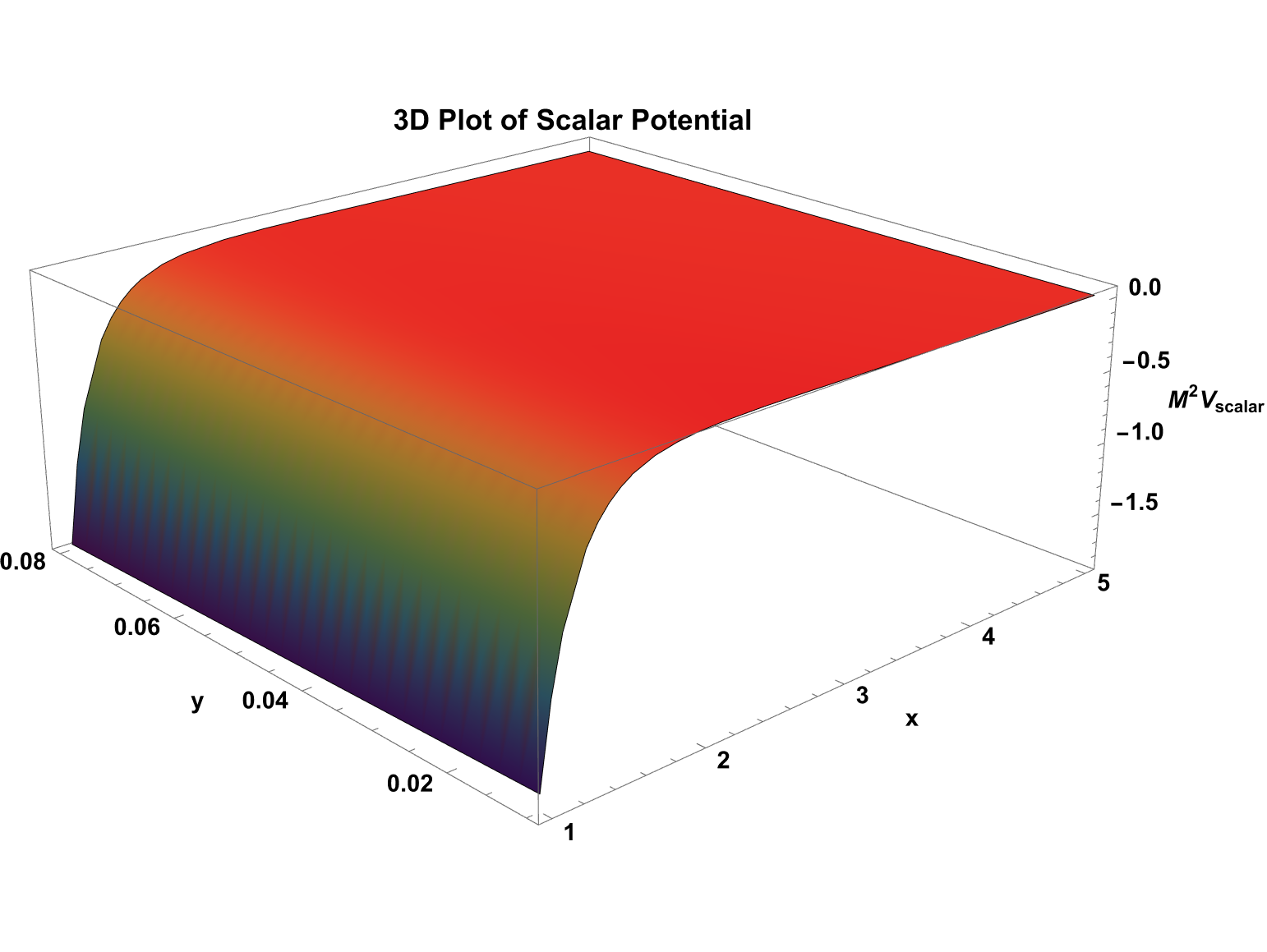}\quad\quad\quad
\includegraphics[width=0.4\linewidth]{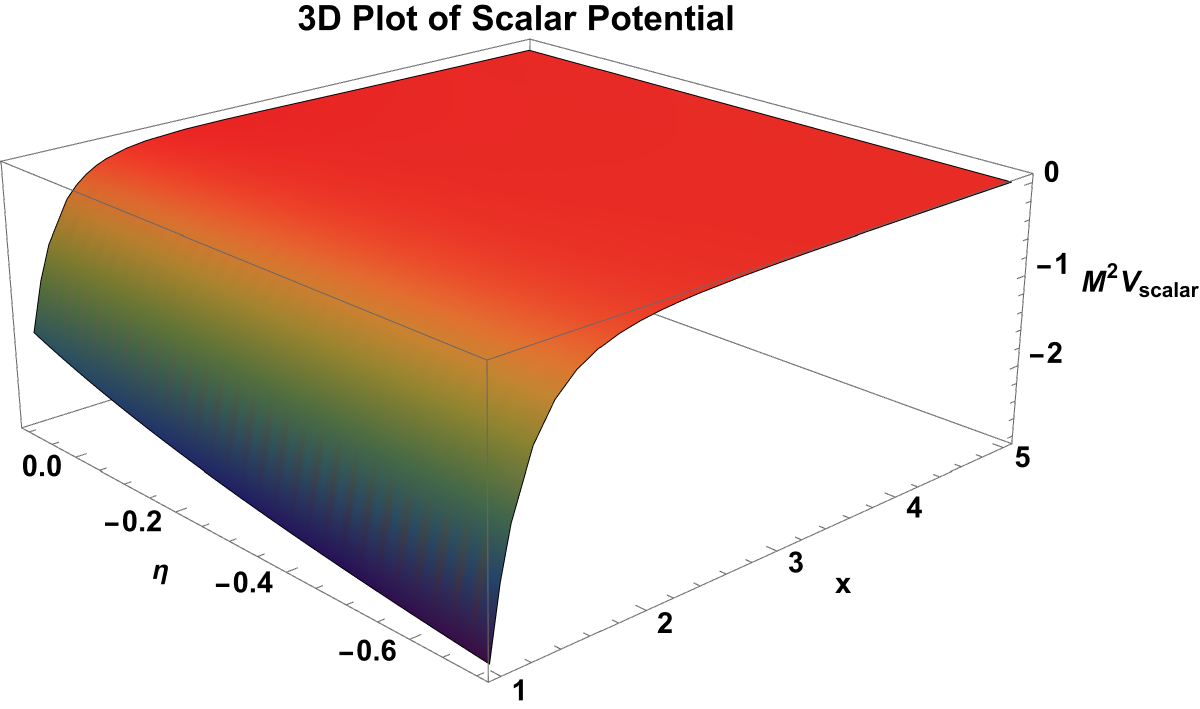}\\
(a) $\eta=0.1$ \hspace{6cm} (b) $y=0.01$
\caption{\footnotesize Three-dimensional visualization of normalized scalar potential $M^2\,V_\text{scalar}$ for monopole mode $\ell=0$, revealing complex parameter interdependencies across the $(x,y,\eta)$ parameter space.}
\label{fig:3d-plot}
\end{figure}

Figure \ref{fig:3d-plot} presents comprehensive three-dimensional visualizations of the potential landscape evolution across the parameter space. Panel (a) demonstrates how the potential surface varies with both radial coordinate $x$ and QF parameter $y$ for fixed LV strength $\eta=0.1$, revealing intricate topological structures that encode the wave scattering properties. Panel (b) shows the complementary view with fixed QF parameter $y=0.01$, illustrating how LV parameter $\eta$ systematically modifies the potential surface morphology. These visualizations provide crucial insights for understanding QNM behavior and stability characteristics in the KRG-QF framework \cite{sec2is30}.

\begin{figure}[ht!]
\centering
\includegraphics[width=0.4\linewidth]{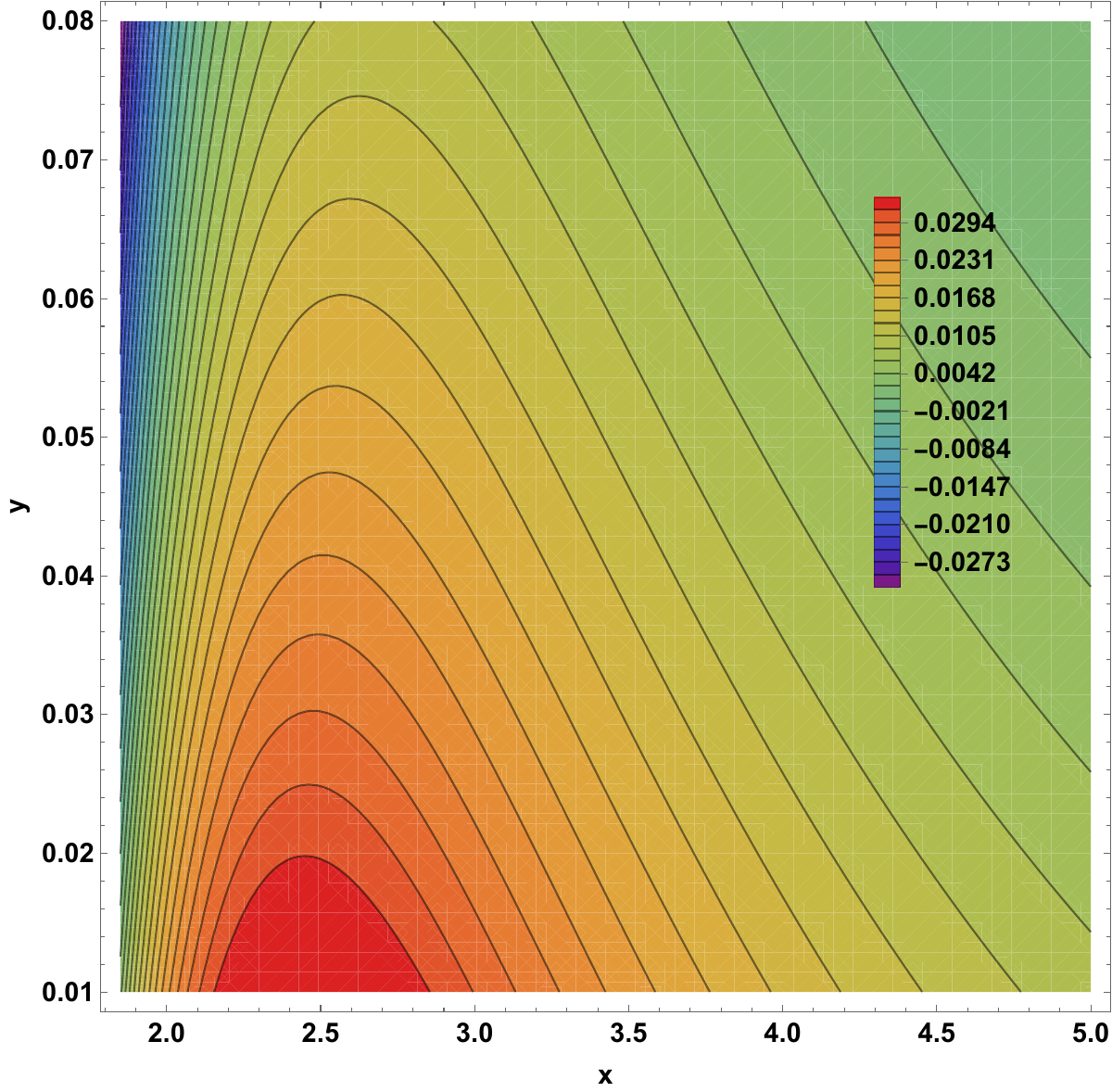}\quad\quad\quad
\includegraphics[width=0.4\linewidth]{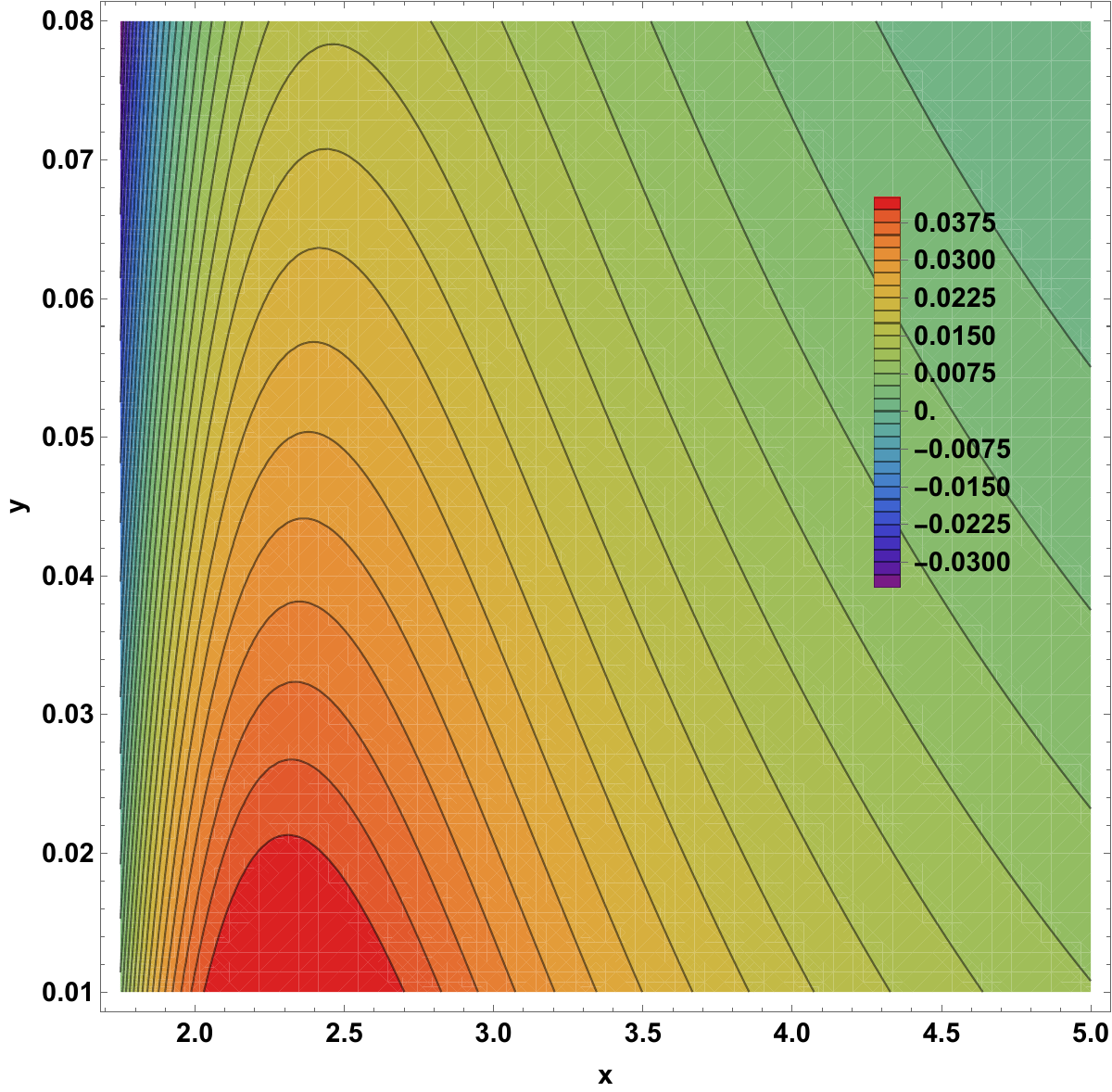}\\
(a) $\eta=0.1$ \hspace{6cm} (b) $\eta=0.15$
\caption{\footnotesize Contour analysis of scalar potential $M^2\,V_{\text{scalar}}$ in the $(x,y)$ plane for monopole mode $\ell=0$, demonstrating LV parameter influence on potential profile topology. Color gradient from red (high) to blue (low) values.}
\label{fig:contour}
\end{figure}

Figure \ref{fig:contour} provides detailed contour analysis revealing the subtle but systematic modifications induced by varying LV parameter $\eta$. The progression from $\eta=0.1$ to $\eta=0.15$ demonstrates how even modest changes in LV strength produce measurable shifts in the potential contour structure. The enhanced potential magnitudes (red regions) with increasing $\eta$ indicate stronger wave reflection and modified transmission coefficients, directly affecting observable QNM spectra and stability timescales. These contour maps establish precise observational targets for discriminating between different parameter combinations through gravitational wave and electromagnetic perturbation measurements \cite{sec2is31}.

The scalar perturbation analysis reveals that KRG-QF modifications introduce rich wave dynamics that significantly deviate from GR predictions. The systematic dependence of the effective potential parameters might establish observational signatures to constrain both LV and quintessence effects through future measurements of BH QNM spectra, stability characteristics, and wave propagation properties. Namely, these results could  provide fundamental theoretical foundations for interpreting future gravitational wave observations and testing modified gravity theories in the strong-field regime.


\section{EM Perturbations in KRG-QF BH Spacetime}\label{isec6}

EM perturbations constitute a fundamental class of field fluctuations that provide crucial insights into BH stability, spectroscopic properties, and electromagnetic field interactions within strong gravitational environments. These perturbations play essential roles in understanding accretion disk physics, relativistic jet formation, plasma dynamics near BH horizons, and serve as valuable analogs for gravitational wave behavior in QNM analysis \cite{sec2is32,sec2is33}. In the context of modified gravity theories, EM perturbations offer particularly sensitive probes for detecting deviations from GR predictions and constraining alternative theoretical frameworks through precision measurements of electromagnetic wave propagation characteristics.

Within our KRG-QF framework, the EM perturbation analysis reveals how the combined effects of LV through parameter $\eta$ and quintessence matter via parameters $(\mathrm{C}, w)$ systematically modify electromagnetic wave dynamics compared to standard Schwarzschild geometry. The resulting perturbative potential incorporates both angular momentum barriers and geometric modifications arising from spacetime curvature alterations, enabling comprehensive characterization of electromagnetic field behavior in modified gravitational environments \cite{sec2is34,sec2is35}.

The dynamics of EM perturbations in curved spacetime are governed by Maxwell's equations, which in covariant form take the expression \cite{sec2is36,sec2is37}:
\begin{equation}
\frac{1}{\sqrt{-g}}\left[F_{\alpha \beta }\,g^{\alpha \nu}\,g^{\beta \mu }\sqrt{-g}\Psi\right]_{,\mu}=0,  \label{em1}
\end{equation}
where $F_{\alpha \beta }=\partial _{\alpha }A_{\beta }-\partial _{\beta}A_{\nu}$ represents the EM field tensor, $A_{\mu}$ denotes the vector potential, and $\Psi$ characterizes the electromagnetic field perturbation amplitude.

Following established procedures for analyzing EM perturbations in spherically symmetric spacetimes \cite{sec2is36,sec2is37}, the radial wave equation can be reduced to the standard Schrödinger-like form:
\begin{equation}
\frac{\partial^2 \psi_\text{em}(r_*)}{\partial r^2_{*}}+\left(\omega^2-V_\text{em}\right)\,\psi_\text{em}(r_*)=0,\label{em2}
\end{equation}
where $r_*$ represents the tortoise coordinate, $\omega$ denotes the perturbation frequency, and $V_\text{em}$ corresponds to the effective EM perturbation potential.

For our KRG-QF BH spacetime, the EM perturbative potential assumes the form:
\begin{eqnarray}
V_\text{em}(r)=\frac{\ell\,(\ell+1)}{r^2}\,A=\left(\frac{1}{1-\eta}-\frac{2\,M}{r}-\frac{\mathrm{C}}{r^{3\,w+1}}\right)\,\frac{\ell\,(\ell+1)}{r^2},\quad \ell\geq 1.\label{em3}
\end{eqnarray}

This expression demonstrates the fundamental structure of EM perturbations in modified gravity, where the potential factorizes into the metric function $A(r)$ encoding spacetime geometry modifications and the angular momentum barrier $\ell(\ell+1)/r^2$ characterizing multipole contributions. The requirement $\ell\geq 1$ reflects the vector nature of electromagnetic fields, which lack monopole components \cite{sec2is38}.

The EM potential explicitly reveals how both LV effects through parameter $\eta$ and QF contributions via parameters $(\mathrm{C}, w)$ systematically modify electromagnetic wave propagation compared to standard GR. The LV parameter $\eta$ appears through the $(1-\eta)^{-1}$ factor, introducing systematic modifications to the effective gravitational coupling, while QF parameters contribute through the $r^{-(3w+1)}$ term that encodes exotic matter effects on spacetime curvature and wave dynamics.

For the specific case of phantom-like quintessence with $w=-2/3$, the EM potential simplifies to:
\begin{eqnarray}
V_\text{em}(r)=\left(\frac{1}{1-\eta}-\frac{2\,M}{r}-\mathrm{C}\,r\right)\,\frac{\ell\,(\ell+1)}{r^2}.\label{em4}
\end{eqnarray}

Introducing dimensionless variables $x=r/M$ and $y=M\,\mathrm{C}$, the normalized potential becomes:
\begin{eqnarray}
M^2\,V_\text{em}=\left(\frac{1}{1-\eta}-\frac{2}{x}-x\,y\right)\,\frac{\ell\,(\ell+1)}{x^2}.\label{em5}
\end{eqnarray}

\begin{figure}[ht!]
\centering
\includegraphics[width=0.4\linewidth]{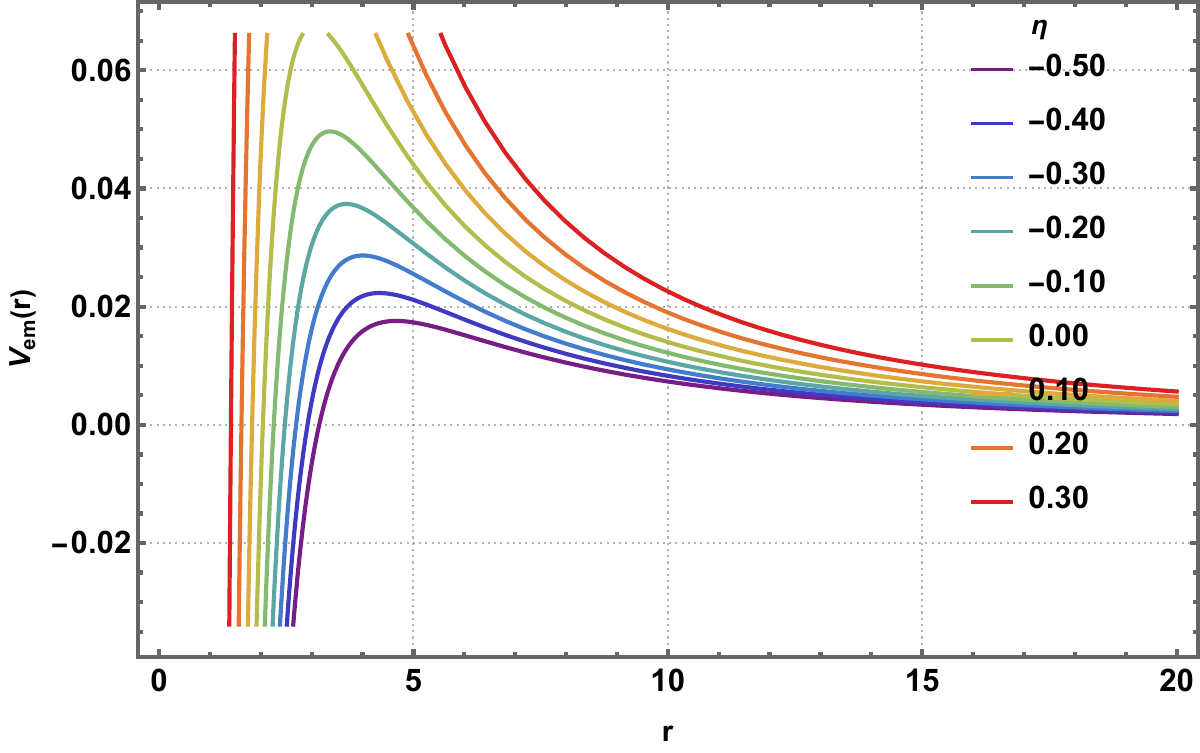}\quad\quad
\includegraphics[width=0.4\linewidth]{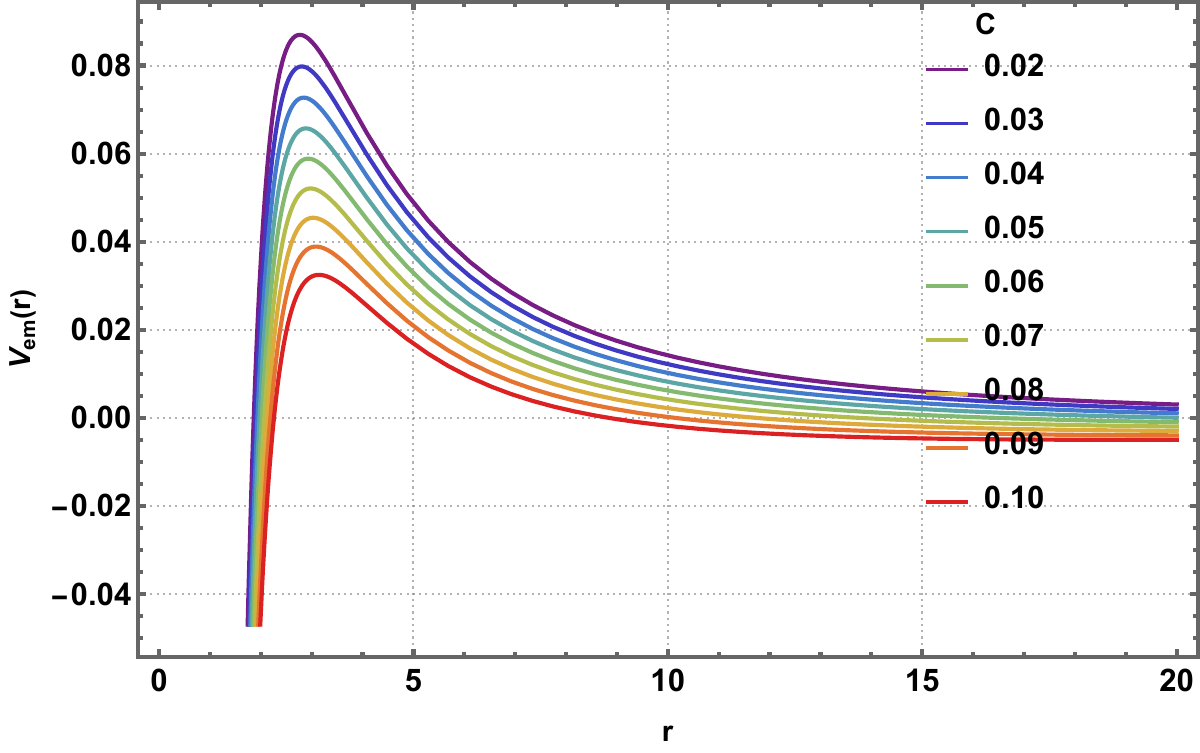}\\
(a) $\mathrm{C}=0.01$ \hspace{6cm} (b) $\eta=0.1$
\caption{\footnotesize EM perturbation potential $V_\text{em}(r)$ behavior for dipole mode $\ell=1$, demonstrating systematic parameter dependence on LV strength $\eta$ and QF normalization $\mathrm{C}$. Parameters: $M=1$, $w=-2/3$.}
\label{fig:em-potential}
\end{figure}

Figure \ref{fig:em-potential} provides fundamental insights into EM wave propagation modifications within the KRG-QF framework. The left panel demonstrates that increasing LV parameter $\eta$ systematically enhances the potential magnitude across all radial coordinates, indicating stronger electromagnetic wave scattering and modified QNM frequencies compared to GR predictions. This enhancement reflects the fundamental alteration of spacetime geometry through spontaneous Lorentz symmetry breaking. Conversely, the right panel reveals that increasing QF parameter $\mathrm{C}$ reduces the potential strength, demonstrating how quintessence matter contributes anti-gravitational effects that facilitate electromagnetic wave propagation and reduce scattering coefficients \cite{sec2is39}.

The complementary behavior between LV and QF effects establishes a rich parameter space where electromagnetic wave characteristics can be systematically tuned through the underlying theoretical parameters. These modifications directly affect observable quantities including QNM frequencies, damping rates, and electromagnetic wave transmission and reflection coefficients, providing potential observational signatures for discriminating between GR and KRG-QF scenarios.

\begin{figure}[ht!]
\centering
\includegraphics[width=0.4\linewidth]{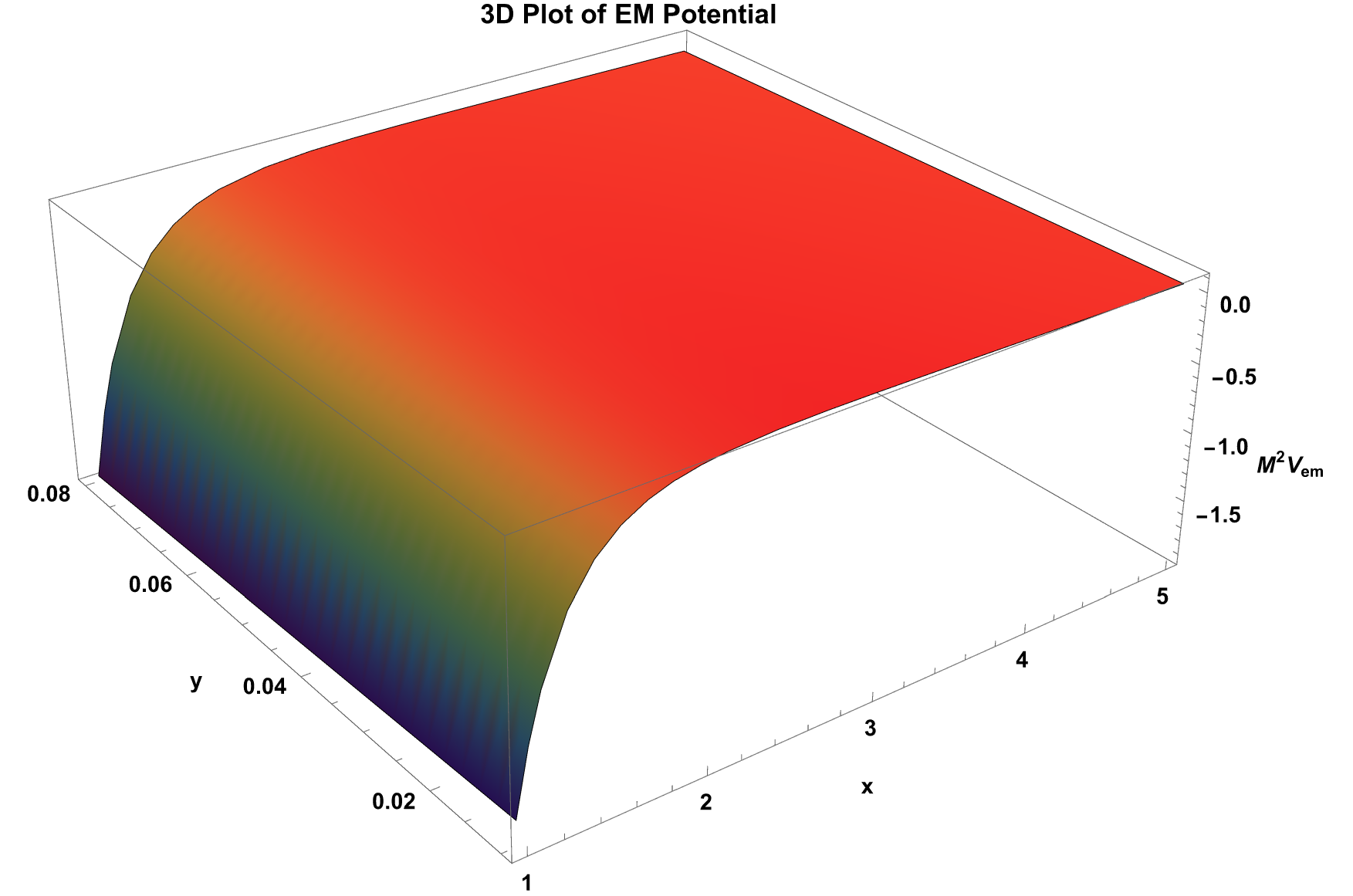}\quad\quad\quad
\includegraphics[width=0.4\linewidth]{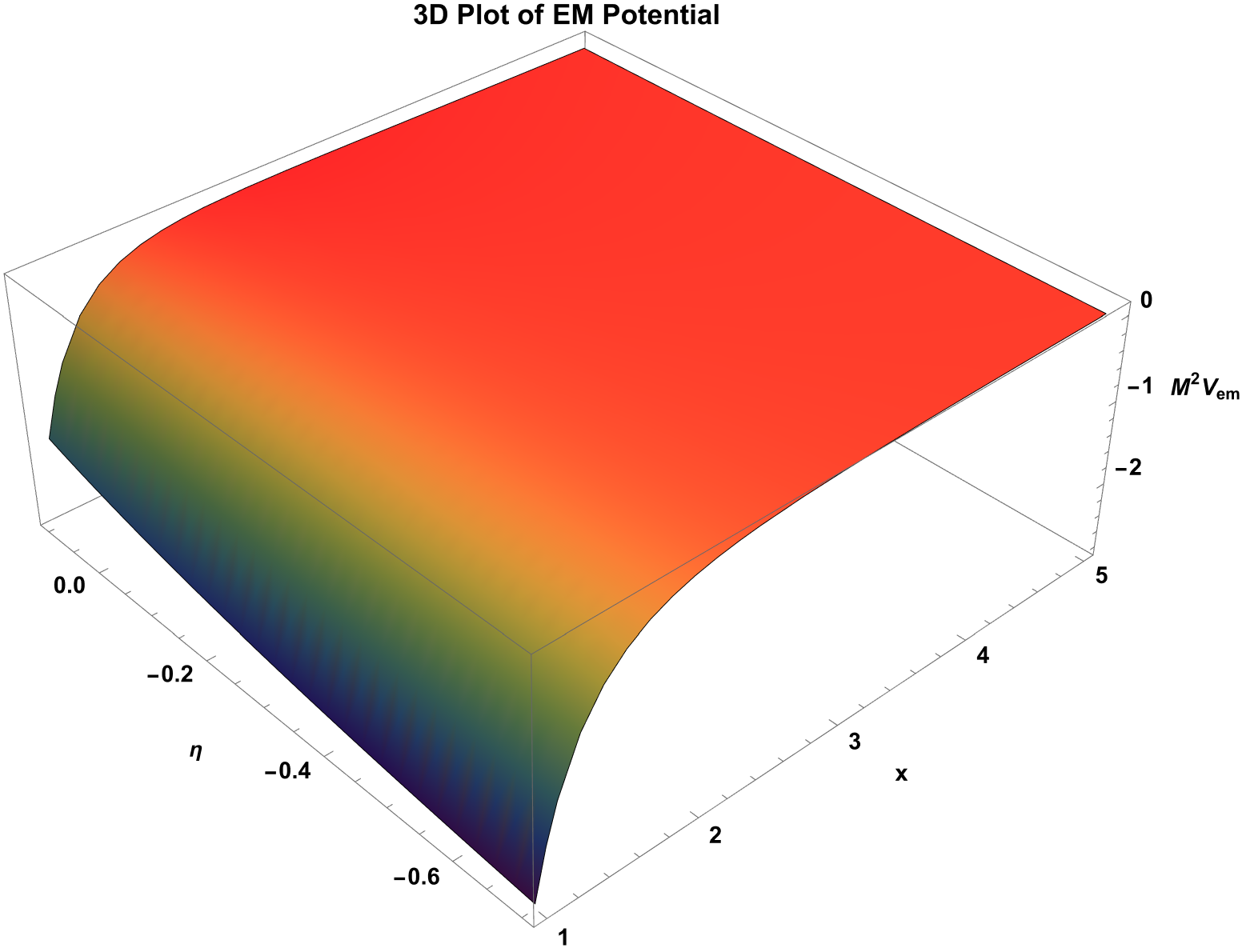}\\
(a) $\eta=0.1$ \hspace{6cm} (b) $y=0.01$
\caption{\footnotesize Three-dimensional visualization of normalized EM potential $M^2\,V_\text{em}$ for dipole mode $\ell=1$, revealing complex parameter interdependencies across the $(x,y,\eta)$ parameter space and systematic potential landscape modifications.}
\label{fig:3d-plot-em}
\end{figure}

Figure \ref{fig:3d-plot-em} presents comprehensive three-dimensional visualizations that illuminate the complex parameter-dependent structure of the EM potential landscape. Panel (a) demonstrates how the potential surface evolves with both radial coordinate $x$ and QF parameter $y$ for fixed LV strength $\eta=0.1$, revealing smooth topological structures that govern electromagnetic wave scattering properties. The systematic variation patterns encode crucial information about wave transmission characteristics and QNM behavior. Panel (b) provides the complementary perspective with fixed QF parameter $y=0.01$, illustrating how LV parameter $\eta$ produces systematic vertical shifts in the potential surface morphology while preserving the fundamental geometric structure.

These three-dimensional representations provide essential insights for understanding electromagnetic wave dynamics in modified gravity theories and establish theoretical foundations for predicting observable electromagnetic signatures in astrophysical environments where KRG-QF effects might be significant \cite{sec2is40}.

\begin{figure}[ht!]
\centering
\includegraphics[width=0.4\linewidth]{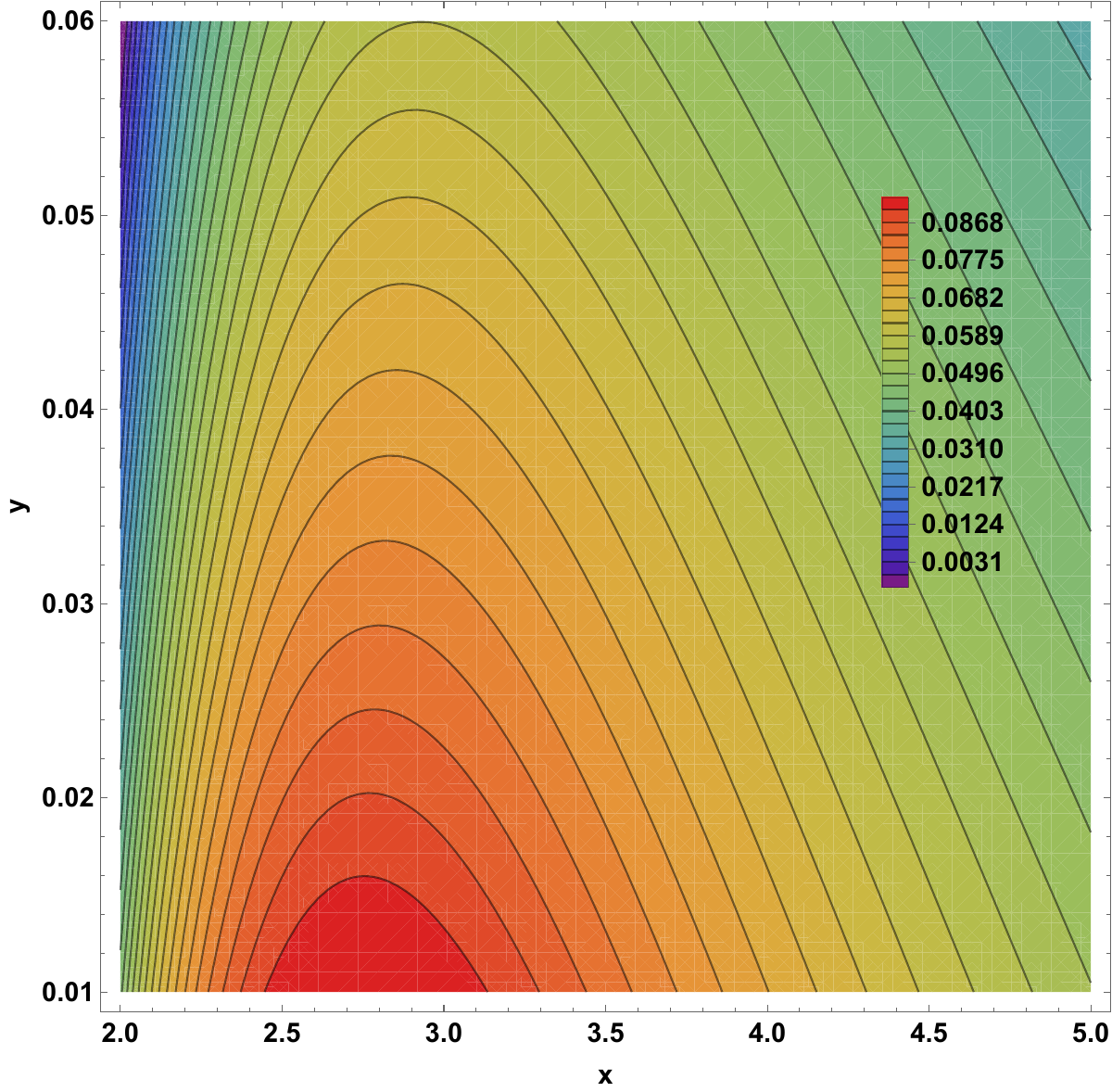}\quad\quad\quad
\includegraphics[width=0.4\linewidth]{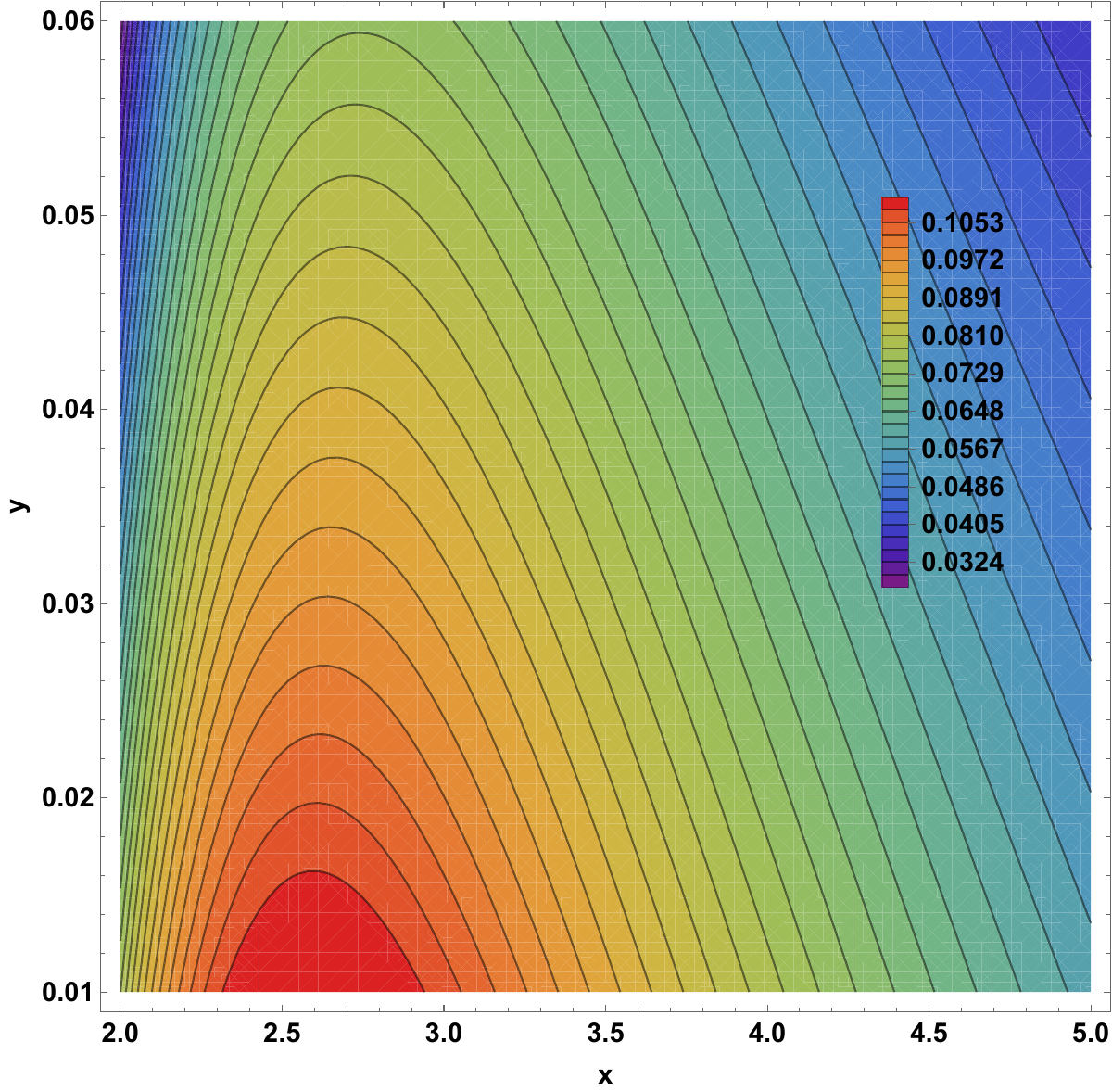}\\
(a) $\eta=0.1$ \hspace{6cm} (b) $\eta=0.15$
\caption{\footnotesize Contour analysis of EM potential $M^2\,V_{\text{em}}$ in the $(x,y)$ plane for dipole mode $\ell=1$, demonstrating systematic LV parameter influence on electromagnetic wave propagation characteristics through potential topology modifications.}
\label{fig:contour-em}
\end{figure}

Figure \ref{fig:contour-em} provides detailed contour analysis that reveals the systematic modifications induced by varying LV parameter $\eta$ on electromagnetic wave propagation characteristics. The progression from $\eta=0.1$ to $\eta=0.15$ demonstrates measurable shifts in the potential contour structure, with enhanced magnitudes corresponding to stronger wave reflection and modified transmission properties. The contour evolution patterns establish precise theoretical predictions for electromagnetic QNM frequencies and damping rates that could be tested through observations of electromagnetic radiation from accreting BH systems or magnetospheric phenomena \cite{sec2is41}.

The EM perturbation analysis demonstrates that KRG-QF modifications introduce remarkable deviations from GR predictions in electromagnetic wave dynamics, establishing theoretical frameworks for testing modified gravity theories through electromagnetic observations and providing essential tools for constraining fundamental physics parameters through precision measurements of electromagnetic phenomena in strong gravitational fields \cite{sec2is42,sec2is43,sec2is44}.

\section{Gravitational Lensing in KRG-QF Spacetime via GBT Analysis}\label{isec7}

Gravitational lensing represents one of the most sensitive probes for testing modified gravity theories and understanding the fundamental nature of spacetime geometry in strong gravitational fields. The weak lensing regime, characterized by small deflection angles, provides valuable insights into how deviations from GR manifest in observable phenomena through precise measurements of light ray trajectories around massive compact objects \cite{sec2is45,sec2is46}. In the context of our KRG-QF framework, gravitational lensing analysis reveals the intricate interplay between LV effects and quintessence matter contributions in determining photon propagation characteristics.

This section presents a comprehensive investigation of weak gravitational lensing properties in KRG BH geometries embedded within QF environments through systematic application of the GBT to the emergent optical geometry. The GBT approach provides a robust mathematical framework for computing deflection angles that naturally incorporates the geometric modifications introduced by both LV parameters and exotic matter contributions \cite{sec2is47,sec2is48}. This methodology enables precise characterization of how KRG-QF spacetime modifications deviate from standard GR predictions in observable lensing phenomena.

Following established GBT methodology \cite{sec2is49}, we initiate the analysis by deriving the optical metric from the null geodesic condition $ds^2 = 0$ applied to our KRG-QF spacetime metric from Eq.~(\ref{aa1}). The resulting two-dimensional optical geometry assumes the form:

\begin{equation}
dt^2 = \gamma_{ij} dx^i dx^j = \frac{1}{A^2(r)} dr^2 + \frac{r^2}{A(r)} d\Omega^2,
\end{equation}

where $\gamma_{ij}$ characterizes the optical manifold structure, and our investigation focuses on the equatorial plane ($\theta = \pi/2$) without loss of generality. Implementing the tortoise coordinate transformation $dr^* = dr/A(r)$, the optical metric simplifies to:

\begin{equation}
dt^2 = dr^{*2} + \tilde{A}^2(r^*) d\phi^2,
\end{equation}

with $\tilde{A}(r^*) = r/\sqrt{A(r)}$ defining the effective lensing geometry that encodes both LV and QF modifications.

The fundamental requirement for GBT application necessitates rigorous computation of the Gaussian curvature $\mathcal{K}$ for the optical manifold. For our KRG-QF metric function $A(r) = \frac{1}{1-\eta} - \frac{2M}{r} - \frac{C}{r^{3w+1}}$, the systematic curvature calculation yields:

\begin{equation}
\mathcal{K} = \frac{1}{2r}\frac{d}{dr}\left(\frac{1}{\sqrt{A(r)}}\frac{dA(r)}{dr}\right) = \frac{A''(r)}{2A^{3/2}(r)} - \frac{3[A'(r)]^2}{8A^{5/2}(r)},
\end{equation}

where the metric function derivatives are:

\begin{equation}
A'(r) = \frac{2M}{r^2} + \frac{C(3w+1)}{r^{3w+2}}, \quad A''(r) = -\frac{4M}{r^3} - \frac{C(3w+1)(3w+2)}{r^{3w+3}}.
\end{equation}

These expressions explicitly demonstrate how both LV parameter $\eta$ and QF characteristics $(\mathrm{C}, w)$ contribute to the curvature structure of the optical manifold, directly influencing photon deflection properties.

Applying the GBT framework to the region $\tilde{D}$ bounded by the light trajectory $C_1$ and a circular arc $C_R$, with vanishing geodesic curvature along the null path ($\kappa(C_1) = 0$) and unit Euler characteristic $\chi(\tilde{D}) = 1$, the theorem establishes:

\begin{equation}
\iint_{\tilde{D}} \mathcal{K} dS + \int_{C_R} \kappa dt = 2\pi.
\end{equation}

In the asymptotic limit $R \to \infty$, employing the weak field approximation with straight-line trajectory $r(\phi) = b/\sin\phi$ where $b$ denotes the impact parameter, the surface element becomes $dS = r A^{-3/2}(r) dr d\phi$. The deflection angle emerges through systematic evaluation of:

\begin{equation}
\Theta = -\int_0^{\pi} \int_{b/\sin\phi}^{\infty} \mathcal{K} \frac{r}{A^{3/2}(r)} dr d\phi.
\end{equation}

For analytical tractability, we express the KRG-QF metric function in factorized form:
 
\begin{equation}
A(r) = \frac{1}{1-\eta}\left[1 - \frac{2M(1-\eta)}{r} - \frac{C(1-\eta)}{r^{3w+1}}\right].
\end{equation}

Employing asymptotic expansion techniques for large $r$ and performing systematic integration, the deflection angle assumes the comprehensive form \cite{sec2is50}:

\begin{multline}
\Theta \simeq \frac{4M}{b}(1-\eta) + \frac{15\pi M^2}{4b^2}(1-\eta)^2 
+ \frac{2C(3w+1)}{b^{3w+1}}(1-\eta)^{3/2}
+ \frac{2MC(3w+1)}{b^{2+3w}}(1-\eta)^{5/2} + \mathcal{O}\left(\frac{M^3}{b^3}\right).
\end{multline}

This expression demonstrates proper consistency with established limits: when $\eta \to 0$ and $C \to 0$, the leading term reduces to the classical Schwarzschild result $\Theta_{\text{Schwarzschild}} = \frac{4M}{b}$. The KRG field modifications manifest through systematic $(1-\eta)$ enhancement factors, while QF contributions appear as power-law corrections with exponents determined by the quintessence state parameter $w$ \cite{sec2is51}.

\begin{figure*}[ht!]
\begin{center}
\subfigure[Deflection angle evolution with LV parameter: $w = -2/3$, $C = 0.5$]{
\includegraphics[width=0.85\textwidth]{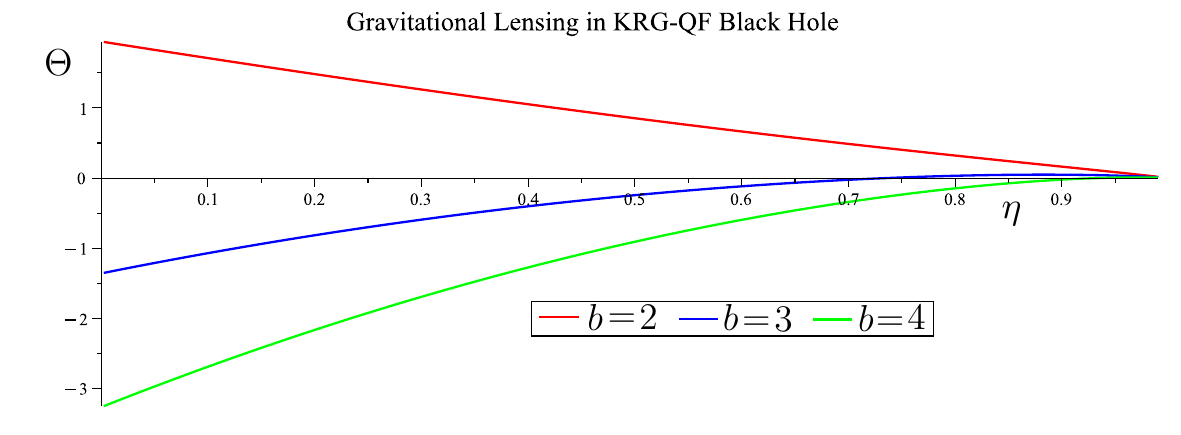}
\label{fig_deflection_a}
}
\quad
\subfigure[Impact parameter dependence analysis: $w = -2/3$, $C = 0.5$]{
\includegraphics[width=0.85\textwidth]{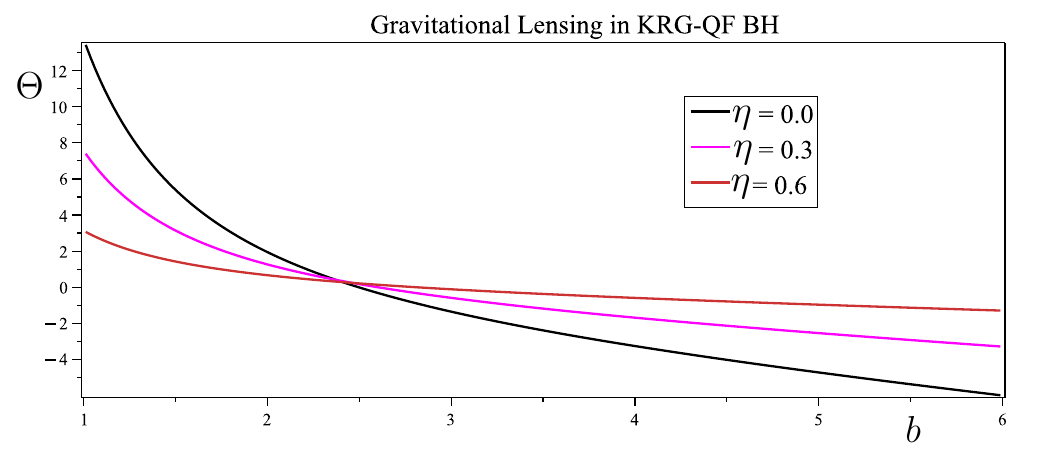}
\label{fig_deflection_b}
}

\caption{\footnotesize Deflection angle $\Theta$ variations across parameter space with quintessence configuration $w = -2/3$ and normalization strength $C = 0.5$. Panel (a) exhibits deflection angle evolution as a function of LV parameter $\eta \in [0, 0.99]$ for discrete impact parameter values $b \in \{2, 3, 4\}M$, revealing enhancement of gravitational focusing with increasing Lorentz symmetry breaking. Panel (b) demonstrates the inverse relationship between deflection magnitude and impact parameter $b \in [1, 6]M$ across varying LV strengths $\eta \in \{0.0, 0.3, 0.6\}$, illustrating the characteristic $b^{-1}$ scaling modified by KRG field contributions and QF corrections.}
\label{fig:deflection_analysis}
\end{center}
\end{figure*}

Figure~\ref{fig:deflection_analysis} provides systematic validation of the theoretical deflection angle formula through comprehensive parameter space exploration. The mathematical structure employs the complete deflection expression with fixed parameters $M = 1$, $w = -2/3$, and $C = 0.5$, enabling detailed investigation of the complex interplay between geometric modifications and exotic matter contributions in KRG-QF spacetimes.

Panel (a) demonstrates the fundamental relationship between LV parameter $\eta$ and deflection amplification across the physical domain $\eta \in [0, 0.99]$ for impact parameters $b \in \{2, 3, 4\}M$. The analysis reveals that increasing Lorentz symmetry breaking systematically enhances gravitational focusing efficiency through the multiplicative $(1-\eta)$ factors embedded throughout the deflection expression. The distinct curves corresponding to different impact parameters exhibit consistent monotonic behavior, with smaller impact parameters producing proportionally larger deflection angles according to the characteristic inverse relationship $\Theta \propto b^{-1}$. Particularly significant is the observation that deflection enhancement becomes increasingly pronounced as $\eta$ approaches unity, indicating potential divergent behavior in extreme LV regimes where $(1-\eta) \to 0$, suggesting fundamental limitations of weak field approximations in this parameter domain \cite{sec2is52}.

Panel (b) provides complementary analysis through systematic examination of impact parameter dependence $b \in [1, 6]M$ across discrete LV configurations $\eta \in \{0.0, 0.3, 0.6\}$. The results demonstrate the expected hyperbolic decay characteristic of gravitational lensing phenomena, with baseline Schwarzschild behavior ($\eta = 0.0$) establishing the reference deflection profile. The progressive enhancement observed for $\eta = 0.3$ and $\eta = 0.6$ configurations quantitatively validates theoretical predictions that KRG field modifications systematically amplify gravitational focusing across all impact parameter scales. The mathematical structure reveals that both leading-order Schwarzschild-like terms and higher-order QF corrections experience multiplicative enhancement through powers of $(1-\eta)$, producing cumulative amplification effects that become increasingly significant for strong LV parameters.

At large impact parameters, the consistent trend of all curves moving towards reduced deflection validates the expected asymptotic behavior from weak field gravitational lensing theory. Meanwhile, increased curve separation at small impact parameters suggests that strong-field corrections are more sensitive to spacetime alterations \cite{sec2is53}. These findings provide a solid theoretical basis for utilizing gravitational lensing observations to restrict the parameters of the KRG-QF model and evaluate modified gravity theories via precise astronomical measurements of light deflection around BHs.

\section{Thermodynamic Properties and Phase Transitions in KRG-QF BH Systems}\label{isec8}

The thermodynamic analysis of BH systems provides fundamental insights into the microscopic structure of spacetime, quantum gravitational effects, and the deep connections between geometry and thermodynamics in modified gravity theories. In the context of KRG-QF frameworks, thermodynamic investigations reveal how LV effects and quintessence matter collectively influence thermal properties, stability characteristics, and phase transition phenomena \cite{sec2is54,sec2is55,sec2is56}. This section presents a comprehensive thermodynamic analysis of our KRG BH solution embedded within QF environments, characterized by the metric function in Eq.~(\ref{bb2}), providing crucial insights into thermal stability, equilibrium properties, and potential observational signatures of modified gravitational dynamics.

The thermodynamic framework for KRG-QF BH systems incorporates both classical thermodynamic principles and the modifications introduced by LV and exotic matter effects. These modifications manifest in altered temperature-entropy relationships, modified stability criteria, and novel phase transition behaviors that significantly deviate from standard Schwarzschild BH thermodynamics \cite{sec2is57,sec2is58}. Understanding these thermodynamic properties is essential for predicting observational signatures in thermal emission spectra and constraining theoretical parameters through precision measurements of BH thermal characteristics.

The event horizon location is determined by solving $A(r_h) = 0$, establishing a fundamental relationship between quintessence parameters $(C, w)$ and the LV parameter $\eta$. This condition yields the modified mass-radius relationship:

\begin{equation}
M = \frac{r_h}{2}\left(\frac{1}{1 - \eta} - \frac{C}{r_h^{3w+1}}\right),
\end{equation}

where the gravitational mass exhibits explicit dependence on both KRG field parameter $\eta$ and quintessence characteristics. This expression demonstrates how Lorentz symmetry breaking and dark energy-like components fundamentally alter the standard Schwarzschild mass-radius correspondence, introducing parameter-dependent corrections that directly affect all subsequent thermodynamic quantities.

\begin{figure}[ht!]
    \centering
    \includegraphics[width=0.4\linewidth]{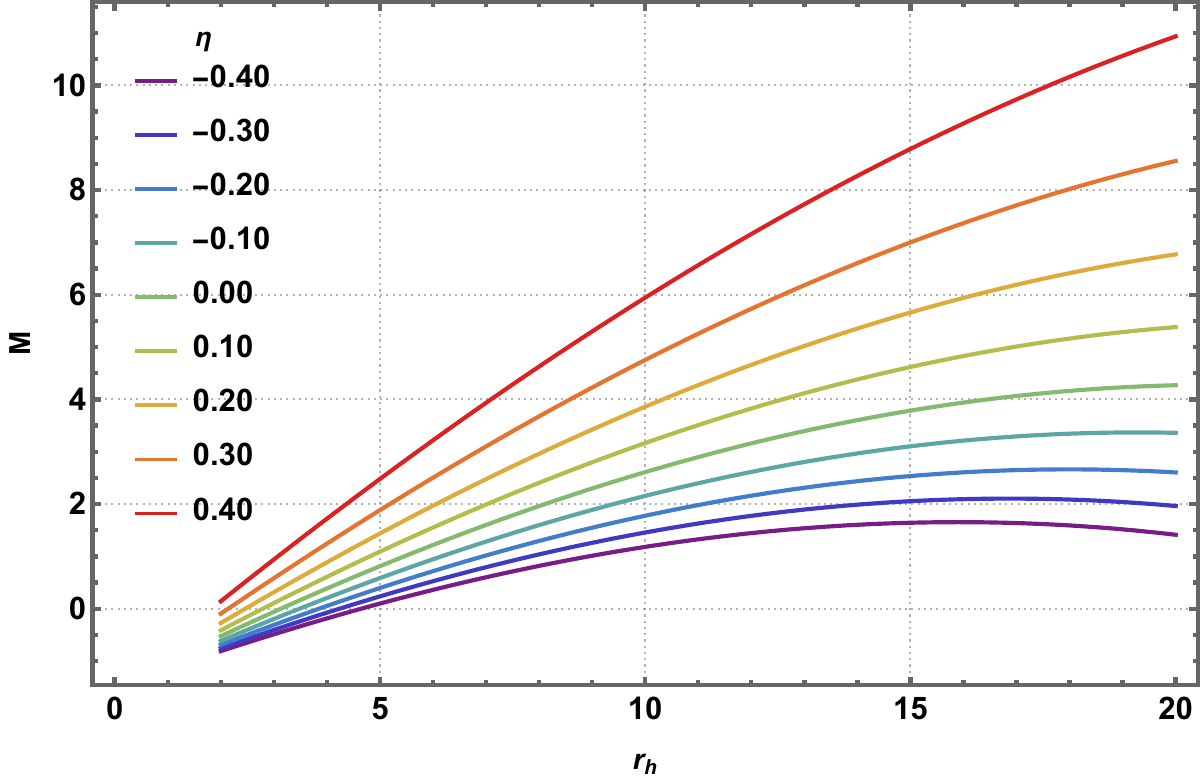}\qquad
    \includegraphics[width=0.4\linewidth]{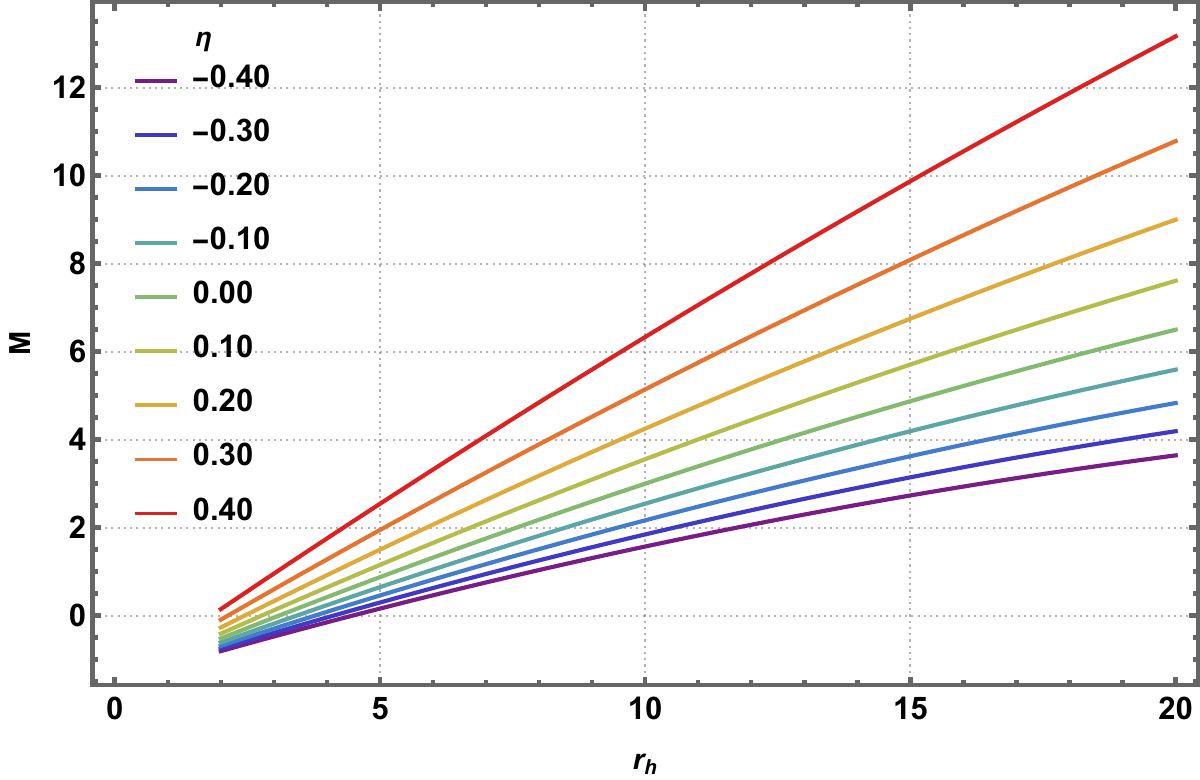}\\
    (a) $w=-3/4$ \hspace{6cm} (b) $w=-2/3$\\
    \includegraphics[width=0.4\linewidth]{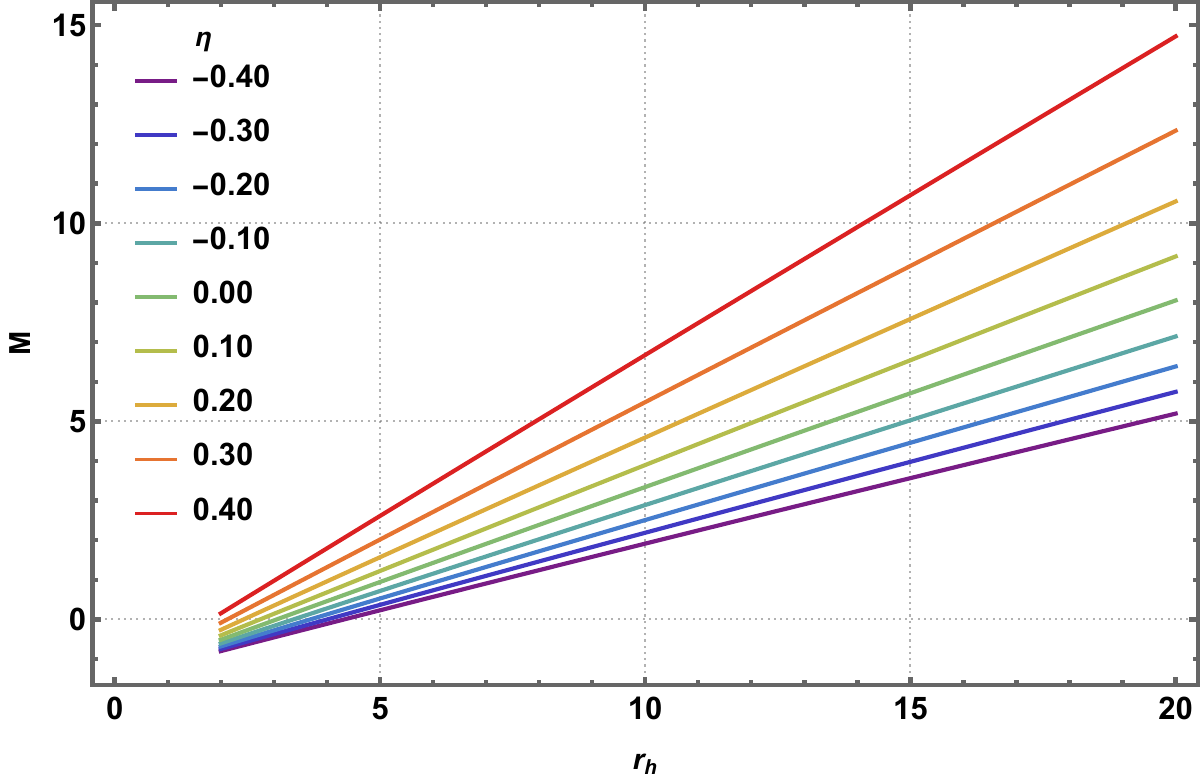}\qquad
     \includegraphics[width=0.4\linewidth]{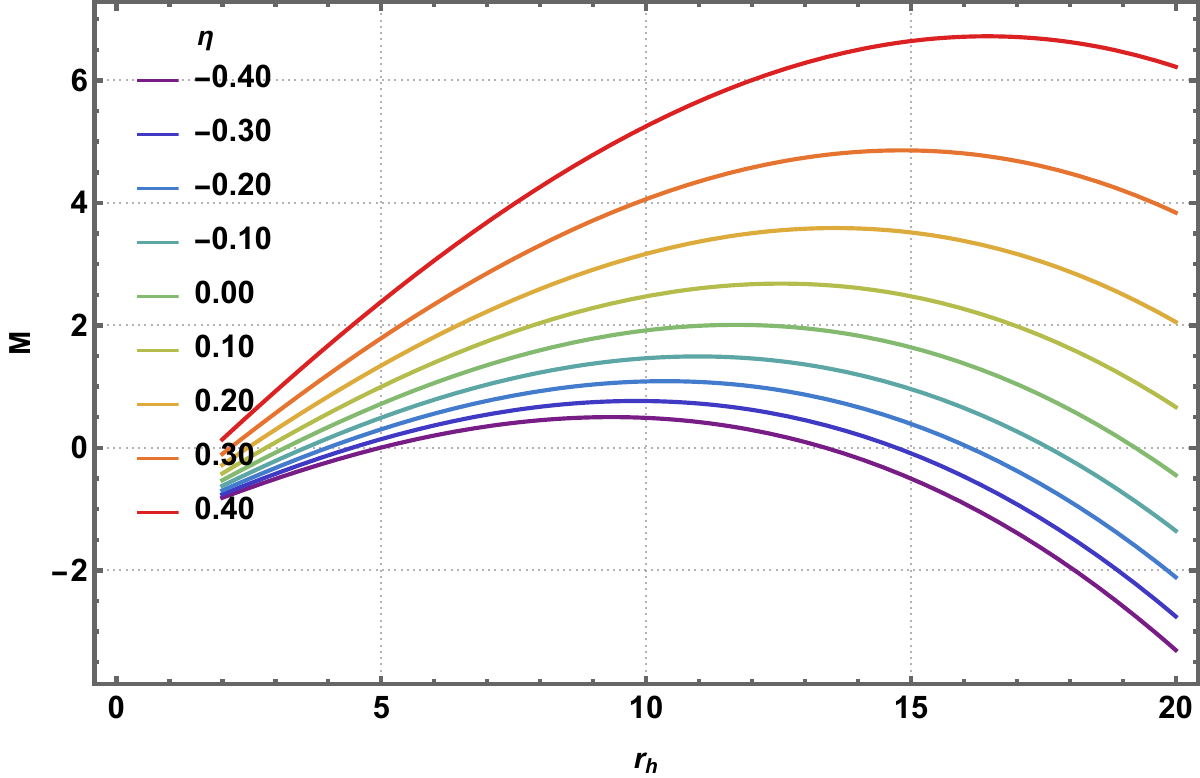}\\
    (c) $w=-1/2$ \hspace{6cm} (d) $w=-5/6$ 
    \caption{Black hole mass $M$ as a function of event horizon radius $r_h$ across different quintessence state parameters, demonstrating the systematic modification of mass-radius relationships due to combined KRG-QF effects. Fixed parameter: $\mathrm{C}=0.01$. The diverse scaling behaviors reveal how varying dark energy equations of state fundamentally alter the gravitational mass structure in modified spacetime geometries.}
    \label{fig:bh-mass}
\end{figure}

Figure \ref{fig:bh-mass} illustrates the fundamental modification of BH mass-radius relationships across different quintessence regimes. The systematic parameter dependence reveals how various dark energy equations of state, characterized by different $w$ values, produce distinct mass scaling behaviors that deviate significantly from standard Schwarzschild predictions. The phantom-like regime ($w=-2/3$) in panel (b) exhibits particularly pronounced deviations, while the intermediate states in panels (a) and (c) show progressive modifications that interpolate between standard and extreme behaviors.

The BH entropy follows the universal Bekenstein-Hawking area law \cite{sec2is59,sec2is60}:

\begin{equation}
S = \frac{A_h}{4} = \pi r_h^2,
\end{equation}

where $A_h$ represents the event horizon area. Remarkably, despite the extensive modifications introduced by KRG-QF dynamics, the fundamental area law for BH entropy remains intact, suggesting a robust geometric foundation for thermodynamic properties that transcends specific gravitational theories. This universality provides important theoretical constraints on the microscopic degrees of freedom underlying BH entropy in modified gravity frameworks.

The Hawking temperature, computed through surface gravity analysis \cite{sec2is61,sec2is62}, yields:

\begin{equation}
T_\text{Haw} = \frac{\kappa}{2\pi} = \frac{1}{4\pi} \left.\frac{dA(r)}{dr}\right|_{r = r_h} = \frac{3r_h^{(-3w-2)}wC}{4\pi} + \frac{1}{4(1-\eta)\pi r_h}.
\label{iztemp}
\end{equation}

This temperature expression explicitly reveals how both LV effects through $(1-\eta)^{-1}$ factors and QF contributions via $w$-dependent terms systematically modify thermal emission characteristics compared to standard GR predictions.

\begin{figure}[ht!]
    \centering
    \includegraphics[width=0.4\linewidth]{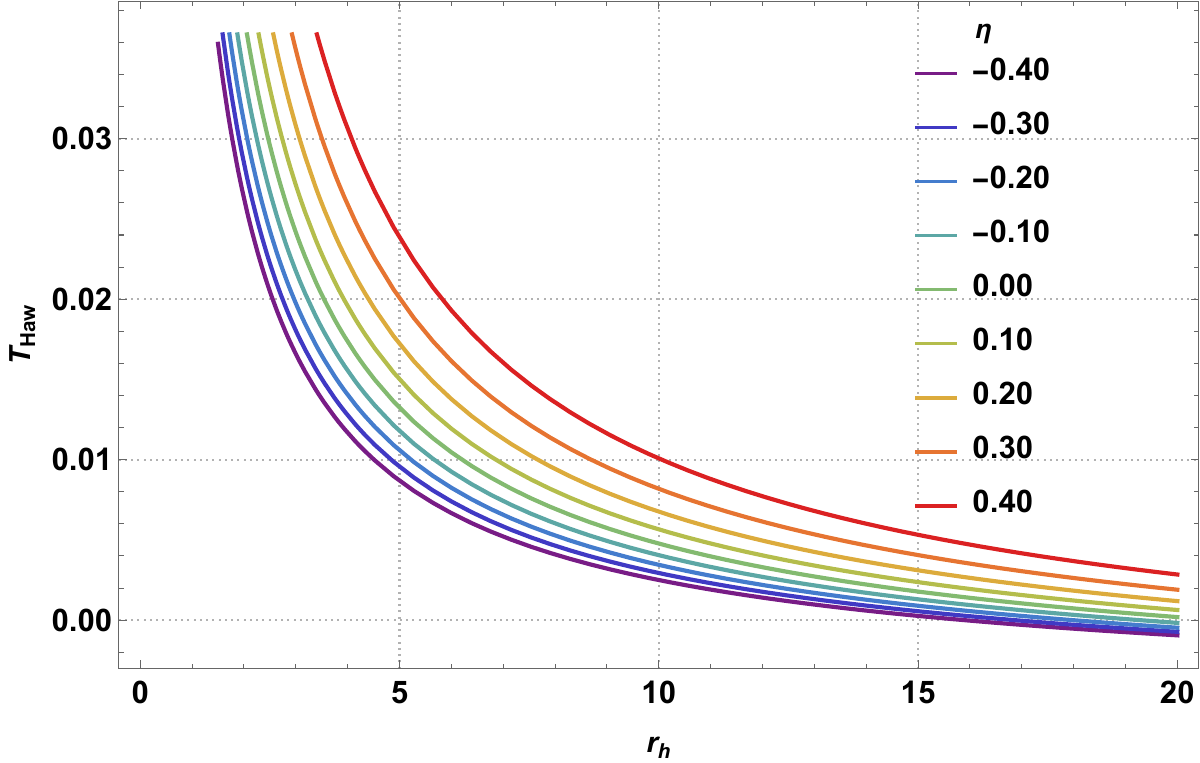}\qquad
    \includegraphics[width=0.4\linewidth]{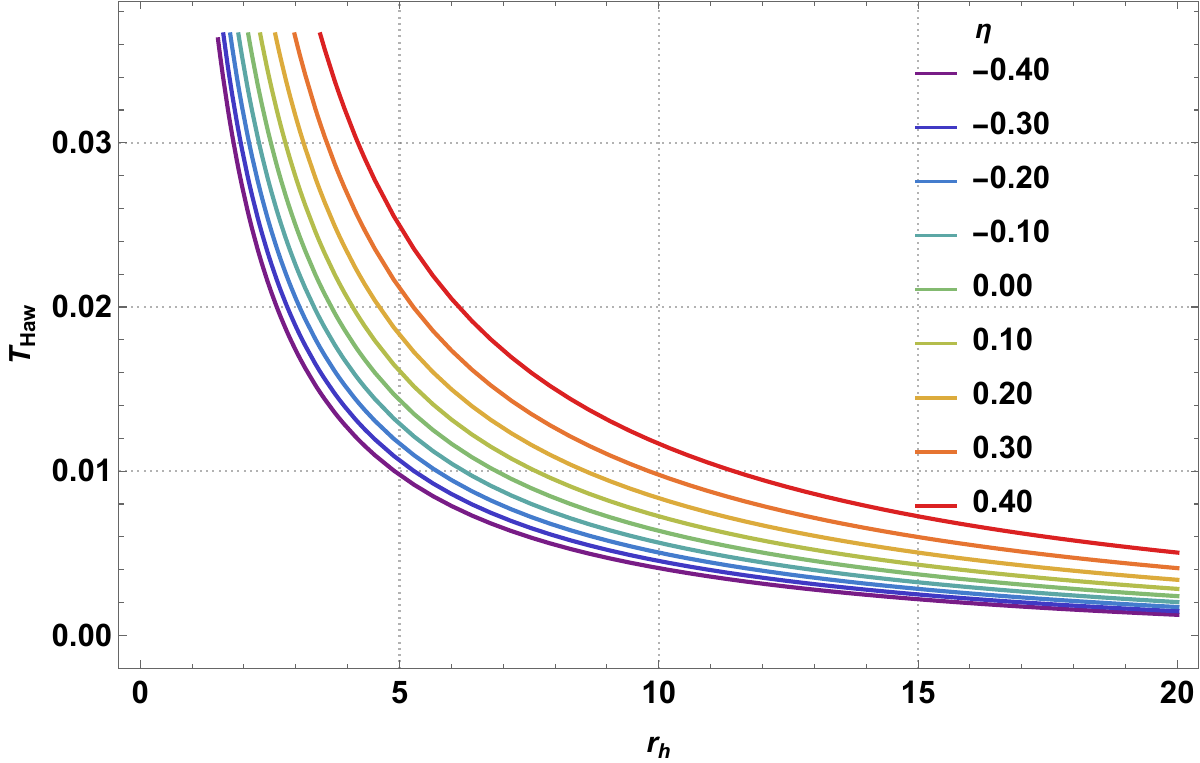}\\
    (a) $w=-3/4$ \hspace{6cm} (b) $w=-2/3$\\
    \includegraphics[width=0.4\linewidth]{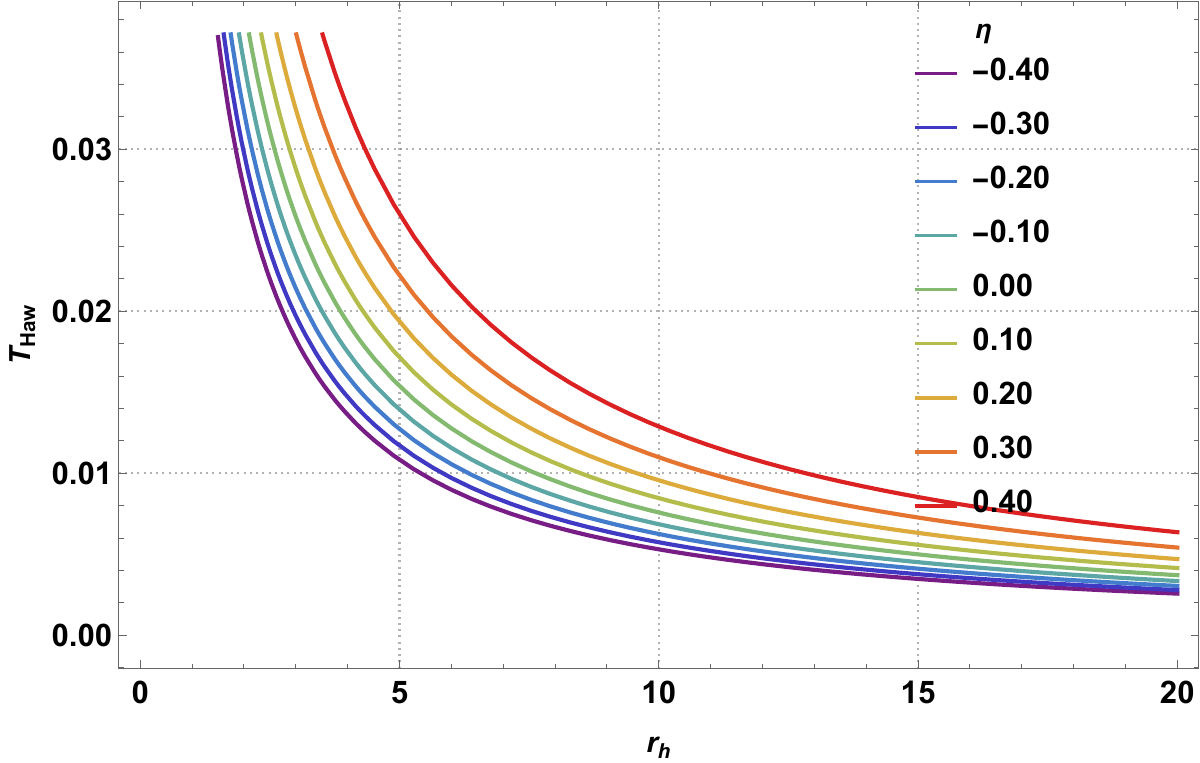}\qquad
     \includegraphics[width=0.4\linewidth]{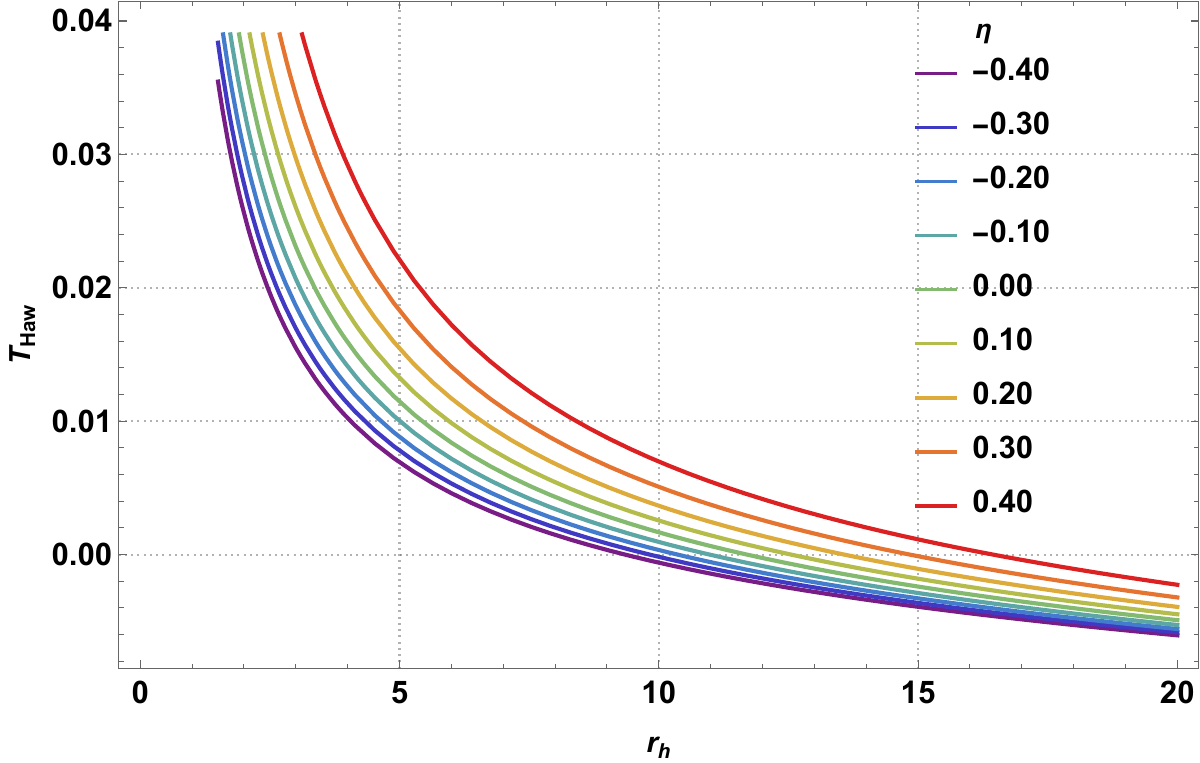}\\
    (c) $w=-1/2$ \hspace{6cm} (d) $w=-5/6$ 
    \caption{Hawking temperature evolution as a function of event horizon radius for different quintessence state parameters, revealing systematic modifications to thermal emission characteristics. Fixed parameter: $\mathrm{C}=0.01$. The distinct temperature profiles demonstrate how varying dark energy equations of state produce markedly different thermal behaviors, with implications for observational signatures in BH thermal radiation spectra.}
    \label{fig:temperature-profiles}
\end{figure}

Figure \ref{fig:temperature-profiles} demonstrates the rich temperature evolution patterns across different quintessence configurations. The systematic variations in thermal emission characteristics provide crucial insights into how dark energy equations of state influence BH thermodynamics. Panel (b) shows the phantom-like regime exhibiting enhanced temperature variations, while panels (a), (c), and (d) reveal progressive modifications that could serve as observational discriminators between different theoretical frameworks.

\begin{figure*}[ht!]
\begin{center}
\subfigure[Low QF regime: Schwarzschild regime corresponds to $\eta = 0$]{
\includegraphics[width=0.3\textwidth]{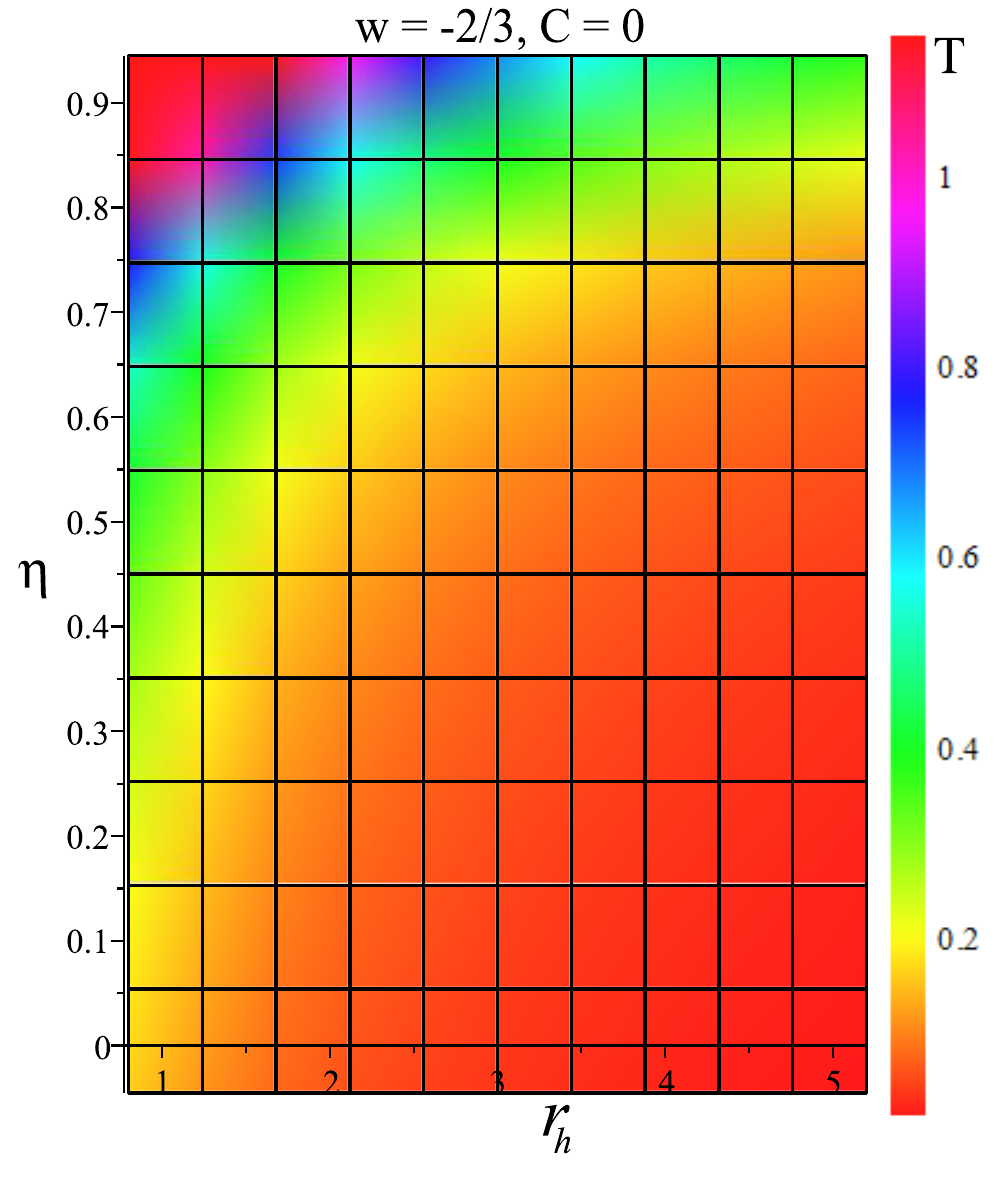}
\label{figtempa_beta2}
}
\quad
\subfigure[Moderate QF regime: $w = -2/3$, $C = 0.5$]{
\includegraphics[width=0.3\textwidth]{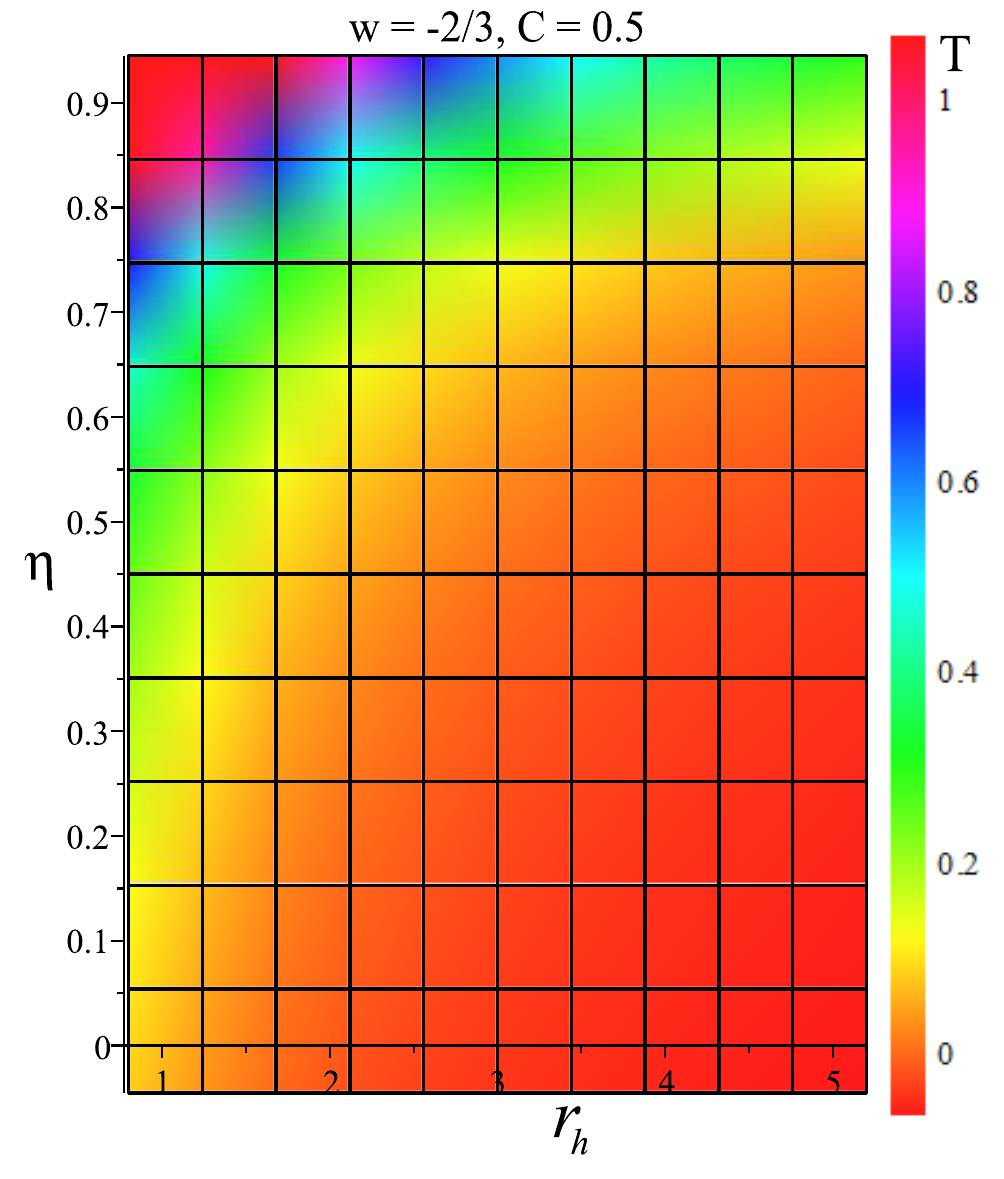}
\label{figtempb_beta2}
}
\quad
\subfigure[Strong QF regime: $w = -2/3$, $C =1$]{
\includegraphics[width=0.3\textwidth]{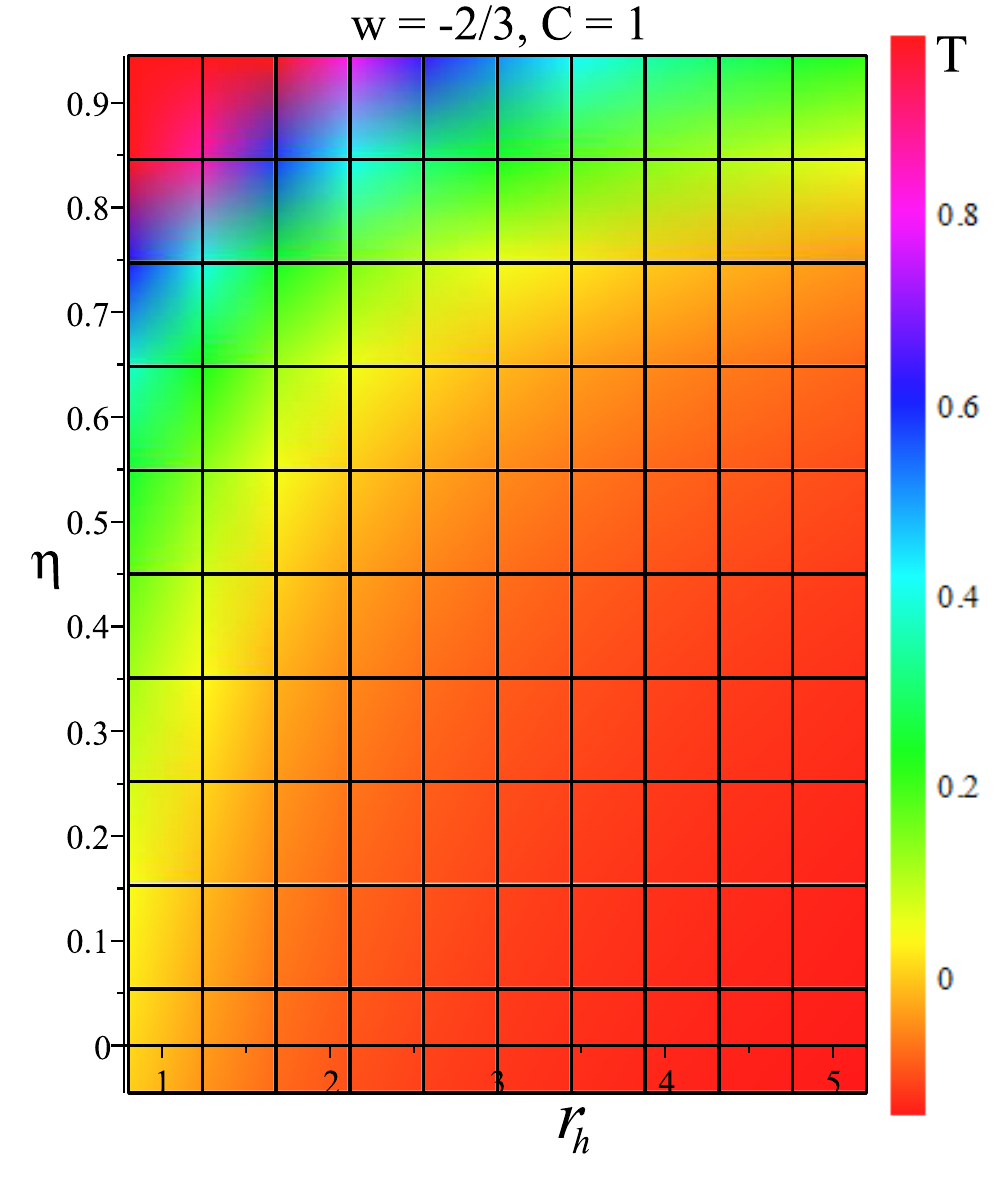}
\label{figtempc_beta2}
}

\quad

\subfigure[Low QF regime: $w = -1/3$, $C = 1$]{
\includegraphics[width=0.3\textwidth]{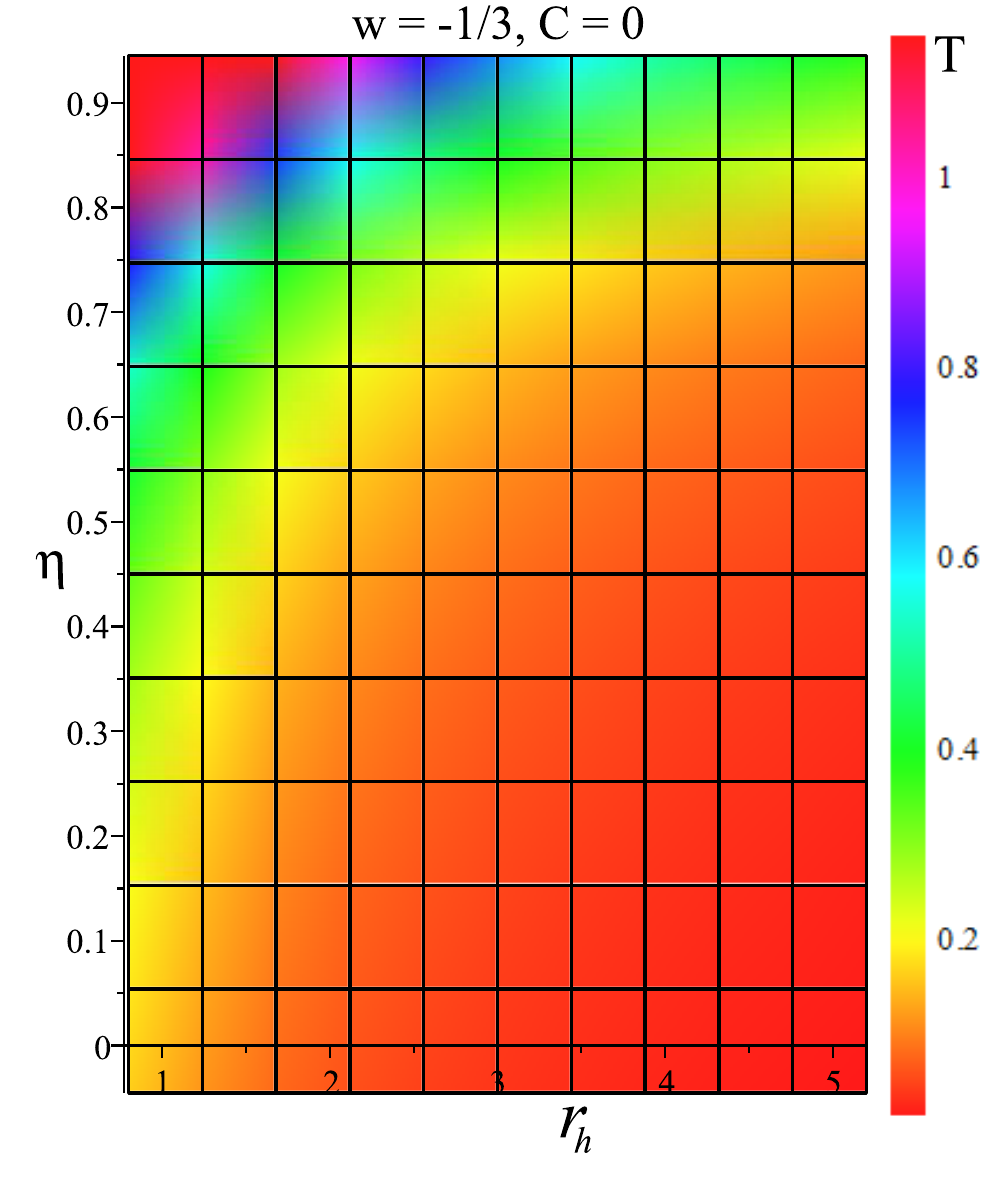}
\label{figtempa_betane2}
}
\quad
\subfigure[Moderate QF regime: $w = -1/3$, $C = 0.5$]{
\includegraphics[width=0.3\textwidth]{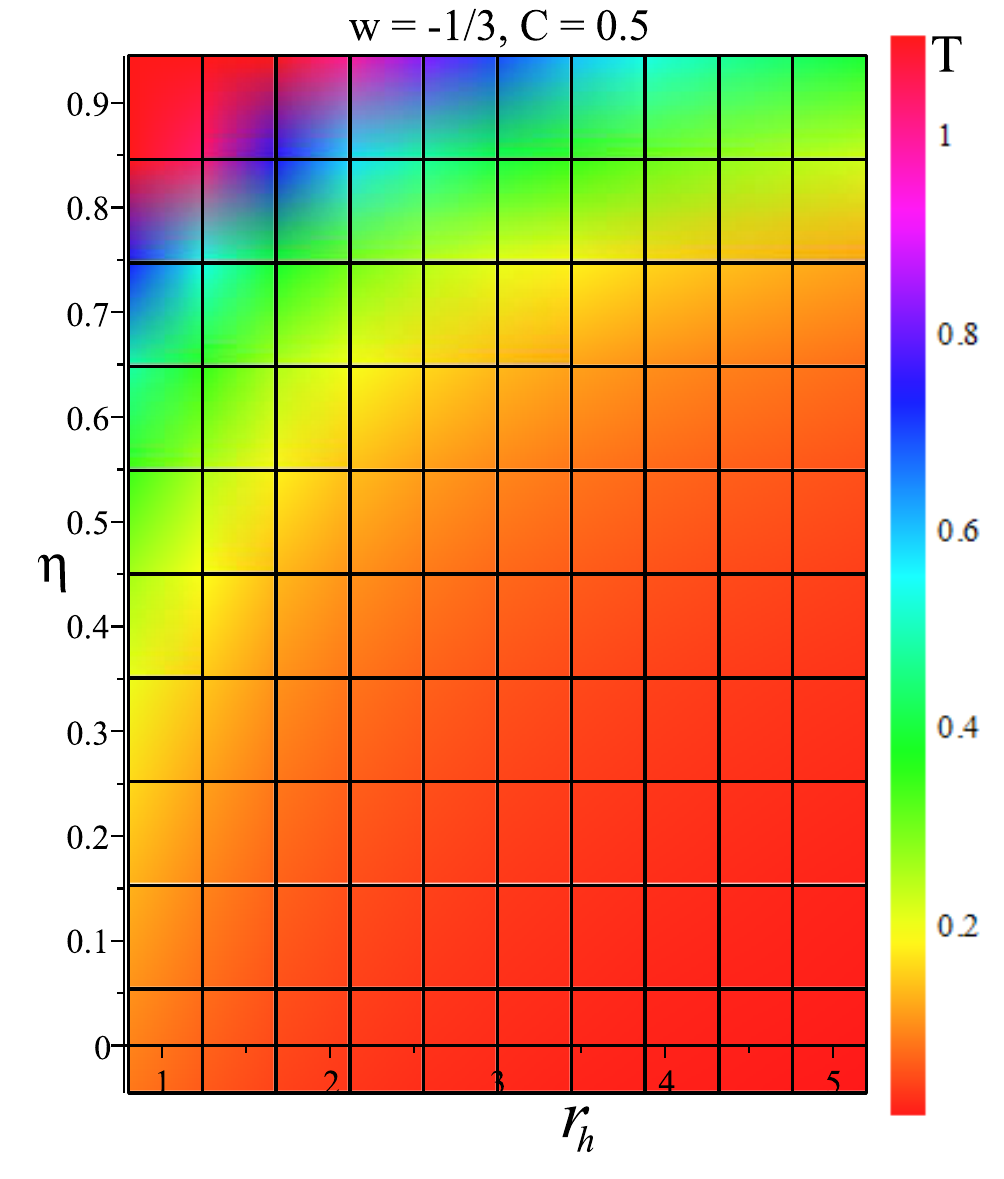}
\label{figtempb_betane2}
}
\quad
\subfigure[Cosmological constant regime: $w = -1$, $C = 0.5$]{
\includegraphics[width=0.3\textwidth]{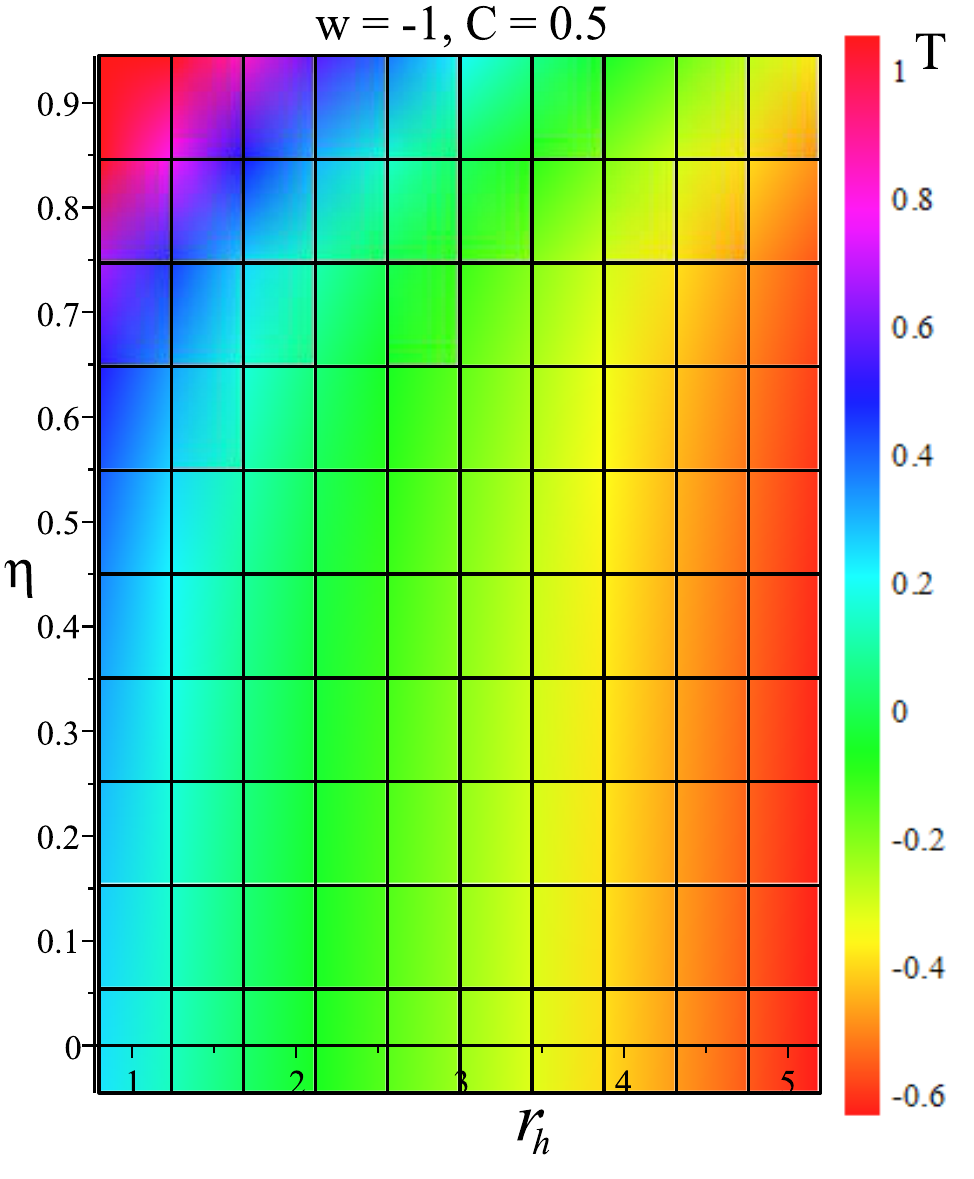}
\label{figtempc_betane2}
}

\caption{Hawking temperature density field analysis across diverse KRG-QF parameter configurations. The six panels systematically explore thermodynamic behavior from standard Schwarzschild baseline (panel a) through varying QF intensities and state parameters. Panels (a-c) examine phantom-like quintessence ($w = -2/3$) with increasing QF strength, while panels (d-e) investigate intermediate dark energy behavior ($w = -1/3$). Panel (f) represents the cosmological constant limit ($w = -1$). Color gradients encode temperature magnitude variations from positive high-temperature zones (red) through zero-temperature regions (green) to negative-temperature zones (blue), revealing complex thermal topologies, thermodynamic instabilities, and parameter-dependent phase structures in modified gravitational systems. The mass parameter was chosen as $M=1$.}
\label{figthermalanalysis}
\end{center}
\end{figure*}

Figure \ref{figthermalanalysis} presents thermodynamic density field analysis across representative KRG-QF parameter configurations. Panel (a) establishes the reference baseline for standard Schwarzschild regime, displaying characteristic smooth positive temperature gradients consistent with classical BH thermodynamics. Panels (b) and (c) demonstrate the progressive intensification of thermal complexity as QF coupling strength increases within the phantom-like regime ($w = -2/3$), with enhanced temperature variations and emerging non-linear structures while maintaining predominantly positive temperatures. Panel (d) explores strong QF coupling in the intermediate dark energy regime ($w = -1/3$), revealing distinct thermodynamic behavior with clear positive-negative temperature boundaries. Panel (e) shows moderate QF coupling in the same regime, exhibiting similar but less pronounced thermal transitions. Panel (f) represents the cosmological constant limit ($w = -1$), exhibiting extensive negative temperature regions that signal potential thermodynamic instabilities and breakdown of the semiclassical description. The appearance of negative Hawking temperatures in certain parameter regimes indicates regions where the standard entropy-energy relationships become unstable, possibly marking the validity limits of the effective field theory approach and suggesting the need for quantum gravitational corrections. The parameter progression illuminates how different combinations of LV and quintessence effects generate diverse thermodynamic phase structures, including potentially unphysical negative temperature zones that significantly deviate from conventional BH thermal behavior \cite{sec2is64}.

\begin{figure}[ht!]
    \centering
    \includegraphics[width=0.4\linewidth]{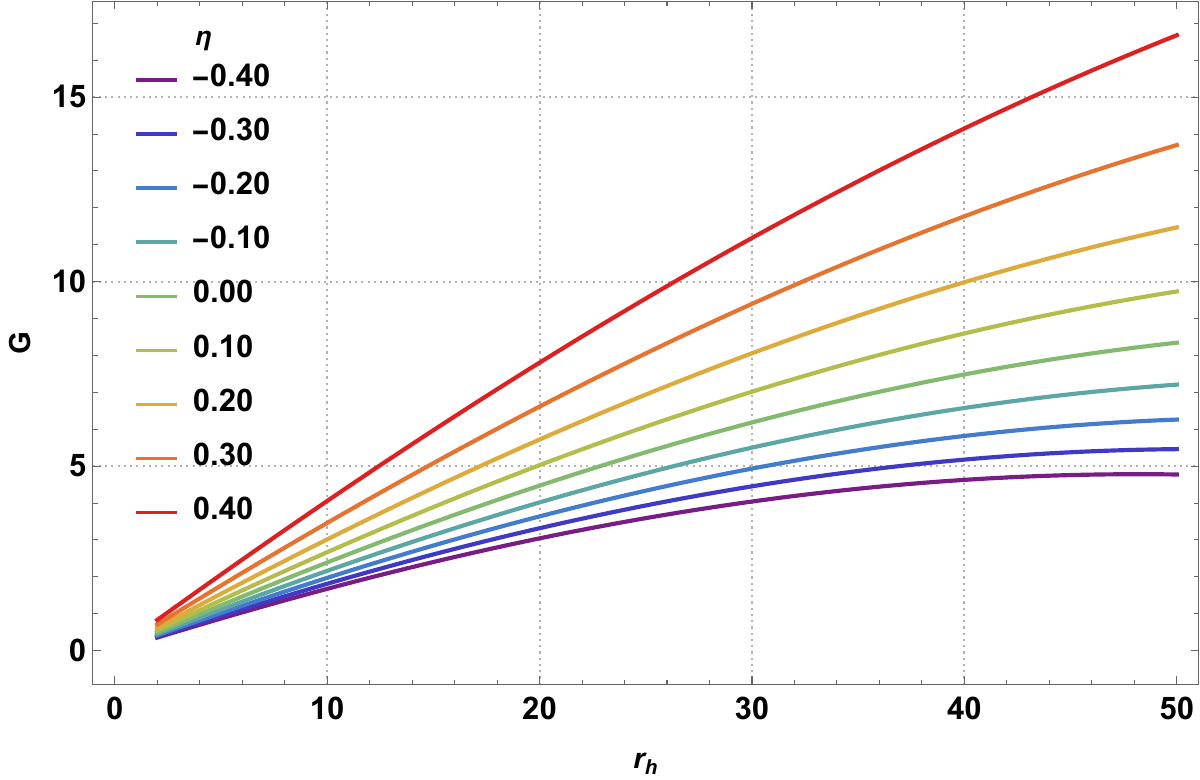}\qquad
    \includegraphics[width=0.4\linewidth]{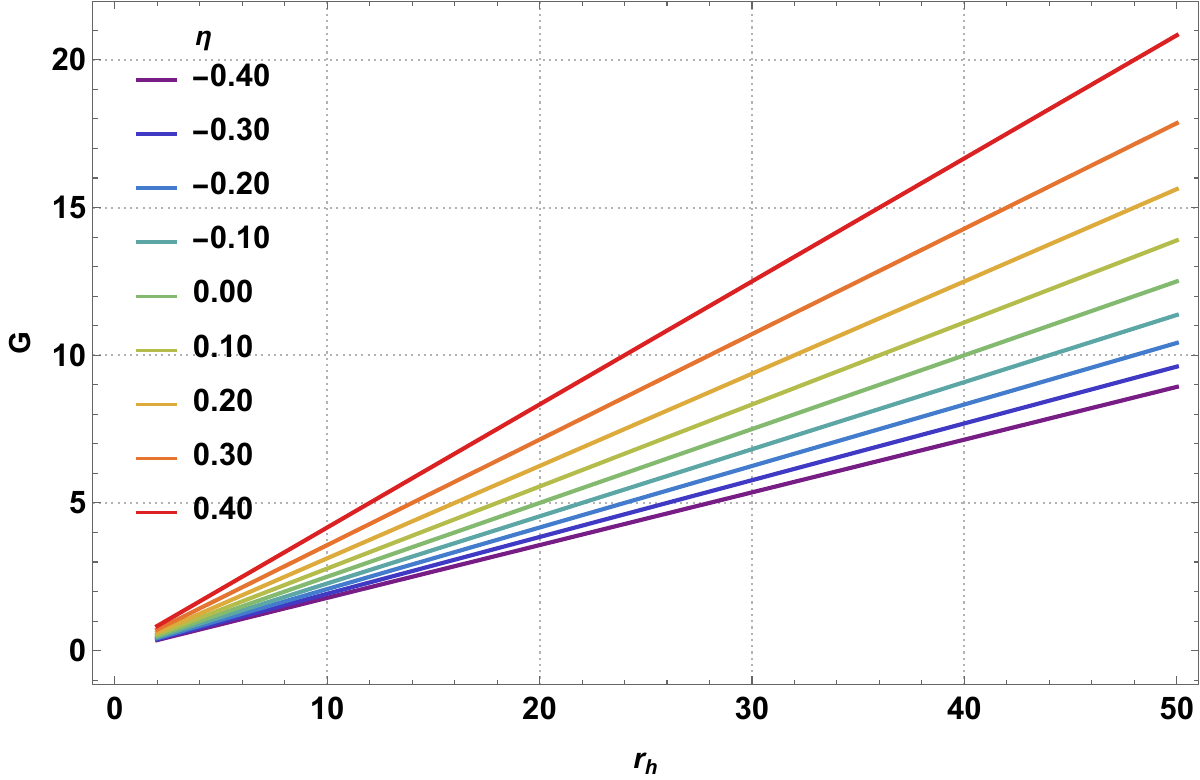}\\
     (a) $w=-3/4$ \hspace{6cm} (b) $w=-2/3$ \\
     \includegraphics[width=0.4\linewidth]{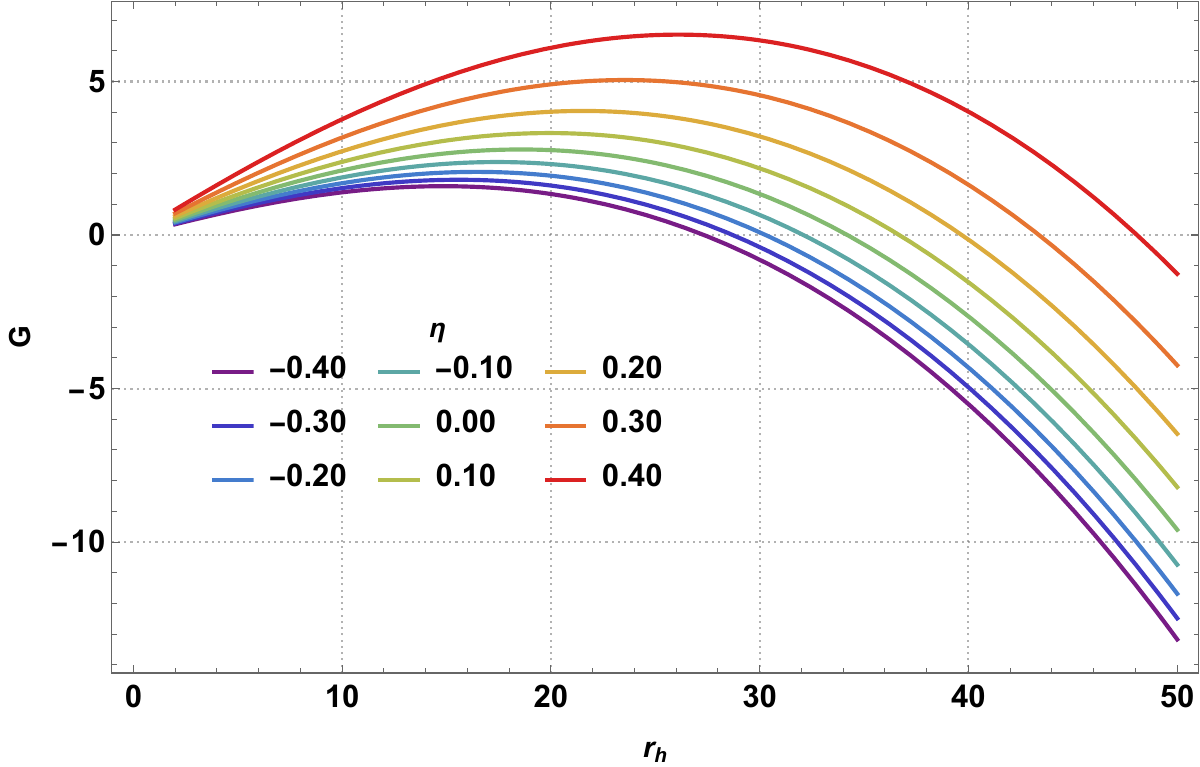}\\
    (c) $w=-5/6$
    \caption{Gibbs free energy evolution with respect to event horizon radius across different quintessence state parameters, demonstrating systematic modifications to thermodynamic equilibrium conditions. Fixed parameter: $\mathrm{C}=0.01$. The monotonic growth patterns with varying slopes encode critical information about system responsiveness to thermal perturbations and establish thermodynamic stability characteristics across different dark energy regimes.}
    \label{fig:gibbs-energy}
\end{figure}

The Gibbs free energy analysis provides fundamental insights into thermodynamic equilibrium and stability conditions \cite{sec2is67,sec2is68}:

\begin{equation}
G = \frac{r_h^{(-3w)}(3w+2)C}{4} + \frac{r_h}{4(1-\eta)}.
\end{equation}

Figure \ref{fig:gibbs-energy} demonstrates the systematic evolution of Gibbs free energy across different quintessence regimes, revealing how varying dark energy equations of state influence thermodynamic stability. The monotonic increase across all parameter regimes suggests global thermodynamic stability, while slope variations encode critical information about system responsiveness to thermal perturbations. The quintessence contribution introduces power-law corrections that become increasingly significant at large horizon radii.

The specific heat capacity represents the most sensitive thermodynamic diagnostic for detecting critical phenomena \cite{sec2is69,sec2is70}:

\begin{equation}
C_P = T_\text{Haw}\, \left( \frac{dS}{dT_\text{Haw}} \right)_P = \frac{2\pi\left(3Cr_h^2 w(-1+\eta)-r_h^{(3+3w)}\right)}{3w(3w+2)(1-\eta)C-r_h^{(3w+1)}}.
\end{equation}

\begin{figure}[ht!]
    \centering
    \includegraphics[width=0.4\linewidth]{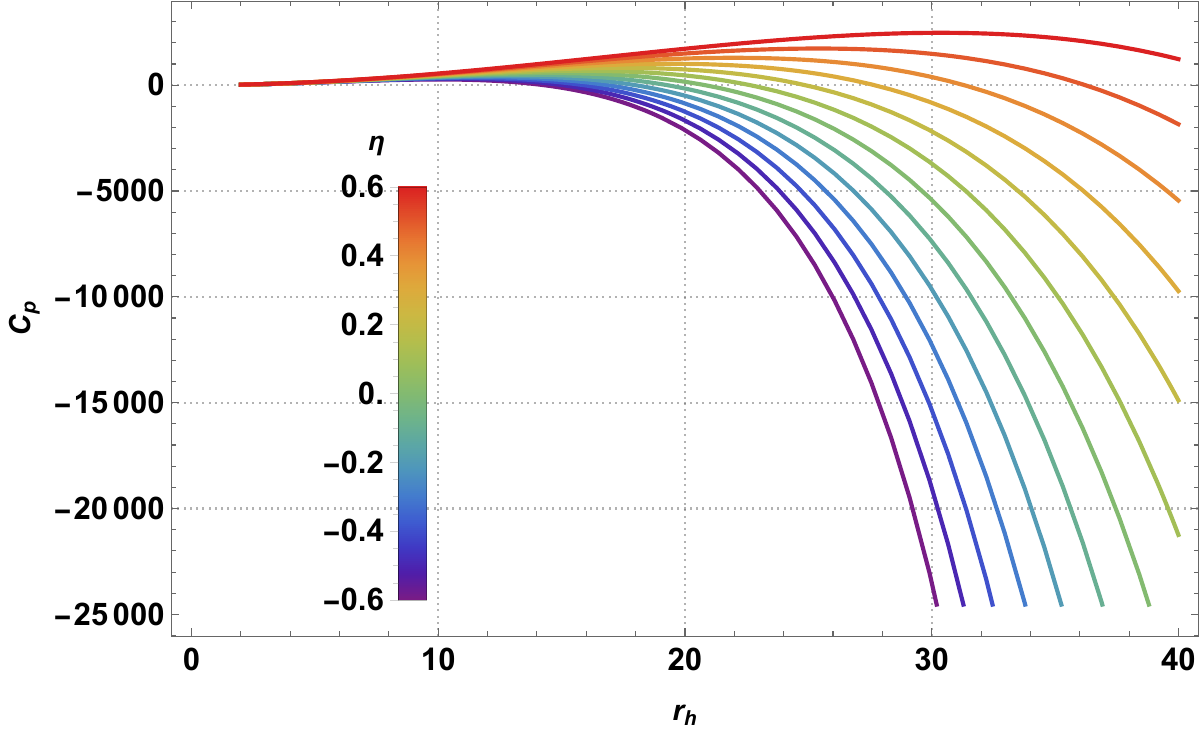}\qquad
    \includegraphics[width=0.4\linewidth]{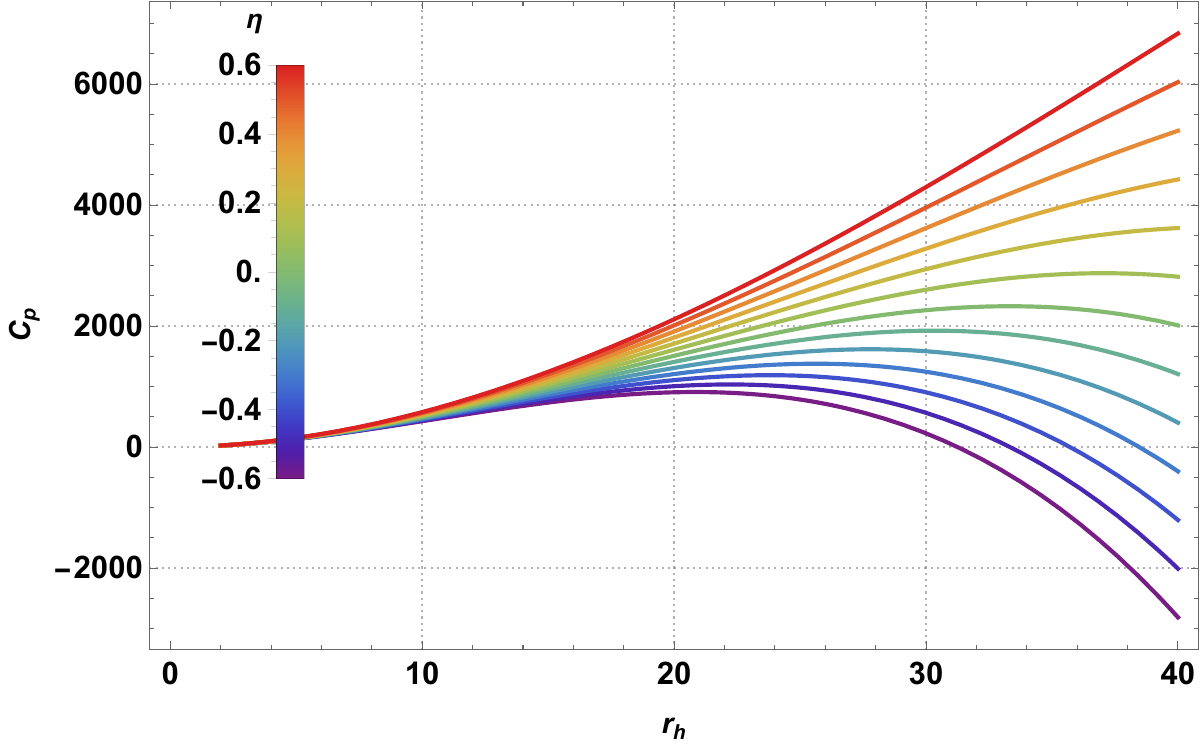}\\
    (a) $w=-3/4$ \hspace{6cm} (b) $w=-2/3$\\
    \includegraphics[width=0.4\linewidth]{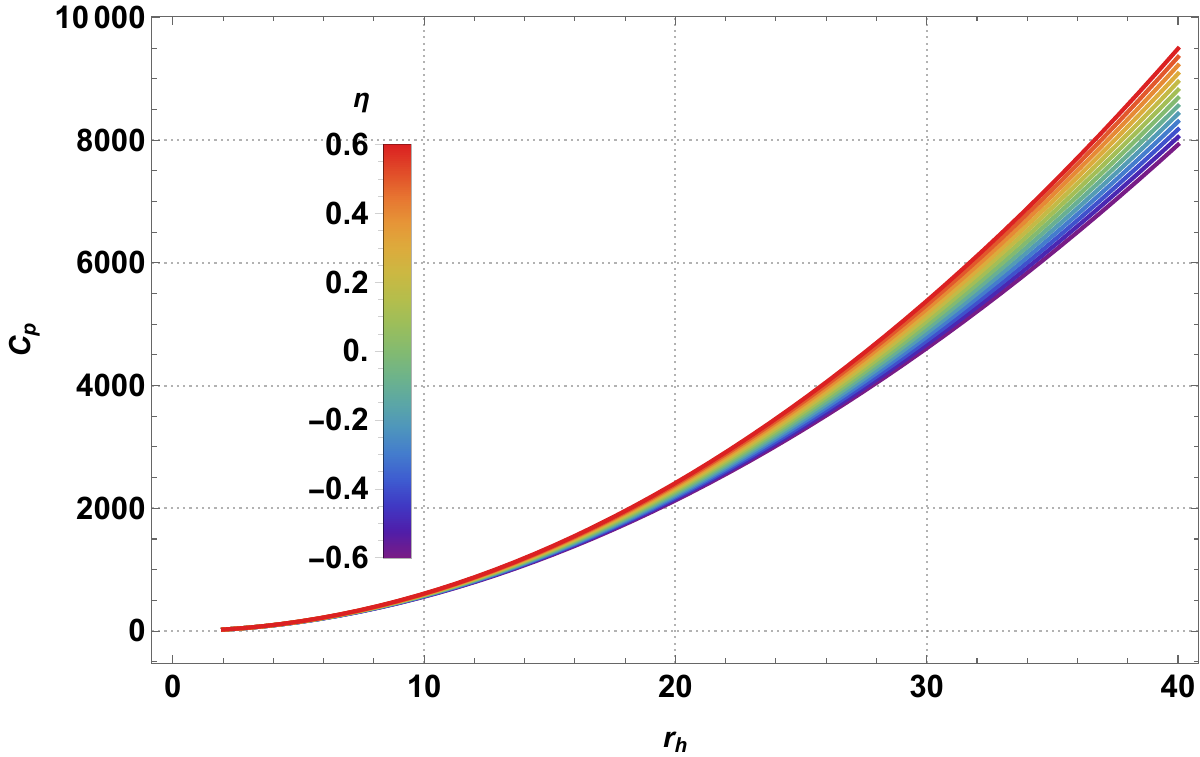}\qquad
     \includegraphics[width=0.4\linewidth]{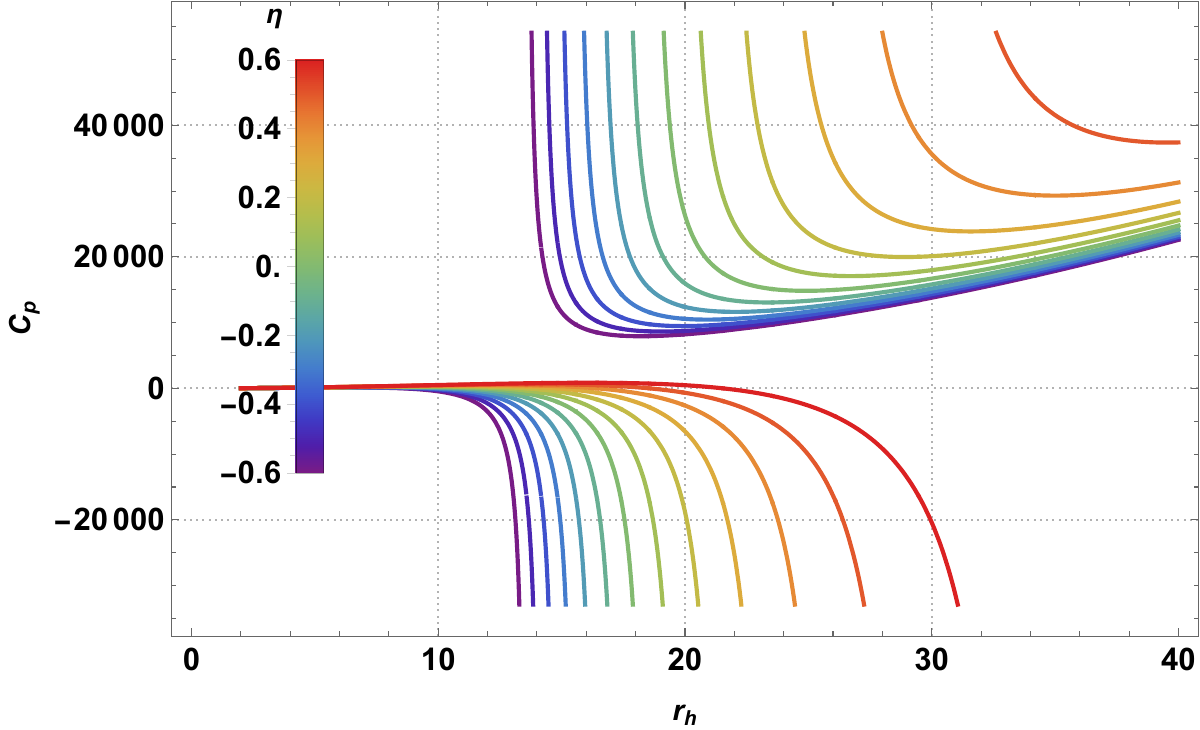}\\
    (c) $w=-1/2$ \hspace{6cm} (d) $w=-5/6$ 
    \caption{Specific heat capacity evolution across different quintessence state parameters, revealing complex critical phenomena and thermodynamic instability regions. Fixed parameter: $\mathrm{C}=0.01$. The divergent structures and sign inversions indicate critical transitions where thermal response becomes singular, characteristic of second-order phase transitions. The systematic parameter dependence demonstrates how quintessence fundamentally restructures stability landscapes in modified gravitational frameworks.}
    \label{fig:heat-capacity-analysis}
\end{figure}

Figure \ref{fig:heat-capacity-analysis} reveals remarkable thermodynamic complexity through divergent structures indicating critical transitions where thermal response becomes singular. The systematic variations across different $w$ values demonstrate how quintessence parameters fundamentally alter stability characteristics. Panel (b) shows particularly pronounced critical behavior in the phantom-like regime, while the other panels reveal progressive modifications that could provide observational signatures for distinguishing between different dark energy scenarios through precision thermal measurements.

Our thermodynamic investigation indicates that quintessence-coupled KRG BH systems display complex phase structures with varying stability zones and critical thermodynamic transitions. These results imply potential observational signatures in thermal emission spectra of astrophysical BH systems, providing essential foundations for testing KRG-QF frameworks through precision thermal measurements and constraining fundamental physics parameters through observations of BH thermal characteristics \cite{sec2is71,sec2is72}.


\section{Conclusions}

In this study, we presented a detailed analysis of BH solutions within the KRG framework coupled with QF environments, systematically exploring the rich phenomenology arising from the interplay between LV effects and exotic matter contributions. Our study encompassed multiple fundamental aspects of modified gravity, from spacetime geometry and geodesic dynamics to observational signatures and thermodynamic properties, establishing a robust theoretical foundation for testing KRG-QF scenarios against observational data.

We began our analysis in Section \ref{isec2} by constructing the KRG-QF BH spacetime through systematic solution of the modified Einstein field equations. The resulting metric function, given by Eq.~(\ref{bb2}) as $A(r)=\frac{1}{1-\eta}-\frac{2\,M}{r}-\frac{\mathrm{C}}{r^{3\,w+1}}$, explicitly demonstrated how LV parameter $\eta$ and QF characteristics $(\mathrm{C}, w)$ fundamentally alter spacetime geometry compared to standard Schwarzschild solutions. Our curvature analysis through Ricci and Kretschmann scalars, summarized in Table \ref{tableiz1}, revealed that while the central singularity persists in the KRG-QF framework, the approach to singular behavior is systematically modified by both theoretical parameters. Figure \ref{hor12} illustrated the emergence of dual horizon structures with event and cosmological horizons, whose locations depend sensitively on the parameter combinations, as shown in Figure \ref{hor123}.

The geodesic analysis presented in Section \ref{isec3} provided crucial insights into particle dynamics within KRG-QF spacetimes. Our investigation of null geodesics revealed systematic modifications to photon trajectories, with the effective potential given by Eq.~(\ref{c1}) demonstrating enhanced gravitational effects from LV and opposing influences from QF contributions. Figure \ref{fig:potential-1} clearly illustrated these complementary trends, while Figure \ref{fig:parametric} showed the complex orbital patterns emerging from varying LV strength. The photon sphere analysis yielded critical impact parameters modified by both theoretical frameworks, with expressions like Eq.~(\ref{c7}) for specific QF states. For massive particles, our ISCO analysis revealed dramatic modifications from GR predictions, with Table \ref{tab:ISCO} documenting ISCO radius increases exceeding order-of-magnitude changes for moderate parameter combinations. Figure \ref{figa92} demonstrated the systematic parameter dependence that could provide observational discrimination between theoretical frameworks \cite{isz01,isz02}.

Our BH shadow investigation in Section \ref{isec4} established direct connections between theoretical parameters and observable quantities. The analytical shadow radius expressions presented in Tables \ref{table:2} and \ref{table:3} provided comprehensive characterization across different QF states, while numerical results in Table \ref{table1a} documented systematic deviations from GR predictions. Figure \ref{figa7} revealed the complementary parameter dependencies that enable observational constraints on both LV and QF effects simultaneously. The shadow morphology analysis through celestial coordinates, visualized in Figure \ref{figa21}, demonstrated clear observational signatures that could be detected by current and future high-resolution imaging campaigns.

The perturbation analyses in Sections \ref{isec5} and \ref{isec6} provided fundamental insights into stability and wave propagation characteristics. Our scalar field perturbation study yielded effective potentials, exemplified by Eq.~(\ref{ff8}), that systematically encode both LV and QF modifications to wave dynamics. Figure \ref{fig:scalar-potential} demonstrated the parameter-dependent potential modifications, while the three-dimensional visualizations in Figure \ref{fig:3d-plot} and contour analyses in Figure \ref{fig:contour} revealed the complex parameter interdependencies governing wave scattering properties. Similarly, our EM perturbation analysis through Eq.~(\ref{em3}) showed analogous modifications to electromagnetic wave propagation, with Figure \ref{fig:em-potential} illustrating the systematic parameter trends that directly affect QNM spectra and electromagnetic signatures near BH horizons \cite{isz03,isz04}.

The gravitational lensing study in Section \ref{isec7} employed GBT methodology to derive comprehensive deflection angle expressions incorporating both LV and QF effects. Our analytical result demonstrated systematic enhancement factors through $(1-\eta)$ terms and power-law QF corrections with exponents determined by the quintessence state parameter $w$. Figure \ref{fig:deflection_analysis} provided detailed validation of theoretical predictions across the parameter space, revealing the characteristic $b^{-1}$ scaling modified by KRG-QF contributions. These results established precise theoretical foundations for using gravitational lensing observations to constrain modified gravity parameters and test alternative theoretical frameworks.

Our comprehensive thermodynamic analysis in Section \ref{isec8} revealed sophisticated thermal behavior transcending standard BH thermodynamics. The Hawking temperature expression in Eq.~(\ref{iztemp}) explicitly showed how both LV and QF effects systematically modify thermal emission characteristics. The detailed mass-radius relationship analysis in Figure \ref{fig:bh-mass} demonstrated how quintessence state parameters fundamentally alter gravitational mass structures across different dark energy regimes. Figure \ref{fig:temperature-profiles} revealed systematic temperature evolution patterns that provide crucial observational discriminators between different theoretical frameworks. The thermodynamic density field analysis in Figure \ref{figthermalanalysis} demonstrated how diverse KRG-QF parameter combinations generate distinct thermal landscape topologies across different quintessence regimes, from standard Schwarzschild baseline through phantom-like and intermediate dark energy scenarios to the cosmological constant limit, revealing parameter-dependent phase structures and remarkable thermodynamic behaviors that significantly deviate from conventional BH thermal properties in modified gravitational frameworks. Our Gibbs free energy investigation through Figure \ref{fig:gibbs-energy} demonstrated systematic modifications to thermodynamic equilibrium conditions across different quintessence regimes, while the specific heat capacity analysis in Figure \ref{fig:heat-capacity-analysis} revealed critical phenomena and divergent structures indicating thermodynamic instability regions that could manifest in observable thermal emission signatures.

Throughout our investigation, we consistently found that KRG-QF modifications introduce systematic deviations from GR predictions across all physical phenomena examined. The LV parameter $\eta$ generally enhances gravitational effects and modifies characteristic scales, while QF contributions provide opposing influences that depend sensitively on the quintessence state parameter $w$. This complementary behavior establishes rich parameter spaces where observable quantities can be systematically tuned, providing multiple independent channels for constraining theoretical parameters through precision measurements \cite{isz05,isz06}.

The thermodynamic analysis particularly revealed the profound impact of quintessence state parameters on BH thermal properties. Different values of $w$ produce markedly distinct mass-radius relationships, temperature evolution patterns, and heat capacity behaviors, suggesting that precision thermal measurements could serve as powerful discriminators between various dark energy scenarios. The emergence of critical phenomena and potential phase transitions in extreme parameter regimes indicates that KRG-QF BH systems may exhibit fundamentally new thermodynamic behaviors that transcend conventional BH thermodynamics.

Our results carry significant implications for observational astronomy and fundamental physics. The dramatic modifications to ISCO radii, shadow characteristics, thermal emission properties, and thermodynamic phase structures suggest that existing observational data from facilities like the EHT, LIGO/Virgo, and future space-based gravitational wave detectors could already contain signatures of KRG-QF effects. The systematic parameter dependencies we documented across mass-radius relationships, temperature profiles, and heat capacity evolution provide robust theoretical frameworks for interpreting observations and constraining alternative gravity theories through multi-messenger astronomy approaches.

The thermal signatures we identified offer particularly promising observational prospects. The distinct temperature evolution patterns across different quintessence regimes, combined with the systematic modifications to heat capacity and Gibbs free energy characteristics, could enable precise constraints on both LV and dark energy parameters through future high-precision thermal emission measurements from astrophysical BH systems.

Looking toward future plans, several promising research directions emerge from our work. First, the extension to rotating BH solutions within the KRG-QF framework would provide access to additional observational channels through spin-dependent effects on geodesics, shadow morphology, gravitational wave emission, and thermal properties. Second, the incorporation of magnetic fields and plasma environments relevant to astrophysical BH systems would enable more realistic modeling of electromagnetic signatures, accretion disk dynamics, and thermal emission spectra. Third, the development of time-dependent QF scenarios could illuminate dynamical evolution effects and temporal variations in observable signatures, particularly in thermal emission characteristics. Finally, the investigation of critical phenomena and phase transitions we identified in extreme parameter regimes warrants further theoretical exploration to understand their fundamental nature and potential observational manifestations in astrophysical contexts.


\section*{Acknowledgments}

F.A. acknowledges the Inter University Centre for Astronomy and Astrophysics (IUCAA), Pune, India for granting visiting associateship. \.{I}.~S. thanks T\"{U}B\.{I}TAK, ANKOS, and SCOAP3 for funding and acknowledges the networking support from COST Actions CA22113, CA21106, and CA23130.

\end{document}